\documentclass[a4paper,structabstract]{aa}  

\usepackage{graphicx}
\usepackage{txfonts}
\begin{document}
   \title{Soft gamma-ray sources detected by {\it INTEGRAL}
   }


   \author{D. Petry
          \inst{1,2,4},
          V. Beckmann\inst{2,3},
          H. Halloin\inst{3}, 
          A. Strong\inst{1}
          }

   \authorrunning{Petry, Beckmann, Halloin, Strong}
   \titlerunning{Soft gamma-ray sources detected by {\it INTEGRAL}}

   \institute{Max Planck Institute for extraterrestrial Physics (MPE), Giessenbachstr., 85748 Garching, Germany
         \and
         ISDC Data Centre for Astrophysics, Ch. d'Ecogia 16, 1290 Versoix,
         Switzerland
         \and
         APC, UMR 7164, Universit\'{e} Paris 7 Denis Diderot, 
         10 rue Alice Domon et L\'{e}onie Duquet, 75025 Paris Cedex 13, France
	 \and
	 now at European Southern Observatory, Karl-Schwarzschild-Str. 2, 85748 Garching, Germany\\
     \email{dpetry@eso.org}
             }

   \date{Received ; accepted }

 
  \abstract
   {}
   {We aim to exploit the available {\it INTEGRAL}/SPI data to provide time-averaged spectra of
    the brightest soft gamma-ray sources.
  }
   {Employing a maximum-likelihood fit technique for our SPI data analysis,
   we take as input to our source model the source catalog derived by 
   Bouchet et al. (2008) from a SPI all-sky study. 
   We use the first four years of public SPI data and
   extract spectra between 25~keV and 1~MeV
   for the 20 catalog sources detected by Bouchet et al. at 200 - 600~keV 
   with $\ge$~2.5\,$\sigma$.
   In order to verify our analysis, we also extract spectra for the 
   same sources from
   the corresponding {\it INTEGRAL}/ISGRI data. We fit adequate spectral models to the
   energy range 25-1000~keV for SPI and 25-600~keV for ISGRI. 
   We use our spectra from the Crab (which is among the 20 sources studied here) and an empty
   location in a crowded field to derive an estimation of the systematic errors. 
   }
   {The agreement between our SPI and ISGRI measurements is good if we normalise them
    on the Crab spectrum.  Our SPI flux measurements also agree well with those by Bouchet et al. (2008).
    All 20 sources in our sample  are detected independently in the bands 25-100~keV and 100-200~keV. 
    At 200-600~keV we detect eight sources, at 600-1000~keV we detect two sources.
    Our spectra agree well with the results from previous publications where available.
    For six of the 14 XRBs in our sample we find evidence for a hard powerlaw-component which becomes
    dominant above the cutoff energy of the thermal Comptonization component. 
    In two of these cases, our study provides the first indication of such emission. For the others, our results
    confirm previous studies. 
    Our spectrum of the Crab, integrated over 1.3~Ms, shows a significant
    flux in all points and is well described by a powerlaw with a break near 100~keV and spectral indices 2.11
    and 2.20.     
    }
   {}


   \maketitle
%

\section{Introduction}

   Soft gamma-rays have to be observed with satellite-based instruments, and so far
   there have been only three major missions taking data 
   in the range from 100 keV to a few MeV with adequate sensitivity: {\it GRANAT} with the SIGMA
   instrument (Revnivtsev et al. \cite{revnivtsev04}), {\it CGRO} with 
   OSSE (Johnson et al. \cite{johnson93}) and COMPTEL 
   (Sch\"{o}nfelder et al. \cite{schoenfelder00}), 
   and {\it INTEGRAL} with IBIS (Ubertini et al. \cite{ubertini03}), 
   and SPI (Vedrenne et al. \cite{vedrenne03}).

   On {\it INTEGRAL}, the only soft gamma-ray observatory operational today, both the 
   high-energy instruments, IBIS  and SPI, are capable of measuring point source spectra. They 
   complement each other in that IBIS  provides angular resolution and good
   sensitivity at energies below a few 100 keV while SPI has superior energy resolution 
   and high-energy sensitivity and permits the study of 
   extended diffuse emission with its large field of view.

   Sources of soft gamma radiation with photon energies between
   100 keV and several MeV are astrophysically interesting
   because they permit us to study matter at supra-thermal energies. Essentially all
   known point-like emitters of such radiation consist of compact objects such as 
   black holes or neutron stars which convert the gravitational energy of their surrounding matter
   or their own extreme rotational and magnetic energy into translational kinetic energy and then in turn partially 
   into high-energy photons. The black holes
   come either as stellar black holes in X-ray binaries or as
   supermassive black holes in the centres of galaxies. 
   All other sources of soft gamma-rays are spatially extended or unresolved and high in number-density.
   They form a continuum which seems, 
   in the case of our Galaxy, to extend over the bulge and a significant part of the disk
   (Bouchet et al. \cite{bouchet05}, Revnivtsev et al. \cite{revnivtsev06}, Krivonos et al. \cite{krivonos07}).   

   In the most recent all-sky study of {\it INTEGRAL}/SPI soft gamma-ray data, Bouchet et al. (\cite{bouchet08}) 
   have presented a detailed analysis
   of both diffuse and point source emission in the energy range from 25 keV to 600 keV 
   based on all SPI data from February 2003 up to May 2006. They conclude that the diffuse component
   becomes more prominent with increasing energy reaching 32~\% in the central radian above 100 keV.
   The diffuse emission is bright over up to $\pm$45$^\circ$ in galactic longitude and $\pm$10$^\circ$ in galactic
   latitude depending on energy.
   
   Bouchet et al. (\cite{bouchet08}), {\it B08} in the following, 
   also present separate catalogs of 
   point sources for the four energy ranges 25-50 keV, 50-100 keV, 100-200 keV, and 200-600 keV
   which they detect at least at a significance level $\ge$~2.5$\sigma$.
   As the sensitivity of SPI relative to the steeply dropping source spectra 
   decreases with increasing energy and the diffuse emission becomes
   more dominant, the number of detected point sources decreases with increasing energy as well.
   In the energy range 200-600 keV a total of 20 sources are detected.
  
   In this work, we take the B08 catalog for the energy range 200-600 keV and derive
   time-averaged {\it INTEGRAL}/SPI spectra for each of the 20 point sources contained in it. 
   The resulting catalog of spectra of the 20 brightest soft gamma-ray sources will be useful 
   to, among others, modellers of the hard X-ray and soft gamma-ray background caused by unresolved sources of the same source classes as in our catalog, 
   high-energy observers of these sources who have to average over longer time-scales in order to achieve
   detections with acceptable significance, and generally researchers interested in the energy balance of the galaxy
   (most of the 20 sources are galactic).

   We employ a newly developed maximum likelihood fit analysis technique based on 
   the {\it spimodfit} software by Halloin \& Strong (\cite{spimodfit}) 
   to all public {\it INTEGRAL} data available for these sources 
   (i.e. up to November 2006, the status when our dataset was frozen).
   In order to minimise inhomogeneities of our systematic uncertainties, the analysis method is not optimised
   for any individual source, in particular not for the Crab. 

   For comparison and verification of our method, we also derive the spectra for the same sources 
   from the corresponding {\it INTEGRAL}/IBIS/ISGRI data. 

   This paper is meant to serve two purposes: (a) be a supplement to B08 and 
   investigate the average spectra of the brightest known soft gamma-ray sources, and
   (b) demonstrate an analysis technique for {\it INTEGRAL}/SPI data based on {\it spimodfit}. 

   In Sect. \ref{secobs}, the SPI and ISGRI instruments are briefly introduced, and the observations
   and the data selection are described. Section \ref{secanalysis} describes the analysis technique. 
   Section \ref{secresults} presents and discusses the spectra individually while Sect. \ref{secconclusions} 
   summarises the results, discusses them jointly and concludes.


\section{Instruments and observations}
\label{secobs}

This study uses data from the SPI and the ISGRI instruments on {\it INTEGRAL}. The properties of these
instruments are discussed in detail elsewhere (see Vedrenne et al. (\cite{vedrenne03}) and 
Roques et al. (\cite{roques}) for SPI and Ubertini et al. (\cite{ubertini03}) for ISGRI). Here we just 
briefly summarise the main aspects.

The ISGRI instrument is part of the high-sensitivity coded-mask imager IBIS.
It has a fully coded field of view of 9$^\circ \times$9$^\circ$, an angular resolution of 12' FWHM, and
a spectral resolution of 9\% at 100 keV. Its well calibrated energy range is 14~keV to ca. 700~keV. 
The nominal 
continuum sensitivity ($\Delta E = E/2$, 3~$\sigma$ in 10$^5$~s) at 100 keV is 3 mCrab. 

The SPI instrument is the coded-mask imaging high-resolution spectrometer on {\it INTEGRAL}.
Its fully coded field of view has a radius of 8$^\circ$. With its 19 pixels (hexagonal, 
cooled Ge detectors) it achieves an angular resolution of 2.5$^\circ$ FWHM. The average
spectral resolution of the detectors deteriorates only marginally as a function of energy from 
$\Delta{}E \approx$ 1.8~keV below 200~keV to $\Delta{}E \approx$ 4~keV at 3~MeV. 
This means, the relative energy resolution
improves from 0.9~\% at 200~keV to 0.13~\% above 3~MeV.
The calibrated energy range is 25 keV - 8000 keV. 
The nominal continu\-um sensitivity (as defined above for IBIS) at 100 keV is 16 mCrab. 
But above 1 MeV, the continuum sensitivity for point sources becomes inadequate. Only the Crab
is detected at 1-8 MeV, and an electronic noise problem between 1 and 2 MeV results in large
systematic errors in this interesting energy range.We therefore limit this study to the range
25-1000~keV.

Both energy resolution and sensitivity are time dependent: Radiation damage degenerates
the detectors leading to a deterioration of the energy resolution by
a few tenths of a keV (negligible for continuum analysis) over timescales of months. 
This degradation is cured by regular annealings (controlled heatings) of the detectors.
An annealing period lasts typically 12 days during which no science data can be taken. 

While the spectral resolution is hence kept constant for our purposes, the sensitivity of SPI 
remains time dependent 
because the continued exposure
to cosmic radiation leads to increasing background from the activation of all parts of the {\it INTEGRAL} satellite. 
In addition, the
intensity of the cosmic radiation itself is strongly time dependent because of solar
wind variablity (solar cycle) and solar flares. Over the course of the {\it INTEGRAL} mission
so far, this has lead to a nearly steady increase in instrumental background. In November 2006,
the instrumental background had increased by about 80\%  compared to the time shortly after launch
(November 2002). It is expected that this trend will not continue (and possibly be reversed) beyond the end
of the present solar minimum which has been a very prolonged one.

\subsection{Data selection}
\label{sec-selection}

The {\it INTEGRAL} data, as they are delivered to scientists,
are subdivided into time periods of constant pointing, so-called science windows
which typically have a duration of about 30 minutes.
A second relevant time unit is the ``revolution'' or orbit which is important
to the science analysis because the highly excentric orbit of the {\it INTEGRAL} spacecraft
takes the instruments through the Earth's radiation belts every three days (``perigee passage'') 
making it necessary to temporarily switch off many of the {\it INTEGRAL} instruments including ISGRI.
Observation scheduling and public data release is then organised per revolution.

This study is based on all available high-quality public data (at the time the dataset was frozen by us). 
These are the data from revolutions  21 to 500, i.e. from 15 December 2002
to 18 November 2006. From this time period, we select good data for each of the objects 
for which we want to derive spectra. The selection criteria are the following:
\begin{enumerate}
\item The angular distance between the pointing direction and the position of the object
    of interest is required to be less than 10$^\circ$. This is a compromise between maximum
    exposure time and maximum data quality. For SPI the maximum off-axis angle could be further increased,
    but for ISGRI, accepting data beyond 10$^\circ$ radius (which is well
    into ISGRI's partially coded field of view) reduces the accuracy of the spectra
    significantly. Since {\it INTEGRAL} does mainly pointed observations
    based on individual proposals (which have emphasised galactic objects so far), 
    the sky exposure achieved by the {\it INTEGRAL} instruments up to now is far from uniform.
    At low galactic latitudes our dataset therefore contains up to a few thousand 
    science windows while only a few hundred are available for the typical object at higher latitudes.

    In those cases where a very large exposure is available, we can reduce the selection radius to
    8$^\circ$ in order to improve data quality. Furthermore, if several objects of interest form a
    closely spaced group, we use a single dataset centred roughly on the centre of the group.

\item Time periods where the data quality was known to be bad such as times near the radiation belt
    entry or exit of the spacecraft, times shortly after
    annealings, times during which solar flares strongly raised the background, or times during
    which hardware problems occurred, were excluded based on the publicly available database
    of such events included in the OSA software (version 7.0) provided by the {\it INTEGRAL}
    Science Data Centre (ISDC, Courvoisier et al. \cite{thierry03}). This results in the removal
    of about 31~\% of the archival data (based on elapsed time and averaged over the entire mission 
    up to revolution 500).

\item In order to exclude time periods with problems which might have been missed in
     the compilation of the database used in the previous criterion,  quality cuts were
     imposed on a number of ``housekeeping'' parameters using the SPI housekeeping database
     compiled at MPE\footnote{\tiny http://www.mpe.mpg.de/gamma/instruments/integral/spi/www/public\_data}.
     In particular, the cuts limited the values of the
     SPI detector temperature, the veto rate of the SPI Anti-Coincidence Shield (ACS), the rate
     of SPI single interaction events, and the orbital phase. The cuts were derived from 
     inspecting the long term behaviour of these parameters and assuming smoothness of their
     time dependence. The orbital phase was conservatively limited to the range between 0.1 and 0.87
     in order to safely exclude any radiation belt influence. These housekeeping parameter cuts
     removed another 9.5~\% of the archival data (percentage defined as above), mostly at the 
     beginning of the mission.
\end{enumerate}

For detailed analysis of the 20 objects from
the B08 catalog we finally utilise the datasets as presented in Table~\ref{tab-sourcedata}.

\begin{table*}
  \caption[]{The 20 objects investigated in this study and the properties of the corresponding datasets.} 
  \label{tab-sourcedata}
  \begin{tabular}{|l|l|l|l|c|c|c|c|}
    \hline
 Object Name$^a$& Object Type$^b$ & RA, DEC (FK5)&galactic l, b& elapsed & \multicolumn{1}{c|}{effective}& var.$^d$ &  Data\\
         &            &                       &              & time & \multicolumn{1}{c|}{SPI exposure$^c$} &       &   Set\\ 
         &           & \multicolumn{1}{c|}{$[^\circ]$}    & \multicolumn{1}{c|}{$[^\circ]$}    & $[$Ms$]$ &  \multicolumn{1}{c|}{$[$10$^6$cm$^2$s$]$} & $[$ks$]$ & \\
\hline                                                                                                                     
Crab              & PWN       & 083.6332, +22.0145 & 184.5575, -05.7843 & 1.33 & 22.7 &  -                         &   1 \\
Vela Pulsar       & PWN       & 128.8361, -45.1764 & 263.5520, -02.7873 & 3.48 & 68.9 &  -                         &   2 \\
NGC 4151          & Seyfert 1.5 & 182.6364, +39.4054 & 155.0765, +75.0637 & 0.085 & 5.2 &  -                         &  3 \\
NGC 4945          & Seyfert 2 & 196.3587,  -49.4708  & 305.2681, +13.3374 & 0.53 &  9.1  & -                       &  4 \\
Cen A             & Radio Gal., Seyf.2 & 201.3651, -43.01911 & 309.5159, +19.4173 & 0.56 & 11.5  & -               &  5 \\
XTE J1550-564     & LMXB, BHC, $\mu$Q&237.7446, -56.4767& 325.8822, -01.8271 & 2.56 & 42.4 & -                        &  6 \\
4U 1630-47        & LMXB, BHC  & 248.5017, -47.3942   & 336.9081, +00.2519 & 3.97 & 154  & -                        &  7 \\
Swift J1656.3-3302& Blazar    & 254.0690, -33.0359   & 350.5993 +06.3581  & 10.1 & 151  & -                        &  10\\
OAO 1657-415      & HMXB, Pulsar & 255.1996, -41.6731   & 344.3538, +00.3111 & 3.97 & 110 &  -                        &  7 \\
GX 339-4          & LMXB, BHC, $\mu$Q & 255.7062, -48.7897   & 338.9394, -04.3267 & 3.36 & 123 &  -                        &  8 \\
4U 1700-377       & HMXB, NS  & 255.9866, -37.8441   & 347.7544, +02.1735 & 3.97 & 73.5&  90, -                    &  7 \\
IGR J17091-3624   & LMXB?, BHC   & 257.275,  -36.410    & 349.519,  +02.215  & 7.08 & 214 &  -                        &  9 \\
GX 354-0          & LMXB, NS  & 262.9892, -33.8347   & 354.3022, -00.1501  & 7.08 & 130 &  -                       &  9 \\
1E 1740.7-2942    & LMXB?, BHC, $\mu$Q& 265.9785, -29.7452   & 359.1160, -00.1057 & 7.08 & 134 &  -                        &  9 \\
IGR J17464-3213   & LMXB?, BHC, $\mu$Q  & 266.5650, -32.2335   & 357.2552, -01.8330  & 7.08 & 114 &  90, -                   &  9 \\
GRS 1758-258      & LMXB, BHC, $\mu$Q& 270.3012, -25.7433 & 004.5076, -01.3607 & 10.1 & 403 &  -                        &  10\\
Ginga 1826-24     & LMXB, NS    & 277.3675, -23.7969   & 009.2724, -06.0878 & 10.1 & 239 &  -                        &  10\\
GRS 1915+105      & LMXB, BHC, $\mu$Q& 288.7983, +10.9456 & 045.3656, -00.2194 & 2.83 & 53.1&  90, -                    &  11\\
Cyg X-1           & HMXB, BHC, $\mu$Q  & 299.5903, +35.2016   & 071.3350, +03.0668  & 1.94 & 32.9 &  90, 180                 &  12\\   
Cyg X-3           & HMXB, BHC, $\mu$Q  & 308.1074, +40.9578   & 079.8455, +00.7001  & 2.79 & 50.2 &  90, -                   &  13\\
Control1$^e$      & -              & 259.5000, -40.3500   & 347.7000, -02.1000  & 7.08 & 181 & - & 9 \\ 
    \hline
  \end{tabular}
  \begin{list}{}{}
    \item[$^a$] See Sect. \ref{sec-selection} for the selection criteria and Sect. \ref{secresults} for
     references concerning the source properties.
    \item[$^b$] PWN = pulsar wind nebula, $\mu$Q = microquasar, NS = binary containing a neutron star,
            BHC = containing a black hole candidate, ? = classification uncertain
    \item[$^c$] Calculated at 30 keV and taking into account the SPI response
    \item[$^d$] Assumed variability timescale in the analysis at $E < 144$~keV (first number) and at $E > 144$~keV
         (second number); ``-'' = the source was assumed constant either because it is known to be steady or because
           the variability timescale and/or amplitude is too small for SPI to detect significant variability.
    \item[$^e$] Control position for the assessment of systematic errors in the analysis of crowded fields
                (see Sects. \ref{sec-syserr} and \ref{sec-control1}).
  \end{list}
\end{table*}

\section{Data analysis}
\label{secanalysis}

\subsection{SPI data analysis using {\it spimodfit}}
The spectral data analysis was carried out using the software packages spiselectscw v3.9 and spimodfit v3.0
(both written by Halloin \& Strong \cite{spimodfit}) with additional infrastructure software written by DP. 
The spimodfit package is part of the standard {\it INTEGRAL} data analysis software ({\it OSA}) since version 7.0 .

Using spiselectscw, the data selection criteria described in Sect. \ref{sec-selection} 
were applied to the calibrated and 
pre-binned data from the
SPI database at the Max Planck Institute for extraterrestrial Physics. 
Using spimodfit, a source region model was then fitted to the selected data.
The parameters of this model are in our case dynamically constructed by spimodfit from two main components:
\begin{enumerate}
\item an isotropic background component.
\item the contributions from those point sources which are within 20$^\circ$ of the optical axis of the
     instrument for a given science window.
\end{enumerate}
The isotropic component was modelled as the {\it product of a common scaling factor and a relative sensitivity
for each of the 19 detectors of SPI} (taking into account the two detector failures during the course of the
mission so far). The relative detector sensitivity was permitted to vary on a timescale of 10 revolutions 
(nearly 30 days) as a compromise between assuming no variability at all (resulting in a degraded fit)
and the minimum reasonable variability
timescale of one revolution (approx. 3 days, resulting in larger uncertainties due to the increased number
of fit parameters). It was verified that assuming a shorter timescale than 30 days did 
only change the results within their statistical errors. 

The common background scaling factor was permitted to vary on a short timescale of 2~h (corresponding to 
typically four science windows). This was the conservative choice determined by the actually observed
changes in the overall event rate (which is strongly dominated by background) and the aim to minimise
the number of model parameters in order to improve sensitivity.

The model components stemming from the point sources are redetermined for every pointing: 
The algorithm first determines which sources from the input catalog are within 20$^\circ$ of the  
pointing axis. Then the tabulated detector response (taking into account the absolute time, spacecraft 
orientation, the coded mask pattern, and the general spacecraft and instrument mass model) is used to 
calculate the relative contribution of each source 
to the counts in each detector. The model fit parameter is the common scaling factor of the contributions
in each detector, i.e. the source flux in the time window corresponding to the pointing.
Depending on the variability timescale assumed for the individual source, this flux is then
assumed constant for a certain number of pointings. 

When the model is constructed, all parameters are tested for relevance, and parameters which
do not show an effect on the fit function are excluded.
Using a maximum likelihood fit which is initialised from a $\chi^2$-fit, the model is then
fit to the data and the statistical errors are determined from the covariance matrix.
For more details see Halloin \& Strong (\cite{spimodfit}). 

The entire analysis is carried out independently in each energy bin such that boundary
conditions such as variability timescales and the input point source catalog can in principle
be chosen differently for each energy bin.

In the largest of our datasets (the one containing GRS\,1758+258, see Table \ref{tab-sourcedata}), 
the number of degrees of freedom is 74292 per energy bin, while the number of fit parameters is 2582.
In the smallest dataset (the one containing NGC~4151) there are only 665 degrees of freedom and 38 fit parameters
per energy bin. Thus,
for an entire SPI spectrum with 17 energy bins (see also the next section), 
the total number of degrees of freedom in our datasets is between $\approx$~11000
and $\approx$~1.3 million, the number of fit parameters between 600 and 44000.

The number of fit parameters is dominated by the background scaling factors and relative detector sensitivities.
The source flux for an individual source from the input catalog constitutes only one fit parameter for each energy bin.
 
Unless otherwise noted, the reduced $\chi^2$ values of the model fit in all energy bins for all results presented here
were very close to 1.0 .   

\subsubsection{Energy binning}

Our default energy binning was chosen a priori based on considerations of SPI energy resolution 
and flux sensitivity. It consists of 16 energy bins of logarithmically increasing width between 25 and 1000 keV.
In addition, a narrow energy bin around 511 keV was inserted in order to check for possible annihilation
radiation features. 

As the analysis showed, many of our sources of interest are too weak to be detected significantly above 
200 keV in individual bins of  even this coarse (compared to the SPI energy resolution)
energy binning. So a second wider binning consisting of 
11 bins was introduced for weaker sources. The narrow and the wide binning have the first three bins (up to 48 keV)
in common, then the wider binning roughly combines every two narrow bins up to 500 keV and after that
every 3 narrow bins. The bin boundaries of the narrow binning are
25, 31, 39, 48, 60, 75, 93, 116, 144, 179, 223, 278, 346, 502, 520, 668, 832, and 1000~keV. 
The bin boundaries of the wider binning are
25, 31, 39, 48, 70, 103, 150, 219, 320, 502, 520, and 1000~keV.

\subsubsection{Input catalog construction}
\label{sec-inputcatalog}

\begin{figure*}
  \centering
  \includegraphics[width=18.5cm]{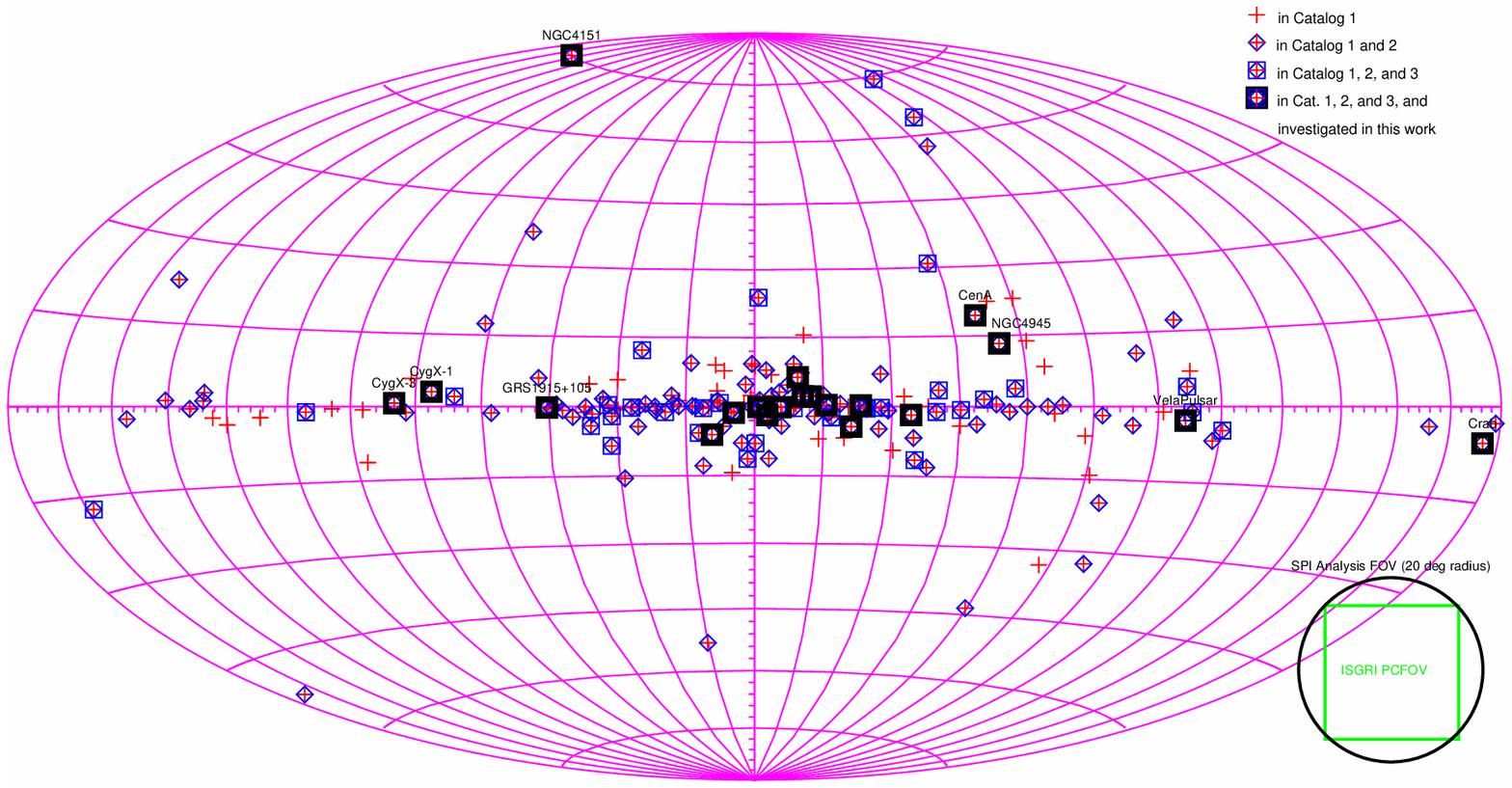}
  \includegraphics[width=18.5cm]{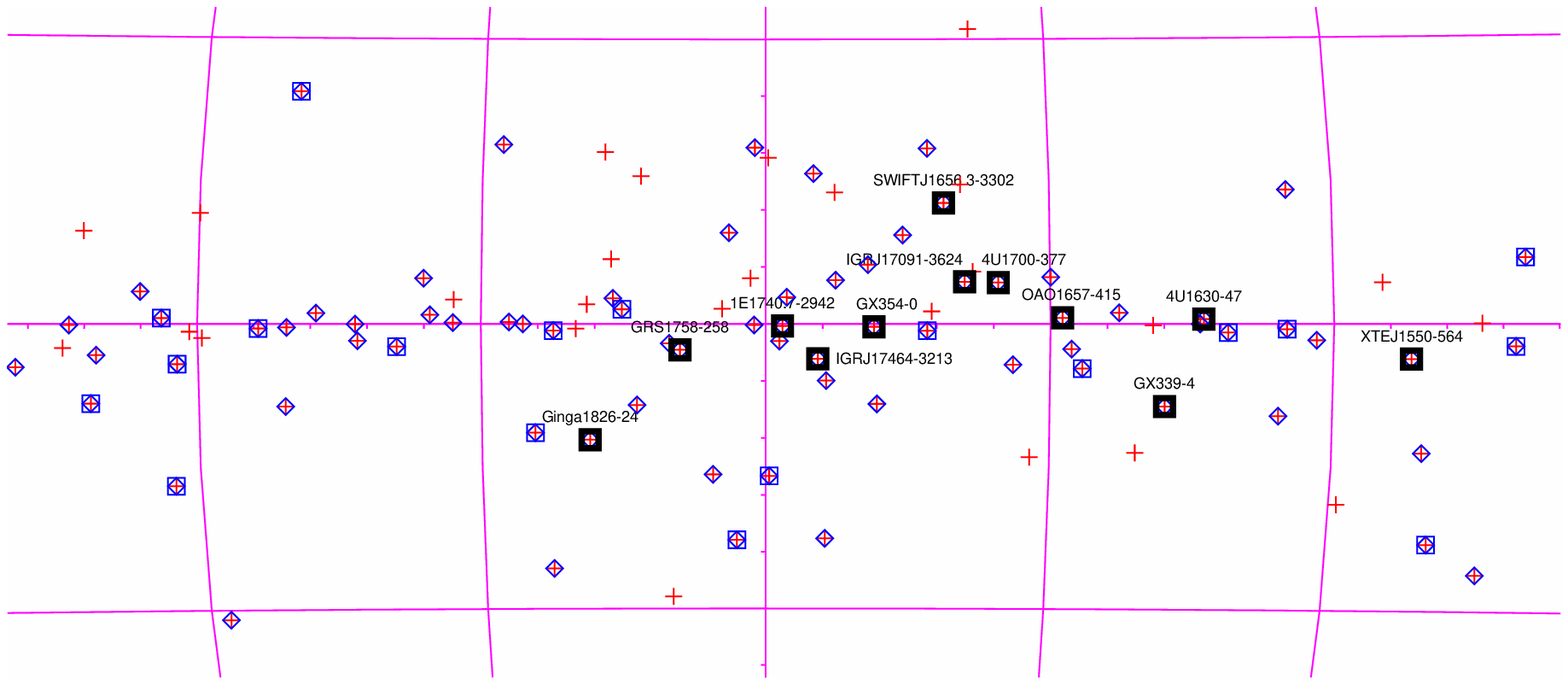}
  \caption{{\bf (a)} (top) The positions of the sources from the input catalogs used in this study for the analysis of the
    SPI data in three separate energy ranges:
    catalog 1 (25-144~keV, crosses, 173 objects, i.e. the entire B08 catalog), 
    catalog 2 (144-346~keV, diamonds, 129 objects), 
    and catalog 3 (346-1000 keV, squares, 52 objects) on an all-sky map in galactic coordinates.
    The 20 objects investigated in this work are marked as thick squares. The grid mesh size
    15$^\circ\times$15$^\circ$. Also shown are the extensions of the SPI analysis field of view and the ISGRI 
    partially coded field of view.
    {\bf (b)} (bottom) The enlarged galactic centre region of (a).
    The catalogs for higher energies are derived by successively removing weak sources from the initial catalog,
    i.e. a lower energy catalog contains all higher-energy ones.
    See Sect. \ref{sec-inputcatalog}.   
}
  \label{fig-catalogs}
\end{figure*}

As mentioned above, we base our input catalog on the one compiled by B08.
This catalog comprises in total 173 sources and states fluxes for them in four energy bands
(25-50 keV, 50-100 keV, 100-200 keV, and 200-600 keV).
At low energies, all sources of the Bouchet catalog have to be included in the input catalog
in order to have a complete model. But as the energy increases, many sources in the catalog 
drop below our flux sensitivity and need no longer be included.
In order to optimise the sensitivity of our analysis, we therefore construct three separate 
catalogs for three energy ranges: (a) 25-144 keV, (b) 144-346 keV, and (c) 346-8000 keV.
For range (a), we use the entire Bouchet catalog. For range (b) we use only those sources 
from the Bouchet catalog which
are detected at 100-200 keV (i.e. a flux is stated in B08, not an upper limit)
and which have at least 3 mCrab at 50-100~keV.
Finally, for range (c) we include only those objects from the Bouchet catalog which
have a flux $>$ 6~mCrab at 100-200 keV or which are marked as variable and 
detected at 50-100 keV. Figure \ref{fig-catalogs} shows the positions of the sources
in these three input catalogs in galactic coordinates.
 
The fraction of the sources from the input catalogs which are actually included in the model
constructed by {\it spimodfit} for an individual dataset, depends on how crowded the field is
and how widely spread the pointings are within the search radius.
In our datasets (see Table \ref{tab-sourcedata}) the number of sources included in the model
ranges from 1 (in energy range (c) for the least crowded fields) to 83
(in energy range (a) of the most extended among our datasets, the one containing GRS~1758-258).

\subsubsection{Time variability}
\label{sec-timevar}

The source variability is an important parameter in the flux extraction process. 
Bright and strongly variable sources can introduce some spurious features in the analysis if
their model variability timescale is chosen too long and
their true variability timescale and amplitude is large enough for SPI to detect a flux change.

We treat those sources as time-variable which are marked as such in the B08 catalog.
These are the following 12 sources (in brackets we give the variability timescale chosen by us 
based on the ISGRI light curves (Courvoisier et al. \cite{isgri-lightcurves}) and the 
sensitivity of SPI):
1A~0535+262 (180~ks), Vela~X-1 (90~ks), GX~301-2 (90~ks), 4U~1700-377 (90~ks), 
IGR~J17464-3213 (90~ks), Sco~X-1 (90~ks), Swift~J1753.5-0127 (2.6 Ms), Aql~X-1 (180~ks),
GRS~1915+105 (90~ks), Cyg~X-1 (90~ks, 180~ks for $E > 144$~keV), EXO 0331 (2.6 Ms),
Cyg~X-3 (90~ks). Cyg~X-1 is the only source which is significantly variable over the
entire energy range. All other variable sources are only treated as signifcantly
variable (given the SPI sensitivity) up to 144 keV. 

The minimum timescales used here were determined from a study of model fit convergence.
We found that our model fits do not converge well if we assume timescales much below 90~ks resulting
in increased statistical errors on all fit parameters.
Particularly in the case of the bright and strongly variable source 4U~1700-377, 
B08 used a variability timescale of $\approx$3~ks 
(the minimum possible due to the way the data is stored). For such sources, dedicated studies
are necessary to further increase the accuracy of spectra for short integration times. 
However, it is our aim here to provide long-term average spectra with a coherent treatment
of all sources. 

For the variable sources, the analysis yields a separate spectrum for each variability time bin,
or in other words a separate light curve for each energy bin.
We derive the average spectrum by applying a  weighted least-squares procedure described
in Alvarez-Gaum\'{e} et al. (\cite{revofpartphys}), page 14, to the light curves from each energy bin.  

\subsubsection{Postprocessing and spectral model fitting}

The analysis is sped up by processing each energy bin on a separate processor reducing the
required computing time by about a factor equal to the number of energy bins, i.e. from several days 
to a few hours for each source data set.

In a final step, the data from each processor are collected, where necessary the time-averaging
of the spectra is performed, and the spectra are converted to the PHA format suitable for
the spectral analysis with XSpec 12.3.1 (Arnaud et al. \cite{xspec})  which is then
used for the fitting of spectral models taking into account 
the SPI response and energy-redistribution.

\subsection{ISGRI data analysis}

The analysis of the {\it INTEGRAL} IBIS/ISGRI data is based on a
cross-correlation procedure between the recorded image on the detector
plane and a decoding array derived from the mask pattern (see Goldwurm
et al. \cite{goldwurm}). Standard spectral extraction has
been applied as provided by the OSA 7.0 software package, using the
same energy binning and science windows as for the SPI data.

\subsection{Statistical and systematic errors}
\label{sec-syserr}

The statistical errors of the photon count rates in each energy bin
are determined both for the ISGRI and the SPI data by the fit software
taking into account proper Poisson count statistics both for background and source signal.

In addition there is a systematic error on the flux calculation stemming from the uncertainty of the calibration
due to the limited amount of calibration data and the propagation of other uncertainties of the calibration process.
We derive the relative error of the calibration of the spectra by assuming a shape for the Crab Nebula
spectrum and independence of energy, i.e. that the uncertainty caused by the error on the calibration
is the same percentage of the measured count rate for each energy bin.
This constant percentage is determined by varying it until the reduced $\chi^2$ from the fit 
of the assumed spectral shape to the Crab spectrum is close to 1.0 for the complete energy range.

For both the ISGRI and the SPI data, we follow the {\it INTEGRAL} Cross-calibration status document by Jourdain et al.
(\cite{jourdain08}) and assume the Crab Nebula spectrum between 25 keV and 1 MeV to be a broken 
powerlaw with break energy fixed at 100 keV (which is close to the break energy we actually measure, 
see Sect. \ref{sec-Crab}). 
With this assumption we derive a calibration error
of our ISGRI and SPI spectra of 0.5\,\% in each energy bin.
Unless otherwise noted, this error has been added in all spectral plots shown in this 
paper. For the fainter sources, it is, however, negligible.

The Crab is a bright source in a relatively quiet field without many other neighbouring hard X-ray sources.
In order to assess the possible systematic errors arising in the analysis of crowded fields in the galactic bulge where
some of our 20 sources of interest are located,
we have added a control position to our input catalog at RA = 259.5$^\circ$, Dec = -40.35$^\circ$ (l = 347.7$^\circ$,
b = -2.1$^\circ$). This position  was randomly chosen in an  empty region (with respect to the B08 catalog)
of the crowded field near GX 354-0. The analysis for this artificial reference source is discussed 
in Sect. \ref{sec-control1}. 

\section{Results}
\label{secresults}

In this section we present our spectral analysis results and discuss the agreement of our 
SPI and ISGRI spectra with each other and with previously published measurements for each object in detail. 
The numerical results are later summarised in tables~\ref{tab-summary1a}, \ref{tab-summary1b} and \ref{tab-summary2} in 
Sect.~\ref{secconclusions}
where we also discuss the agreement with the fluxes given by B08.

Studying data below 25~keV would exceed the scope of this paper. Also,
due to the small field-of-view of the {\it INTEGRAL} soft X-ray instrument
Jem-X, a soft-X-ray dataset truly concurrent with our SPI and ISGRI dataset would have meant shrinking the
SPI and ISGRI exposure by more than a factor two. When we compare to previously published measurements, this lack
of soft X-ray data leads often to poorly constrained low energy features such as cutoffs near our lower energy limit 
(25~keV) or soft photon temperatures in thermal Comptonisation models. Where necessary, we try to overcome this 
by taking the appropriate values from the literature. 

In the following, unless otherwise noted, the normalisation parameters ``Norm'' or ``Norm$_1$'' in the spectral fit results
are given in units of photons~keV$^{-1}$cm$^{-2}$s$^{-1}$. They correspond to the flux at 1~keV. The letter $\Gamma$
designates a powerlaw spectral index. If we need to introduce a second power-law component for high-energy emission,
we use the XSpec model PEGPWRLW with peg energy 200~keV. The normalization of it (``Norm$_2$'') is given as the 
differential flux at 200~keV. 

In the figures in this section we show the spectra unfolded by XSpec with statistical errors (slightly increased
by the calibration errors discussed in Sect.~\ref{sec-syserr}). The solid line always indicates the fitted model 
(as a step function according 
to the energy binning). If the model has several components,
these are also separately shown with dashed lines.

\subsection{Crab Nebula}
\label{sec-Crab}

\begin{figure}
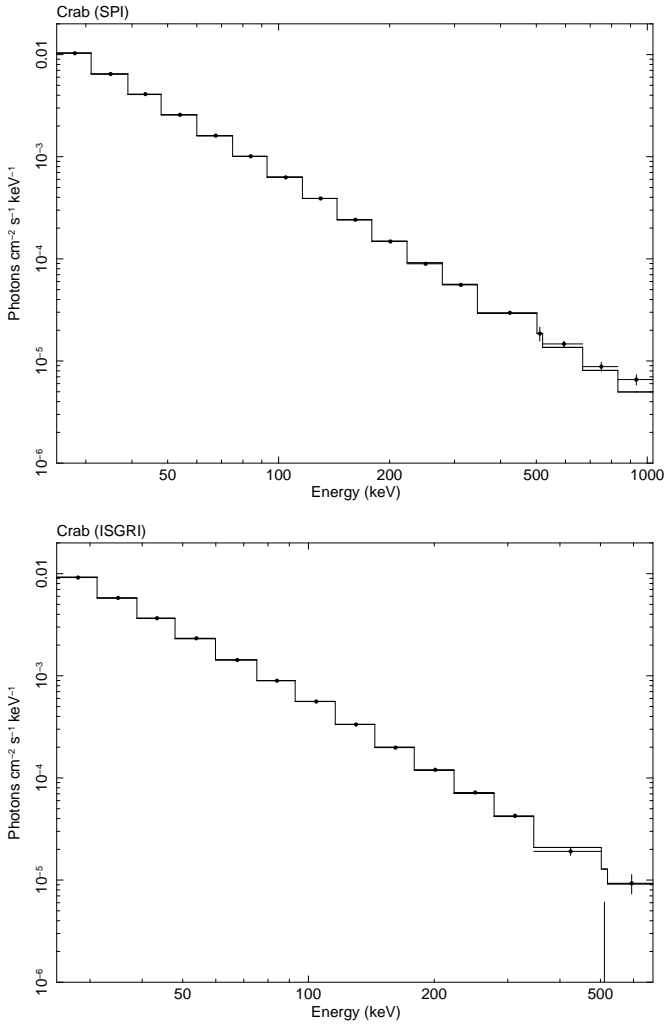

  \centering
  \includegraphics[angle=-90,width=9cm]{1284402a.eps}
  \includegraphics[angle=-90,width=9cm]{1284402b.eps}
  \caption{The unfolded spectra of the Crab Nebula as derived from  dataset 1 (see table \ref{tab-sourcedata}) with statistical errors: 
   {\bf(a)} (top)  SPI spectrum {\bf(b)} (bottom) ISGRI spectrum; fit function is in both cases a broken powerlaw. Parameters of both 
   fits are given in Sect. \ref{sec-Crab}.
}
  \label{fig-crab}
\end{figure}
 
The Crab Nebula is the standard candle of high-energy astronomy. Because of the steady, strong emission
from this pulsar wind nebula, it has always been among the first
objects to be observed by new instruments. For a recent compilation of X-ray data on the Crab see, e.g.,
Kirsch et al. (\cite{kirsch05}). Still, its high energy spectrum is not too well known. Many instruments use
an assumption about the Crab's spectral shape as the basis of their calibration and can hence not
say much about the fine details of the true spectrum. This is also the case for {\it INTEGRAL}/ISGRI.
On the other hand, 
{\it INTEGRAL}/SPI with its high energy resolution could   
measure a Crab spectrum which is less dependent on such assumptions about the spectral shape.

Figure \ref{fig-crab}a shows the result of our analysis of the SPI Crab spectrum.
A good fit is a broken powerlaw (XSpec model ``bknpow'') with the following parameters:

{\setlength{\tabcolsep}{1mm}
\begin{tabular}{rcrcl}
$\Gamma_1$ & = & 2.11 &$\pm$&  0.01\\
$E_{\mathrm{break}}$ & = & (81 &$\pm$&11)~keV\\
$\Gamma_2$ & = & 2.20 &$\pm$&  0.01\\
Norm      & = & 11.6 & $\pm$ &  0.5 \\
\end{tabular}
}

with a reduced $\chi^2$ of 0.8 for 13 degrees of freedom (see also Sect. \ref{sec-syserr}).

The corresponding ISGRI dataset gives the following Crab spectrum
(Fig. \ref{fig-crab}b):
$\Gamma_1$ = 2.119 $\pm$ 0.006,
$E_{\mathrm{break}}$ = (96$\pm$ 4)~keV,
$\Gamma_2$ = 2.36 $\pm$ 0.02 , Norm = 10.7 $\pm$ 0.3, with
a reduced $\chi^2$ of 1.3 for 11 d.o.f. (see  Sect. \ref{sec-syserr}).  

This needs to be compared to our best fit for the SPI data. The shape is in very
good agreement below 100 keV. 
Table \ref{tab-crab} (see Sect. \ref{secconclusions}) 
gives our measurements of the Crab flux in several energy bands.
The model fluxes between 25 and 100 keV ( 0.2287$\pm$0.0004~cm$^{-2}$s$^{-1}$ for SPI and
0.2046$\pm$0.0004~cm$^{-2}$s$^{-1}$ for ISGRI )
differ by a factor 1.11$\pm$0.01 (SPI/ISGRI)
which has to be regarded as a measurement of the systematic difference in absolute flux normalisation. 

Above the fixed break, the ISGRI spectrum is significantly softer
than the SPI spectrum.
However, as the ISGRI response was {\it made} (by the authors of the OSA 7 ISGRI response) 
to give the Crab spectrum a powerlaw
above 100 keV, this deviation tells us that there are probably significant systematic errors in 
the ISGRI calibration at the upper end of its energy range in addition to the calibration 
errors derived by us in Sect. \ref{sec-syserr}.

Our SPI spectrum is in very good agreement with that derived from SPI data
up to 1 MeV by Jourdain et al. (\cite{jourdain08}) whose analysis was not based on {\it spimodfit}. 
Furthermore, the smooth fit of the narrow energy bin around the 511~keV line, 
which is also a strong background line of the instrument, demonstrates that the 
background determination works well.

\subsection{Vela Pulsar}     
\label{sec-Vela}

\begin{figure}
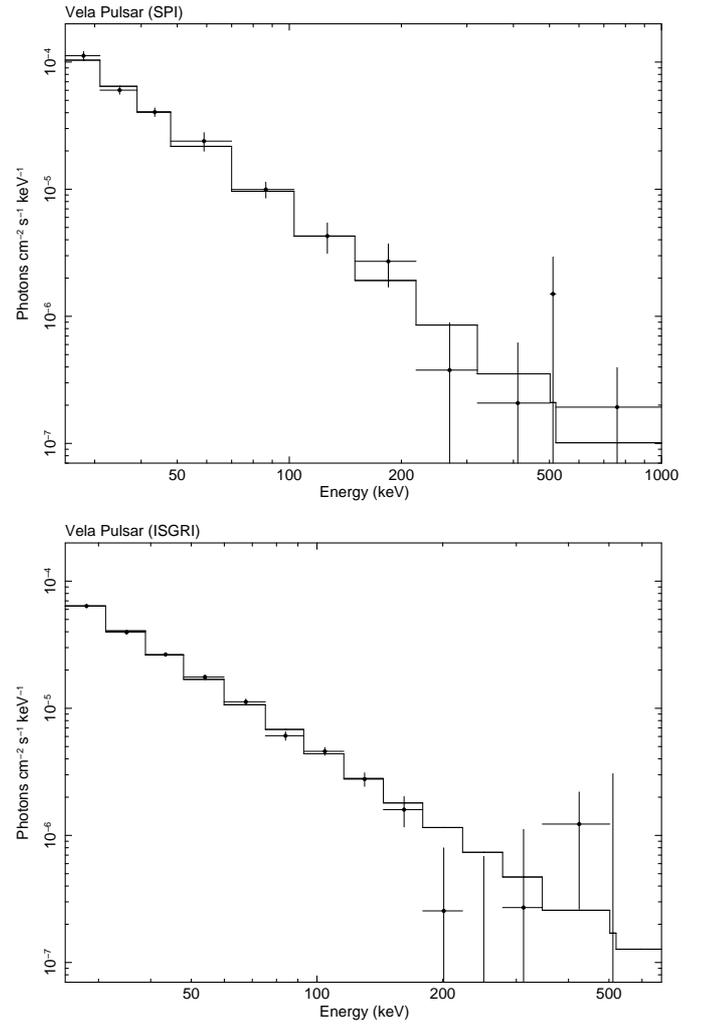

  \centering
  \includegraphics[angle=-90,width=9cm]{1284403a.eps}
  \includegraphics[angle=-90,width=9cm]{1284403b.eps}
  \caption{ The unfolded spectra of the Vela Pulsar and Pulsar Wind Nebula derived from dataset 2 
   (see table \protect\ref{tab-sourcedata}) with statistical errors: 
   {\bf (a)} (top) SPI spectrum, {\bf (b)} (bottom) ISGRI spectrum. Fit function is in both cases 
    a simple powerlaw (parameters are given in Sect. \ref{sec-Vela}).
}
  \label{fig-vela}
\end{figure}

The other pulsar wind nebula/pulsar in our set of sources,
the Vela PWN, is generally regarded as similar to the Crab. However, 
in hard X-rays the object is about two orders of magnitude
fainter even though it is nearly an order of magnitude
closer. It has been studied in detail by all high-resolution soft X-ray observatories
(see e.g. Mangano et al. (\cite{mangano06}) and references therein). 

Figure \ref{fig-vela}a shows the spectrum from our analysis of the SPI Vela Pulsar data.
Because of the weakness of the source, we choose the wider energy binning in the SPI analysis.
A simple powerlaw with $\Gamma$ = 2.13 $\pm$0.10, Norm = 0.12$\pm$0.05 describes the spectrum well.
The reduced $\chi^2$ of the fit is 0.55 for 9 d.o.f.

The SPI spectrum is in good agreement with our corresponding ISGRI spectrum
(Fig. \ref{fig-vela}b) which has the parameters
$\Gamma$ = 2.04$\pm$0.04 , Norm = (0.057$\pm$0.008), with
a reduced $\chi^2$ of 1.4 for 13 d.o.f.

The hard X-ray spectrum of this source was  studied
up to 200 keV by Mangano et al. (\cite{mangano06}) using {\it Beppo-SAX}
data. They find a spectral index of 2.00$\pm$0.05 at 15-200 keV which is in
very good agreement with our measurements.

At energies above 200 keV, our analysis does not show
significant evidence for emission. 
On the other hand, our measurement is well consistent with a continuation
of the spectrum in a way similar to what we observe for the Crab.
Given the nature of the 
source and its morphological similarity to the Crab, this is to be expected. 
OSSE detected the Vela PWN
only marginally between 200 and 760 keV with a spectral index of 1.8$\pm0.3$
(Strickman et al. \cite{strickman96}) 
compatible with a continuation of our spectrum without a break.

\subsection{NGC 4151}        
\label{sec-NGC4151}

\begin{figure}
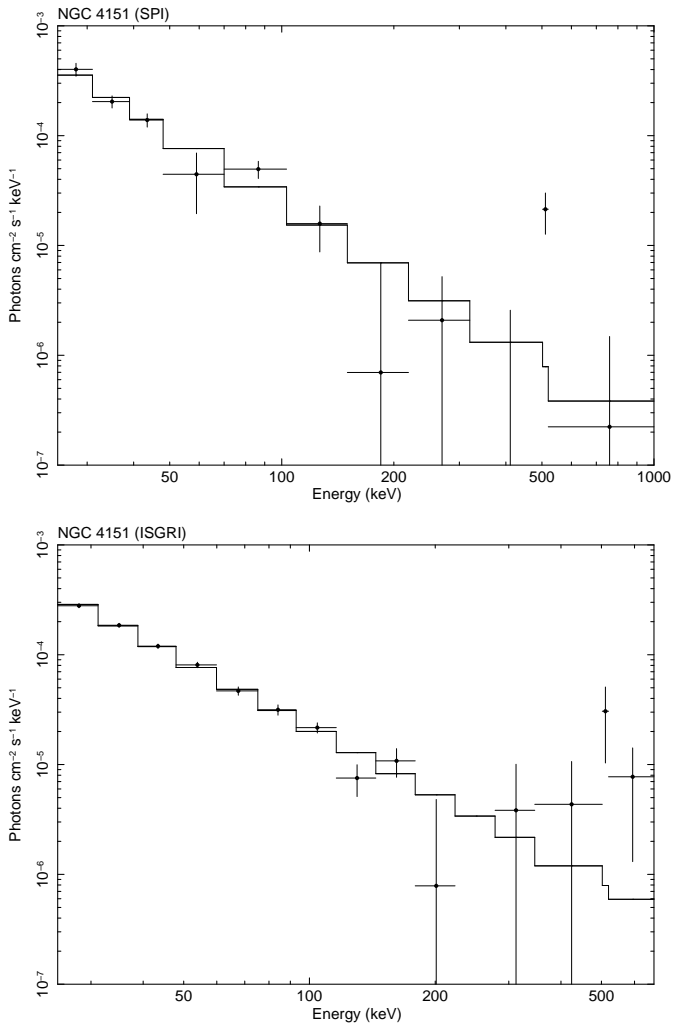

  \centering
  \includegraphics[angle=-90,width=9cm]{1284404a.eps}
  \includegraphics[angle=-90,width=9cm]{1284404b.eps}
  \caption{ The unfolded spectra of NGC~4151 derived from dataset 3 (which has very low exposure, see table \protect\ref{tab-sourcedata}) 
  with statistical errors: {\bf (a)} (top) SPI spectrum, {\bf (b)} (bottom) ISGRI spectrum. The fit function is in both cases a powerlaw 
  (parameters are given in Sect. \ref{sec-NGC4151}).
}
  \label{fig-NGC4151}
\end{figure}
 
The prototypical Seyfert galaxy NGC 4151 is one of the best studied active galaxies.
For a comprehensive review on this object  see Ulrich (\cite{ulrich00}).
Unfortunately, this is our smallest dataset with less than 100~ks  
observation time. We show the resulting spectra here only for completeness.
A detailed study of this source using INTEGRAL data up to 300 keV was published by
Beckmann et al. (\cite{beckmann05}) (they use only ISGRI data and get a larger dataset
by using data taken during SPI annealings).

Figure \ref{fig-NGC4151}a shows the SPI spectrum. The best fit is a simple powerlaw
with $\Gamma$ = 2.10$\pm$0.17, Norm = 0.38$\pm$0.25, and a reduced $\chi^2$
of 1.4 for 9 d.o.f. The excess in the annihilation line bin is marginal (2.4\,$\sigma$ significance).

The ISGRI spectrum (Fig. \ref{fig-NGC4151}b)
gives a compatible fit result with $\Gamma$ = 2.03$\pm$0.06,
Norm = 0.24$\pm$0.05, reduced $\chi^2$=1.4 for 13 d.o.f. 

Fitting a powerlaw with exponential cutoff to the ISGRI data 
as in Beckmann et al. (\cite{beckmann05}) improves the fit slightly ($\chi^2$=1.3)
and results in values well compatible with previously published measurements:
$\Gamma$ = 1.57$\pm$0.24, $E_{\mathrm{cutoff}}$ = 121$\pm$55~keV, Norm = 0.07$\pm$0.04 . 

\subsection{NGC 4945}        
\label{sec-NGC4945}

\begin{figure}
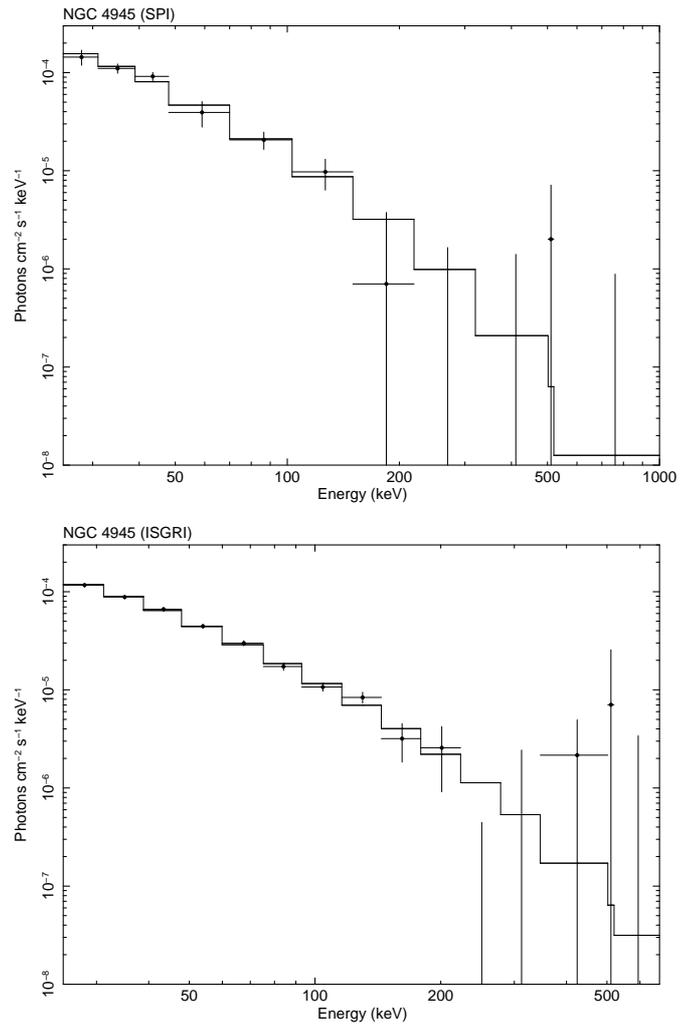

  \centering
  \includegraphics[angle=-90,width=9cm]{1284405a.eps}
  \includegraphics[angle=-90,width=9cm]{1284405b.eps}
  \caption{ The unfolded spectra of NGC~4945 derived from dataset 4 (which has low exposure, see table \protect\ref{tab-sourcedata}) 
   with statistical errors: {\bf (a)} (top) SPI spectrum, {\bf (b)} (bottom) ISGRI spectrum. The fit function is in both cases a powerlaw 
   with exponential cutoff and fixed photo-electric absorption (parameters are given in Sect. \ref{sec-NGC4945}).
}
  \label{fig-NGC4945}
\end{figure}

NGC 4945 is the brightest Seyfert 2 and the second brightest hard X-ray source (after NGC 4151)
of all radio-quiet AGN (Done et al. \cite{done96}).
Its X-ray spectrum has been extensively studied by various authors (see e.g. Itoh et al. (\cite{itoh08})
and references therein). As a Seyfert 2, the source is strongly absorbed at lower energies.
Itoh et al. (\cite{itoh08}) using  {\it Suzaku} measure
a hydrogen column density $N_{\mathrm H}$ of 5.4$\times$10$^{24}$~cm$^{-2}$ and a hard X-ray spectrum with a 
powerlaw index of 1.5$\pm^{0.2}_{0.1}$ and an exponential cutoff at an energy of 150$\pm^{100}_{50}$\,keV
(in agreement with previous measurements).

Our dataset for NGC4945, like the one for NGC 4151, has relatively low exposure.  
The hydrogen column density for this source is by far the highest of all the sources
in our sample. We find that it is still low enough to be negligible for our SPI spectral
modelling but the ISGRI sensitivity below 50~keV is high enough to make the spectral
fits sensitive to the presence of the absorption.
In order to compare better with previous measurements, we therefore include the photoelectric
absorption in the spectral model (Xspec model WABS) and
fit an absorbed powerlaw with exponential cutoff to the SPI and the ISGRI spectrum fixing 
the hydrogen column and the cutoff energy to the values found by Itoh et al. (\cite{itoh08}).

Figure \ref{fig-NGC4945}a shows the resulting SPI spectrum. The best fit parameters are
$\Gamma$ = 1.66$\pm$0.17, Norm = 0.061$\pm$0.039 with a reduced $\chi^2$ of 0.4 for 9 d.o.f. 

The ISGRI spectrum is shown in Fig. \ref{fig-NGC4945}b. Keeping $N_{\mathrm H}$ fixed as for SPI, 
the best fit parameters
are $\Gamma$ = 1.56$\pm$0.05, Norm = 0.033$\pm$0.006  with a reduced $\chi^2$
of 0.6 for 13 d.o.f. 

SPI and ISGRI spectrum are both consistent with each other and with the
spectra measured by Itoh et al. (\cite{itoh08}).

\subsection{Cen A}           
\label{sec-CenA}

\begin{figure}
  \centering
  \includegraphics[angle=-90,width=9cm]{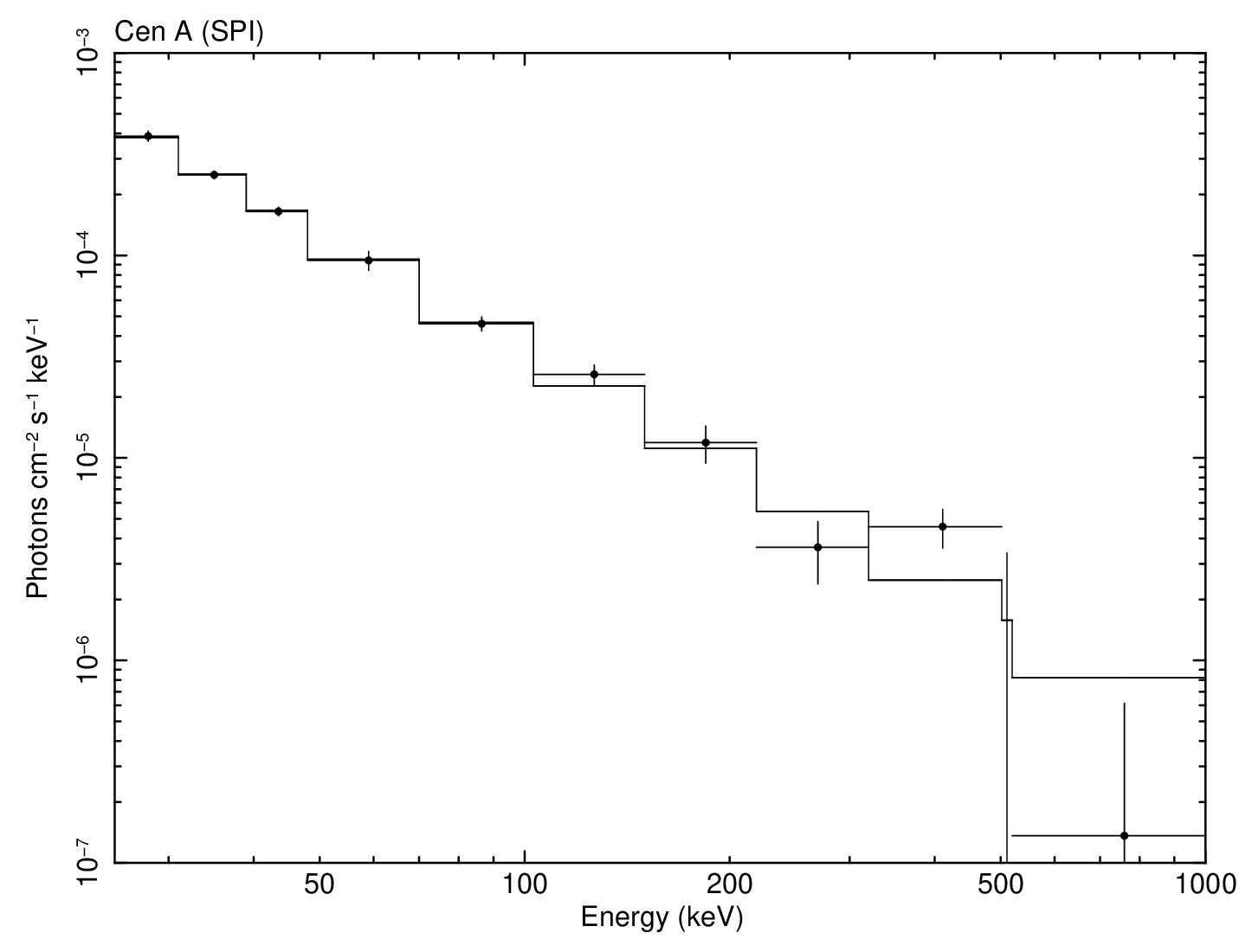}
  \includegraphics[angle=-90,width=9cm]{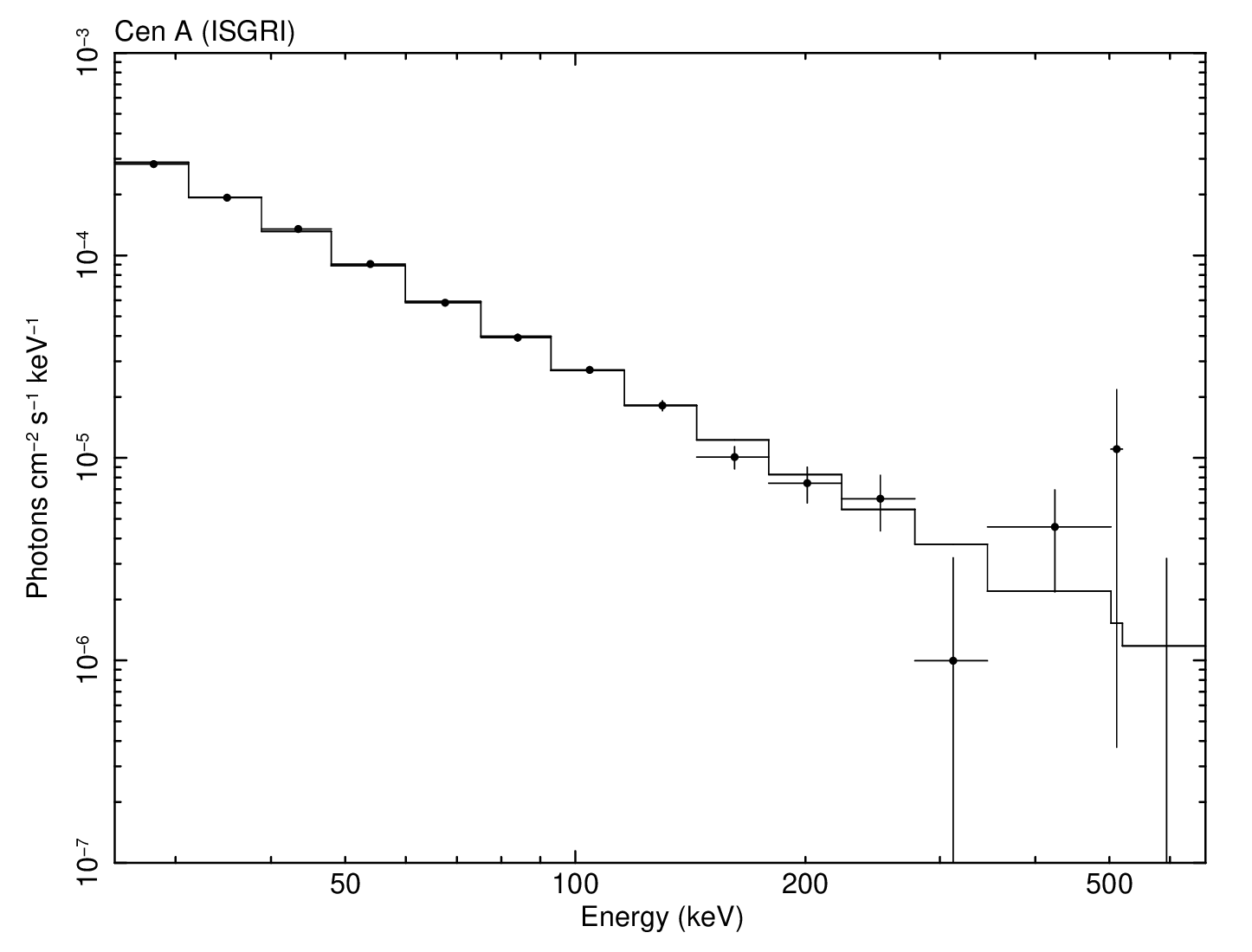}
  \caption{ The unfolded spectra of Cen~A derived from dataset 5 (see table \protect\ref{tab-sourcedata}) with statistical errors: 
   {\bf (a)} (top) SPI spectrum, {\bf (b)} (bottom) ISGRI spectrum. The fit function is in both cases a powerlaw (parameters are 
   given in Sect. \ref{sec-CenA}).
}
  \label{fig-CenA}
\end{figure}

The nearby radio galaxy Cen~A has been observed
by all X-ray observatories (see Rothschild et al. (\cite{rothschild06})
and references therein). It has been classified as a Seyfert 2 (see Dermer \& Gehrels \cite{dermer95}).
The consistently observed hard X-ray spectrum is, in spite of flux variability, a simple
powerlaw of index $\approx$ 1.8 (Rothschild et al. \cite{rothschild06})
which extends to the MeV range where the spectrum steepens (Steinle et al. \cite{steinle98}).

Even though the available SPI exposure for Cen~A is modest, we obtain a clear detection
also beyond 200 keV. The SPI spectrum (Fig. \ref{fig-CenA}a) gives the best fit parameters
$\Gamma$ = 1.89$\pm$0.05, Norm = 0.20$\pm$0.04, and a reduced $\chi^2$ of 1.2 for 9 d.o.f.. 

Similarly, the ISGRI spectrum (Fig. \ref{fig-CenA}b) gives a best fit
with $\Gamma$ = 1.80$\pm$0.02, Norm = 0.117$\pm$0.009 and a reduced $\chi^2$ of
0.9 for 13 d.o.f. 

SPI and ISGRI spectral shape are in excellent agreement with each other and
previous measurements.

\subsection{XTE J1550-564}
\label{sec-XTEJ1550}

\begin{figure}
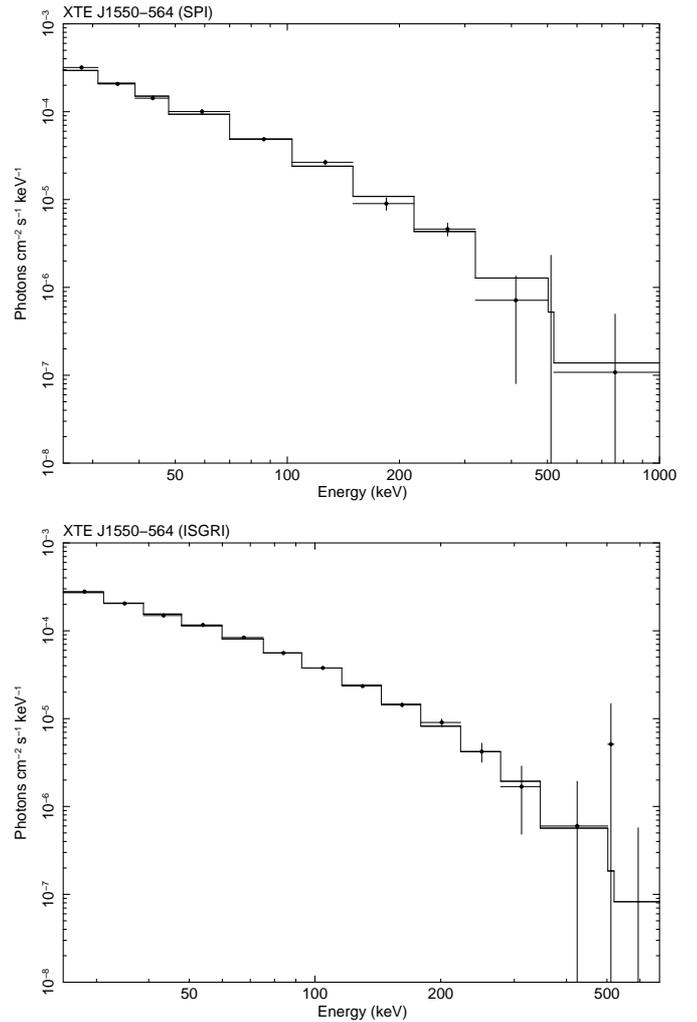

  \centering
  \includegraphics[angle=-90,width=9cm]{1284407a.eps}
  \includegraphics[angle=-90,width=9cm]{1284407b.eps}
  \caption{ The unfolded spectra of XTE J1550-564  derived from dataset 6 (see table \ref{tab-sourcedata}) with 
  statistical errors: {\bf (a)} (top) SPI spectrum, {\bf (b)} (bottom) ISGRI spectrum (increased 
   errors were required, see text). The fit function is in both cases a powerlaw with exponential cutoff (parameters are given
  in Sect. \ref{sec-XTEJ1550}).
}
  \label{fig-XTEJ1550}
\end{figure}

The Low-Mass X-ray Binary XTE J1550-564 has been classified as a microquasar and 
a black hole candidate. Since its discovery in 1998 (Smith \cite{smith98})
it has been subject of several detailed studies and was found to be similar to 
Cyg~X-1, however, with lower luminosity (see e.g. Wu et al. \cite{wu07} and references therein).
Sturner \& Shrader (\cite{sturner05}) studied the source using {\it RXTE} and {\it INTEGRAL} data
and found that it stayed in its low-hard spectral state even when going through modest
outbursts. According to the ISGRI lightcurve of the source
(Courvoisier et al. \cite{isgri-lightcurves}), XTE J1550-564 had only two short major outbursts with INTEGRAL coverage
and was otherwise in a low state. The time-averaged spectrum we are determining here should therefore
be close to the low-hard type, i.e. consistent with thermal Comptonisation which is essentially a powerlaw with
exponential cutoff in our energy range.

We find that a powerlaw with exponential cutoff fits the SPI data reasonably.
From the SPI spectrum (Fig. \ref{fig-XTEJ1550}a) we get the parameters $\Gamma$ = 1.36$\pm$0.09,
$E_{\mathrm{cutoff}}$ = (206$\pm$50)~keV, Norm = 0.031$\pm$0.010 with a reduced $\chi^2$ of 1.6 for 8 d.o.f.  

For the ISGRI spectrum we don't achieve a reduced $\chi^2$ below 2.0 with any common model 
(we tried cutoff powerlaw,
thermal Comptonisation, and black body + powerlaw) although the residuals don't show any clear
trend in the deviations. This can be explained by additional systematic errors in the image
reconstruction caused by the complex source region. Allowing for this by doubling the assumed calibration
errors from 0.5~\% to 1.0~\% (see also Sturner \& Shrader (\cite{sturner05})), 
we get a reduced $\chi^2$ of 1.3 for 12 d.o.f. The
best fit parameters are marginally consistent with the ones obtained from
the SPI data: $\Gamma$ = 1.00$\pm$0.05, $E_{\mathrm{cutoff}}$ = (108$\pm$9)~keV, Norm = 0.009$\pm$0.002. 
This is due to higher photon fluxes measured by ISGRI below 200 keV. Above 200 keV,
the fluxes are consistent (cf. Table \ref{tab-summary2}).   

For comparison with earlier publications, we also fit the thermal Comptonisation model
(COMPTT) by Titarchuk (\cite{titarchuk94}) which is the best fit to the ISGRI and SPI data in
Sturner \& Shrader (\cite{sturner05}). Averaging over their datasets, they obtain a
plasma temperature $kT$ = (48.4$\pm$2.7)~keV,
a plasma optical depth $\tau_p$ = 1.48$\pm$0.07, 
and a soft photon temperature $kT_0$ = (0.5$\pm$0.14)~keV. 
Since we lack the data between 3~keV and 25~keV, $kT_0$ is
not well constrained. We therefore fix it to the value 0.5~keV by Sturner \& Shrader. 
From the SPI data we then obtain $kT$ = (56.9$\pm$7.7)~keV and $\tau_p$ = 1.32$\pm$0.18
(reduced $\chi^2$=1.3 for 9 d.o.f.) in good agreement with Sturner \& Shrader.

\subsection{4U 1630-47}  
\label{sec-4U1630}

\begin{figure}
  \centering
  \includegraphics[angle=-90,width=9cm]{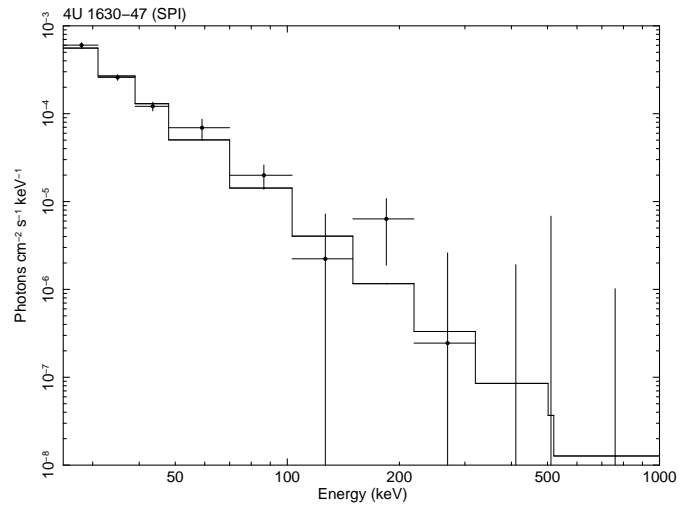}
  \caption{ The unfolded spectrum of 4U~1630-47  derived from the SPI dataset 7 (see table \ref{tab-sourcedata}) with 
  statistical errors. The corresponding ISGRI data is flawed because of source contamination by IGR J16358-472 (see text). 
  The fit function is a powerlaw (parameters are given in Sect. \ref{sec-4U1630}).
}
  \label{fig-4U1630}
\end{figure}

The X-ray transient 4U 1630-47 has been classified as a LMXB, a very likely black hole candidate, and
has also been found to show indications of the presence of jets but is not yet regarded as a microquasar
(see Kubota et al. (\cite{kubota07}) and references therein). During the observation time covered by our dataset,
the source showed frequent but short outbursts. The time-averaged spectrum should therefore be a superposition
of the various states the source can be in, with the low state dominating. 
Tomsick et al. (\cite{tomsick05}) analyse {\it RXTE} and {\it INTEGRAL} data from 4U 1630-47. 
They distinguish five 
different spectral states which they all fit with a two-component model made from
a black body with interstellar absorption and a powerlaw. They find that the para\-meters of this model
vary strongly between the states. The lowest flux state is found to be the one where the emission is dominated by
black body emission ($kT \approx 1.4$~keV) below 15 keV, and above that by a steep powerlaw 
with index $\Gamma \approx$3.
A simple powerlaw is also the best fit to our spectra of this on average relatively weak source.

The SPI spectrum for 4U 1630-47 (Fig. \ref{fig-4U1630}) is best fit by a powerlaw with $\Gamma$ =  
3.3$\pm$0.2 and Norm = 33$\pm$21. The reduced $\chi^2$ is 0.6 for 9 d.o.f.

The ISGRI spectrum may suffer from source contamination from the neighbouring
source IGR J16358-4726, especially at higher energies
(see also Tomsick et al. \cite{tomsick05}).
 We try to account for this approximately by increasing the systematic errors to 5~\%.
Fitting a powerlaw then results in  $\Gamma$ = 2.33$\pm$0.24, Norm = 0.04$\pm$0.04 
with a reduced $\chi^2$ = 1.0 for 13 d.o.f. which is much harder than the SPI spectrum.

Furthermore, the observed fluxes (see Table \ref{tab-summary2}) from SPI and ISGRI
differ strongly at low energies while the SPI measurements agree with those
from B08 who also note that the source may
be confused with IGR~J16358-472.  

We conclude that the neighbouring source IGR J16358-472 may be hampering
the measurement for ISGRI or SPI or both. Since this distortion is
strongly dependent on the instrument characteristics, the effects are 
different for SPI and ISGRI and lead very probably to the observed difference
in the spectra. The smaller differences between our spectra and those
from Bouchet are to be expected since here the same instrument is used.
The fact that the SPI spectrum agrees well with {\it RXTE} measurements
of 4U~1630-47 in a low state by Tomsick et al. (\cite{tomsick05}) argues
for the better quality of the SPI spectrum. Also, our source model for 
the SPI data explicitely includes IGR~J16358-472. The SPI analysis 
should therefore in this case be more immune to contamination than the ISGRI analysis.

\subsection{Swift J1656.3-3302} 
\label{sec-SwiftJ}

\begin{figure}
  \centering
  \includegraphics[angle=-90,width=9cm]{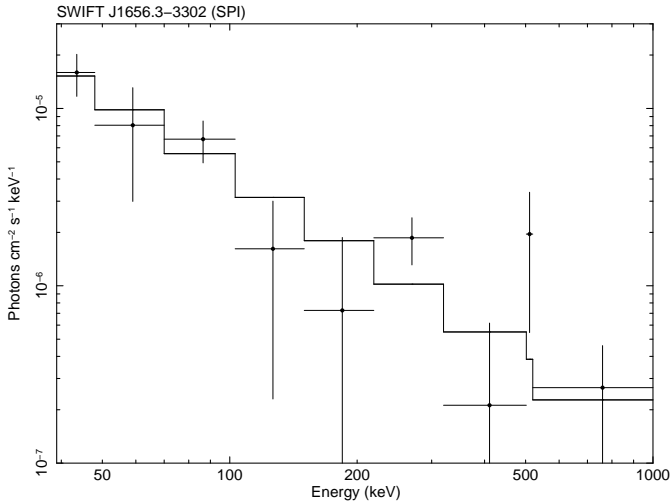}
  \caption{ The unfolded SPI spectrum of Swift J1656.3-3302 derived from dataset 10 
(see table \protect\ref{tab-sourcedata}) with statistical errors. The fit function 
is a powerlaw (parameters are given in Sect. \ref{sec-SwiftJ}).
}
  \label{fig-SwiftJ}
\end{figure}

The recently discovered source Swift~J1656.3-3302 (Okajima et al. \cite{okajima06})
has been identified with a high redshift ($z$ = 2.4) blazar (Masetti et al \cite{masetti08}).
The late discovery of this source is due to its hard spectrum, i.e. its weakness at soft X-rays, 
and to its location near the galactic bulge which makes
for a very complex field of view with many variable and bright neighbouring sources.
Probably because many bright sources are very near the edge of the partially coded field of view when
Swift~J1656.3-3302 is near its centre, our standard modelling technique 
turned out to be inadequate for extracting a SPI spectrum of Swift~J1656.3-3302 from a 
dataset {\it centred} at the source. The quality of the spimodfit fits was never acceptable
in more than a few energy bins.
But when the source was observed by SPI with high exposure and off-axis by at least 5$^\circ$ in the dataset
selected for GRS~1758-258 (dataset 10, see table \protect\ref{tab-sourcedata}), 
the extraction of a useful spectrum with our standard input catalog
and variability information was possible in all energy bins except for the lowest two. 
The corresponding ISGRI spectrum of Swift~J1656.3-3302 from dataset 10, however, was not useful 
since the object was too far outside the ISGRI fully coded field of view.
We therefore present in Fig. \ref{fig-SwiftJ} only the SPI spectrum for this source.
It is well described by a simple powerlaw with $\Gamma$ = 1.49$\pm$0.18 and
Norm = 0.004$\pm$0.005. The reduced $\chi^2$ is 1.0 for 7 d.o.f.

An ISGRI spectrum for  Swift~J1656.3-3302 was measured by Masetti et al. (\cite{masetti08}).
They find a spectral index of 1.64$\pm$0.16 in agreement with our measurement.

\subsection{OAO 1657-415}
\label{sec-OAO1657}

\begin{figure}
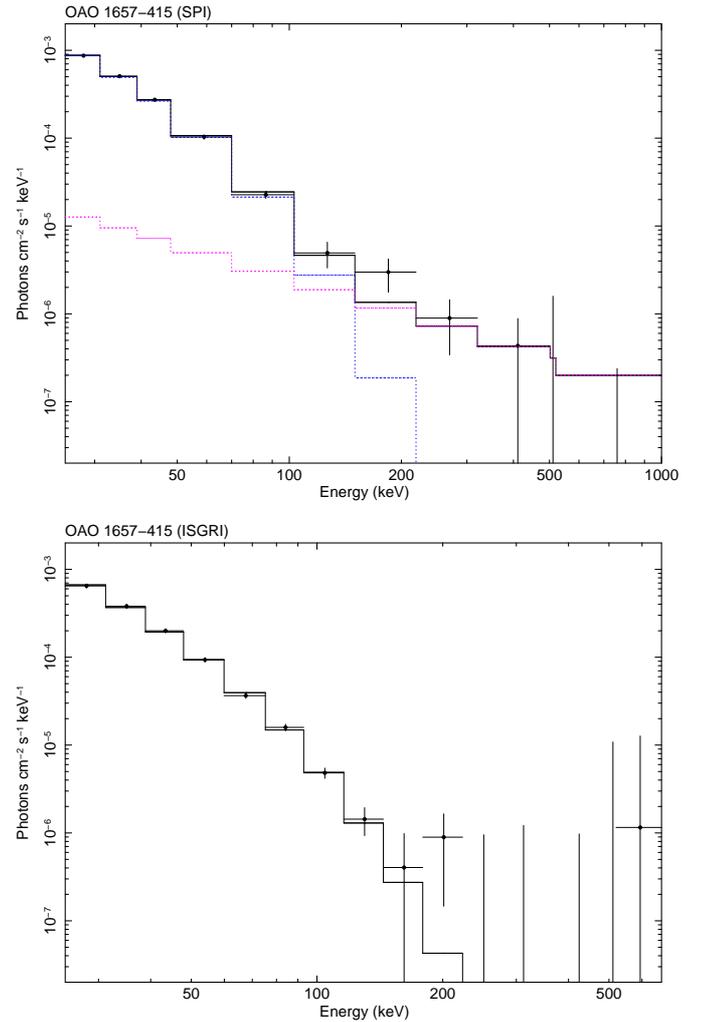

  \centering
  \includegraphics[angle=-90,width=9cm]{1284410a.eps}
  \includegraphics[angle=-90,width=9cm]{1284410b.eps}
  \caption{ The unfolded spectra of OAO 1657-415 derived from dataset 7 (see table \ref{tab-sourcedata}) with 
  statistical errors:
  {\bf(a)} (top) SPI spectrum fit with powerlaw with exponential cutoff and additional powerlaw component 
  {\bf (b)} (bottom) ISGRI spectrum fit with powerlaw with exponential cutoff (with increased errors, see text). 
  The fit parameters are given in Sect. \ref{sec-OAO1657}.
}
  \label{fig-OAO1657}
  \end{figure}

The high mass X-ray binary OAO 1657-415 is a system consisting of
an eclipsing 38~s pulsar and a highly reddened B supergiant with orbital period
10.4~d (see e.g. Audley et al. \cite{audley06} and references therein).
The object has been studied since its discovery in the 1970s by a fair number 
of authors with several X-ray observatories up to about 160~keV, recently also with {\it INTEGRAL}
(Filippova et al. \cite{filippova05}, Barnstedt et al. \cite{barnstedt}). In the range from 3 to 160~keV
they find consistently an absorbed, exceptionally hard spectrum with index $\Gamma \approx$~1.0 which then softens
with an exponential cutoff at energies around 25~keV.

We fit our time-averaged (and therefore also phase-averaged)
SPI spectrum initially with the simplest of these models, the powerlaw with exponential cutoff.
We obtain $\Gamma$ = 1.17$\pm$0.23 and $E_{\mathrm{cutoff}}$ = (23.9$\pm$3.2)~keV with a reduced $\chi^2$ of 1.4
for 8 d.o.f., in good agreement with previous measurements. 

Inspecting the residuals,  we find marginal evidence for a hard tail
diverging from the model above 150 keV. The feature is not excluded by previous studies
since they did not cover this energy range. 
The significance of the hard tail is about 3~$\sigma$ which does not warrant
a much more detailed study.
In order to quantify the possible flux of this feature, we
add a powerlaw to the model keeping $E_{\mathrm{cutoff}}$ fixed. 
This results in a lower reduced $\chi^2$  of 0.7 and best fit parameters
$\Gamma_1$ = 1.23$\pm$0.07,
Norm$_1$ = 0.16$\pm$0.04,
$\Gamma_2$ = 1.27$\pm$0.58 and a flux of the second component 
at 200~keV of $(2.1\pm1.0)\times 10^{-6}$~cm$^{-2}$s$^{-1}$keV$^{-1}$.
Fig. \ref{fig-OAO1657}a shows the fit result with the added powerlaw.

For the ISGRI data, we increase the statistical errors by 5~\%
of the flux in order to get take into account the larger systematic uncertainties 
of the analysis (this is the same dataset as the one for 4U~1630-47 where 
this increase was also necessary).

If we fit the same model with added powerlaw as for the SPI data, we get
 reduced $\chi^2$  of 0.4 for 10 d.o.f. compatible with the SPI fit. 
The parameters of the second powerlaw are essentially unconstrained. The reality 
of the potential hard tail in the SPI data is not ruled out.

If we fit with the powerlaw with exponential cutoff without the second powerlaw, we get
a reduced $\chi^2$  of 0.4  for 12 d.o.f. and best fit parameters
$\Gamma_1$ = 1.4$\pm$0.3,
$E_{\mathrm{cutoff}}$ = 25$\pm$4~keV, Norm = 0.23$\pm$0.21 which is well
compatible with the SPI data. This spectrum in shown in Fig. \ref{fig-OAO1657}b.   

\subsection{GX 339-4}        
\label{sec-GX339-4}

\begin{figure}
  \centering
  \includegraphics[angle=-90,width=9cm]{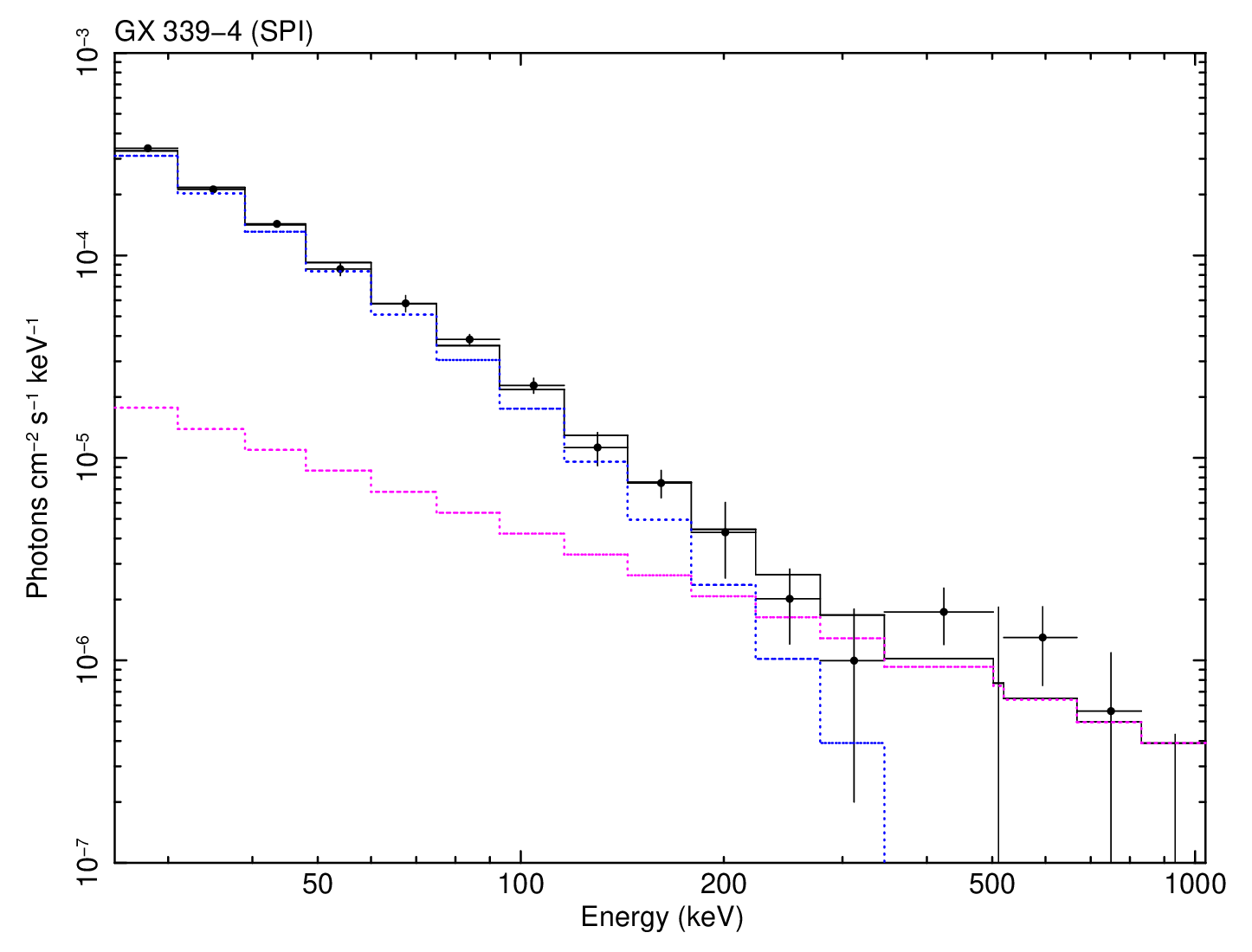}
  \includegraphics[angle=-90,width=9cm]{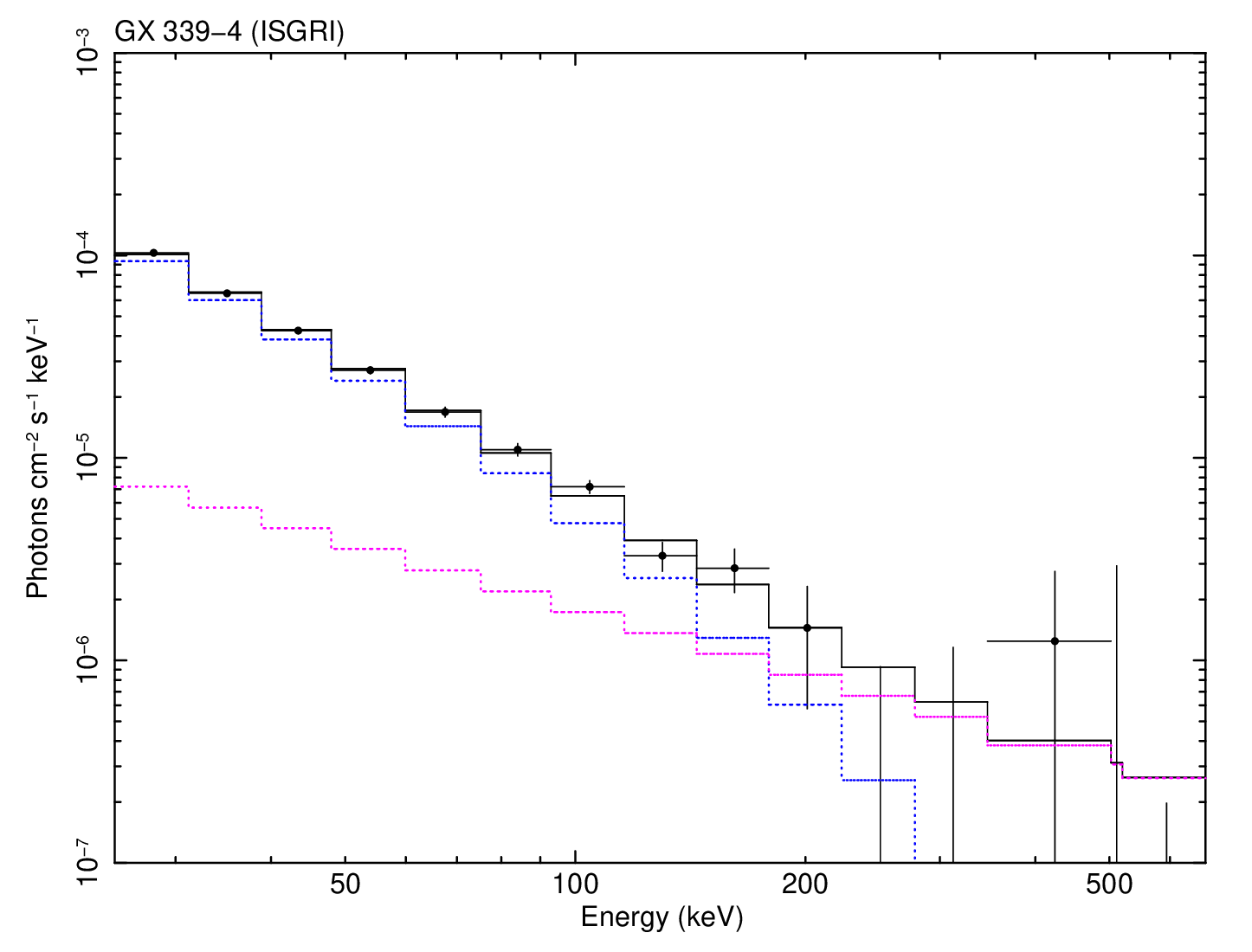}
  \caption{ The unfolded spectra of GX 339-4 derived from dataset 8 (see table \ref{tab-sourcedata}) with 
  statistical errors: 
  {\bf(a)} (middle) SPI spectrum fit with two-component model: powerlaw with exponential cutoff and second 
   powerlaw. {\bf (b)} (bottom) ISGRI spectrum fit with
  the same model as in (a). 
  The fit parameters are given in Sect. \ref{sec-GX339-4}.
}
  \label{fig-GX339-4}
  \end{figure}

The microquasar GX~339-4 in one of the classical black hole X-ray binaries and has been studied by
all X-ray and soft gamma-ray observatories, also with {\it Swift} (Tomsick et al. \cite{tomsick08}) and 
{\it INTEGRAL} (see Joinet et al. \cite{joinet07} who also give a review of the literature).
Like other black hole binaries, the source has two main states, the low-hard state
during which the high-energy spectrum is essentially a powerlaw with index $\Gamma$ = 1.4-2.1
with an exponential cutoff at hundred to a few hundred keV, and the high-soft state during
which $\Gamma$ goes to values $\geq 2.4$ but without cutoff (see again Joinet et al. \cite{joinet07}).
Studies of the soft-gamma ray emission with OSSE have also found emission up to 400 keV
(Grabelsky et al. \cite{grabelsky95}).
Joinet et al. (\cite{joinet07}) found evidence for emission above 200 keV
in the low-hard state at the 5~$\sigma$ level.

Our time-averaged spectrum should be a superposition of all states of the object.
As a first step, we fit a simple powerlaw to the data and obtain
index $\Gamma$ = 2.08$\pm$0.04, Norm = 0.35$\pm$0.05 with a reduced $\chi^2$ of 1.2 for 15 d.o.f.
The residuals are systematically negative from 100-300~keV but seem to be systematically positive 
above 300~keV indicating a hard tail. 
The significance of the emission above 200~keV is 6.4$\sigma$.

The same fit applied to the ISGRI data gives a good reduced $\chi^2$
of 0.6 for 13 d.o.f and best fit parameters  $\Gamma$ = 2.07$\pm$0.04 and
Norm = 0.10$\pm$0.01 confirming the shape of the SPI result up to 200 keV.
Beyond 200~keV, the ISGRI result is not sensitive enough.
As in the SPI spectrum, there is no evidence for a cutoff.
A hard tail cannot be ruled out by the ISGRI data.

In order to improve our fit to the SPI data and to better be able to compare with
previous measurements, we fit a more complex model consisting of two components:
(a) a powerlaw  with exponential cutoff (in our energy range this is
equivalent to a thermal Comptonisation model, one of the main candidate emission
mechanisms for microquasars, see Sect. \ref{sec-XTEJ1550}) 
and (b) an additional powerlaw in order
to describe a potential hard tail which may obscure the observation of a 
cutoff in our spectrum.  

Since we lack the soft X-ray data to better constrain the cutoff energy, we fix it
to 100~keV consistent with previous measurements (if $E_{\mathrm{cutoff}}$ is permitted to
vary freely, we get a best fit value of 88$\pm$56~keV).
Figure \ref{fig-GX339-4}a shows the result of this fit. The best fit parameters
are  $\Gamma_1$ = 1.6$\pm$0.1, Norm$_1$ = 0.09$\pm$0.03, $\Gamma_2$ = 1.0$\pm$0.6,
Norm$_2$ (Flux at 200~keV) = 
$(3.8\pm2.4)\times10^{-6}$~cm$^{-2}$s$^{-1}$keV$^{-1}$
with an improved reduced $\chi^2$ of 0.8 for 13 d.o.f.

The index $\Gamma_1$ of the dominant powerlaw at lower energies is well
consistent with that observed by Tomsick et al. (\cite{tomsick08}) up to
100~keV. The additional powerlaw becomes dominant at about 200~keV. 
This confirms the feature already
noted by Joinet et al. (\cite{joinet07}) who speculated that it could be
produced in a jet. 

The more complex model is also fully compatible with the ISGRI
data. Fitting the two-component model with fixed $E_{\mathrm{cutoff}}$ = 100~keV
and $\Gamma_2$ = 1.0 to the ISGRI data, we obtain a reduced $\chi^2$ of 0.7 for 12 d.o.f.
and best fit parameters $\Gamma_1$ = 1.7$\pm$0.1 (Fig. \ref{fig-GX339-4}b).

\subsection{4U 1700-377}
\label{sec-4U1700}

\begin{figure}
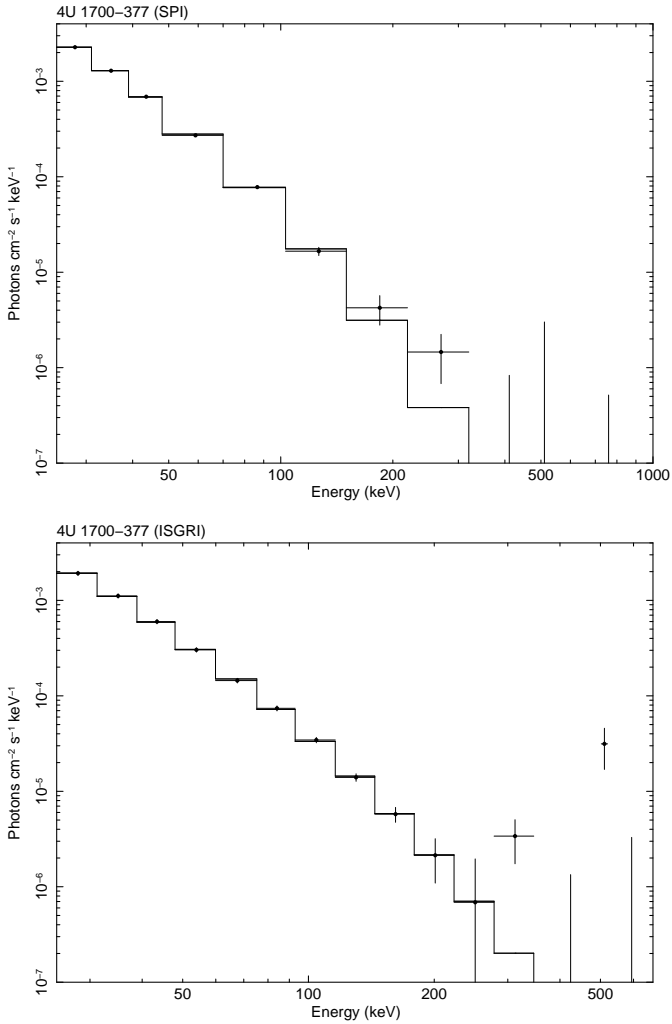

  \centering
  \includegraphics[angle=-90,width=9cm]{1284412a.eps}
  \includegraphics[angle=-90,width=9cm]{1284412b.eps}
  \caption{ The unfolded spectra of 4U 1700-377 derived from dataset 7 (see table \ref{tab-sourcedata}) with 
  statistical errors: {\bf (a)} (top) SPI spectrum, 
  {\bf(b)} (bottom) ISGRI spectrum. Spectral model is in both cases a powerlaw with high energy cutoff. 
  The fit parameters are given in Sect. \ref{sec-4U1700}.
}
  \label{fig-4U1700}
  \end{figure}

The eclipsing, 3.4 day  X-ray binary 4U~1700-377 consists of a hot, bright supergiant
and very probably a neutron star because its overall X-ray
spectrum is similar to that of an accreting pulsar. The system
has been studied in detail by many authors. 
For a review see e.g. van der Meer et al. (\cite{vandermeer05}).
In soft X-rays, the source displays a multitude of line features.
The high energy spectrum has been modelled successfully
by a powerlaw with high energy cutoff with additional photoelectric
absorption for the points below $\approx$ 25~keV. Authors have also employed
more complex models like thermal bremsstrahlung or thermal Comptonisation.
In all cases, even though the source is strongly variable, the 
spectral shape above about 25~keV was found to be independent of the flux state.

Our SPI  spectrum is indeed well described by a powerlaw with high energy
cutoff (Xspec model HIGHECUT).
Figure \ref{fig-4U1700}a shows the SPI spectrum. The best fit parameters are
$\Gamma$ = 2.3$\pm$0.1, Norm = 5.0$\pm$2.4, $E_c$ = (31$\pm$2)~keV,
$E_f$ = (65$\pm$11)~keV. The reduced $\chi^2$ is 0.9 for 7 d.o.f.

This agrees well with the literature: e.g. Maisack et al. (\cite{maisack94})
find $\Gamma$ = 2.55$^{+0.15}_{-0.55}$, $E_c$ = (20.1$^{+11.4}_{-20.1}$)~keV,
$E_f$ = (69.8$^{+9.7}_{-38.3}$)~keV.
But also a simpler power law with exponential cutoff at 48$\pm$4~keV and $\Gamma$ = 2.0$\pm$0.1
describes the data reasonably.

As for the other two sources in dataset 7, the statistical errors for the ISGRI
spectrum of 4U~1700-377 (Fig. \ref{fig-4U1700}b) had to be increased by 5~\% of the flux.
This is certainly at least partially caused by the strong variability of 4U~1700-377
(see also Sect. \ref{sec-timevar}).
A fit with a power law and exponential cutoff gives the best fit parameter values   
$\Gamma$ = 2.3$\pm$0.2, Norm = 7$\pm$3, $E_{\mathrm{cutoff}}$ = (74$\pm$14)~keV,
in agreement with SPI. 
The reduced $\chi^2$ is 0.9 for 12 d.o.f.

\subsection{IGR J17091-3624}
\label{sec-IGRJ17091}

\begin{figure}
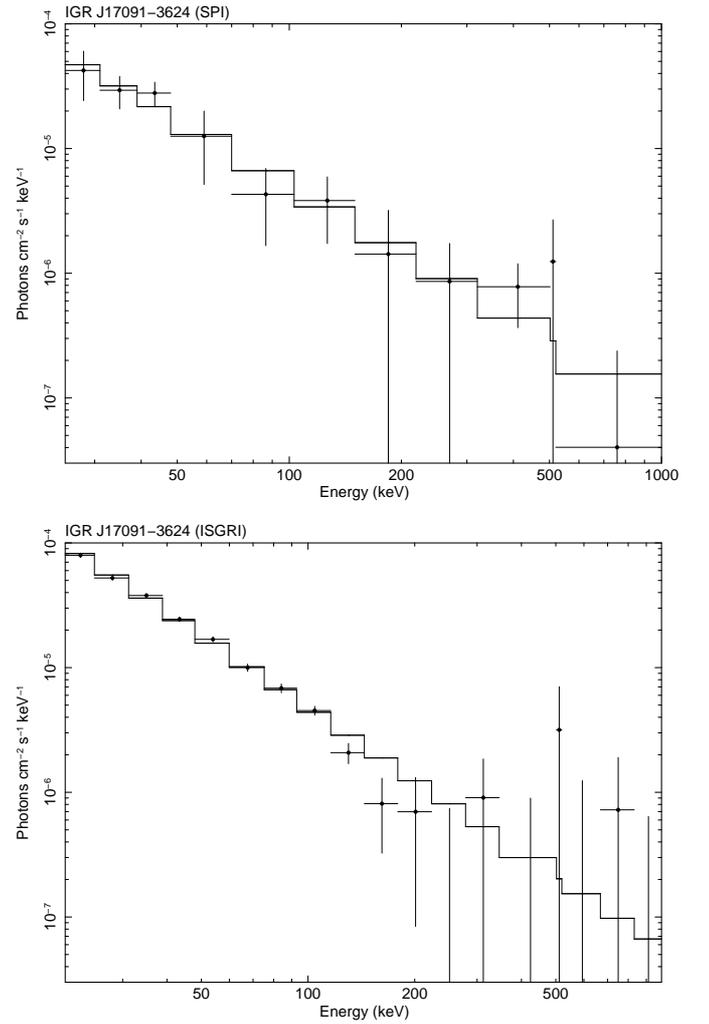

  \centering
  \includegraphics[angle=-90,width=9cm]{1284413a.eps}
  \includegraphics[angle=-90,width=9cm]{1284413b.eps}
  \caption{ The unfolded spectra of IGR J17091-3624 derived from dataset 9 (see table \ref{tab-sourcedata}) with 
  statistical errors: {\bf (a)} (top) SPI spectrum with powerlaw fit, 
  {\bf(b)} (middle) ISGRI spectrum with powerlaw fit. 
The fit parameters are given in Sect. \ref{sec-IGRJ17091}.
}
  \label{fig-IGRJ17091}
  \end{figure}

The object IGR J17091-3624 was discovered in 2003 by {\it INTEGRAL} (Kuulkers et al. 2003)
and subsequently studied by a small number of authors based on archival data 
(Revnivtsev et al. \cite{revnivtsev03}) and {\it INTEGRAL} and {\it RXTE} data
(Capitanio et al. \cite{capitanio06}). Chaty et al. (\cite{chaty08}) recently found that
the most probable candidate for a companion star is neither a giant nor a supergiant. They
conclude that the system is most probably a LMXB.
The source is variable but weak ($\leq$ 10 mCrab above 25 keV). 
The spectrum above a few ten keV is in good approximation
a hard powerlaw with index varying between 1.6 and 2.2, but has also been successfully
modelled with thermal Comptonisation (Capitanio et al. \cite{capitanio06}).

When fitting our time-averaged SPI spectrum of IGR~J17091-3624 (Fig. \ref{fig-IGRJ17091}a), we find that a 
simple powerlaw describes it well. We find as best fit parameters $\Gamma$ = 1.8$\pm$0.2,
Norm = 0.02$\pm$0.01 with a reduced $\chi^2$ of 0.4 for 9 d.o.f.
This is consistent with the previous measurements mentioned above and
with the notion that the source was mostly in the low-hard state during our
observation time. 
 
Fitting a powerlaw to the corresponding ISGRI spectrum (Fig. \ref{fig-IGRJ17091}b)
gives a reduced $\chi^2$ of 1.2 for 16 d.o.f.,$\Gamma$ = 1.93$\pm$0.04, and
Norm = 0.034$\pm$0.005 consistent with the SPI spectrum.

In order to compare our spectra with the ISGRI spectra by Captanio et al. (\cite{capitanio06}),
we also fit a thermal Comptonisation model (Xspec model COMPTT).
We follow Captanio et al. and fix the soft photon temperature $kT_0$ = 0.1~keV.
From our ISGRI spectrum we then obtain a plasma optical depth $\tau_p$ = 1.5$\pm$0.3, 
and a plasma temperature $kT$ = (36$\pm$8)~keV, Norm = 0.0016$\pm$0.0005,
and a reduced $\chi^2$ of 0.5 for 12 d.o.f. (Fig. \ref{fig-IGRJ17091}c).
This is consistent with the values found by Captanio et al. (\cite{capitanio06}).

\subsection{GX 354-0}
\label{sec-GX354-0}

\begin{figure}
  \centering
  \includegraphics[angle=-90,width=9cm]{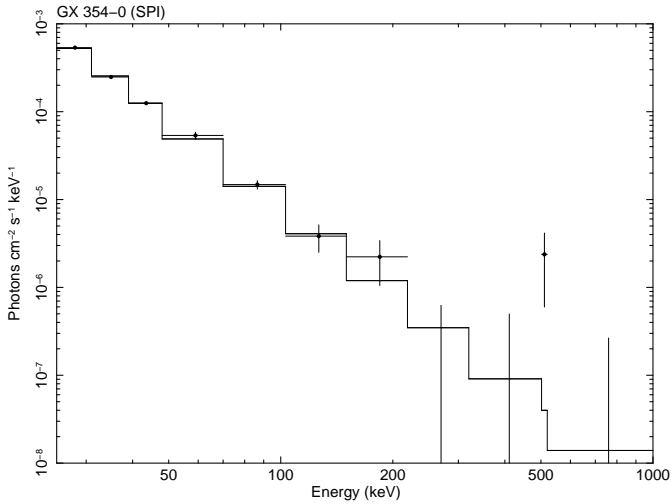}
  \includegraphics[angle=-90,width=9cm]{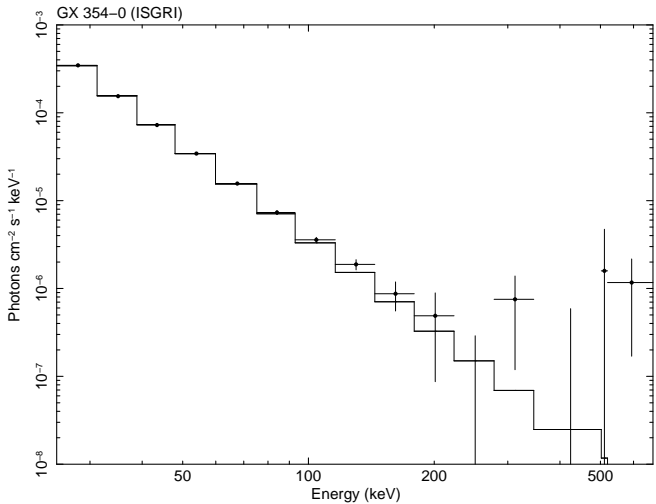}
  \caption{ The unfolded spectra of GX~354-0 derived from dataset 9 (see table \ref{tab-sourcedata}) with 
  statistical: {\bf (a)} (top) SPI spectrum with powerlaw fit, 
  {\bf(b)} (bottom) ISGRI spectrum fit with powerlaw.  
  The fit parameters are given in Sect. \ref{sec-GX354-0}.
  The combined significance of the emission above 200~keV from the SPI and ISGRI data is 3~$\sigma$.
}
  \label{fig-GX354-0}
  \end{figure}

The LMXB GX~354-0, also known as 4U~1728-34, consists of an accreting neutron star and
probably a main sequence star and has been classified as a bursting atoll source
which implies significant spectral variability.
It has been studied by most X-ray observatories including {\it INTEGRAL} 
(see Falanga et al. \cite{falanga06} and references therein).
Above 25 keV, the spectrum has been typically modelled with a thermal Comptonisation
spectrum. However, the studies do not extend much beyond 100 keV.

In our spectral analysis of the time-averaged spectrum, we initially fit a
simple powerlaw. For the SPI data, this already gives a satisfactory fit (Fig. \ref{fig-GX354-0}a)
with a reduced $\chi^2$ of 0.7 for 9 d.o.f. and best fit parameters
$\Gamma$ = 3.26$\pm$0.06 and Norm = 26.4$\pm$5.9 .
The same is true if we fit a powerlaw to the ISGRI data (Fig. \ref{fig-GX354-0}b): We obtain
a reduced $\chi^2$ of 1.0 for 13 d.o.f., and parameters $\Gamma$ = 3.53$\pm$0.02, Norm
= 43.3$\pm$3.0, steeper than the SPI spectrum.

Falanga et al.(\cite{falanga06}) find that a thermal Comptonisation
model (Xspec model COMPTT) describes the data well. 
Lacking soft X-ray information we  take the values for
the soft photon temperature $kT_0$ and the plasma temperature  $kT$ from the recent
measurement by  Falanga et al. (for the low-hard state),
$kT_0$=1.18~keV and $kT$=35~keV, and fix them in our fit.
Applied to the SPI data this results in a reduced $\chi^2$ of 1.9 for 9 d.o.f. and
unevenly scattered residuals suggesting a hard tail. 
However, the sensitivity of our measurement does not permit a more detailed study of this feature.
Also the ISGRI data are not compatible with the pure COMPTT spectrum proposed by Falanga et al. Here we get
a reduced $\chi^2$ of 7.6 for 11 d.o.f. even if we let all model parameters vary freely.

We conclude that our data are not well compatible
with a pure thermal Comptonisation spectrum since it underpredicts the high-energy emission. 
The combined significance of the emission above 200~keV from GX~354-0 is, however, only 3~$\sigma$.
Possibly, the averaging over several spectral states mimics a power law.
    
\subsection{1E 1740.7-2942}  
\label{sec-1E17407}

\begin{figure}
  \centering
  \includegraphics[angle=-90,width=9cm]{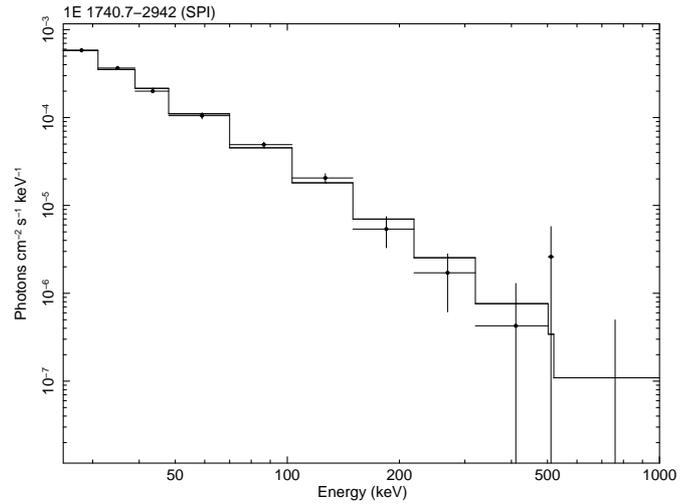}
  \includegraphics[angle=-90,width=9cm]{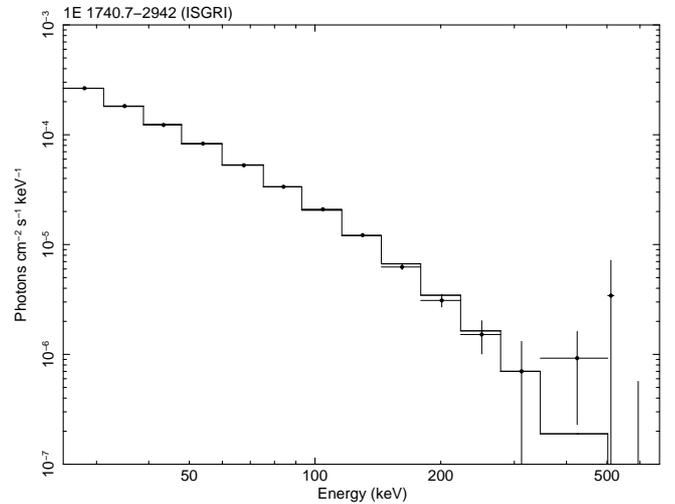}
  \caption{ The unfolded spectra of 1E 1740.7-2942 derived from dataset 9 (see table \ref{tab-sourcedata}) with 
  statistical errors: {\bf (a)} (top) SPI spectrum, 
  {\bf(b)} (bottom) ISGRI spectrum, in both cases fit by a powerlaw with exponential cutoff.
  The fit parameters are given in Sect. \ref{sec-1E17407}.
}
  \label{fig-1E17407}
  \end{figure}

The black hole candidate 1E 1740.7-2942 is a bright persistent source
very near the Galactic Centre and the first source to be classified as a microquasar
(see Bosch-Ramon et al. \cite{bosch06} for a recent review). Its classification
as a LMXB is controversial. 
Like other BHCs, 1E~1740.7-2942 exhibits low-hard and high-soft states.
Using ISGRI data, Del Santo et al. (\cite{delsanto05}) find that the average spectrum above 25~keV
is approximately described by a powerlaw of index $\Gamma$ = 2.3 .
In individual periods they find harder spectra ($\Gamma \approx$~1.5) with
exponential cutoffs around 100~keV which can alternatively be described 
with thermal Comptonisation models.  

Bouchet et al. (\cite{bouchet91}), using {\it GRANAT}/SIGMA data, have reported a transient 
additional component (modelled by a Gaussian with FWHM of 240~keV)  peaking at 480~keV which 
appeared for about a day in the spectrum and then disappeared. This observation has not
yet been confirmed and our study would not pick up on such features since we are averaging
over much longer timescales (unless such events would occur frequently). 

We fit our SPI spectrum with a powerlaw and exponential cutoff (Fig. \ref{fig-1E17407}a) and obtain
a reduced $\chi^2$ of 1.4 for 10 d.o.f. with $\Gamma$ = 2.17$\pm$0.14 in agreement
with the average spectrum by Del Santo et al. (\cite{delsanto05}). The normalisation
is Norm = 0.9$\pm$0.4. The cutoff energy is a badly constrained 434$\pm$400 keV. 

The same model fit to the ISGRI data gives best fit parameters of
$\Gamma$ = 1.44$\pm$0.04, $E_{\mathrm{cutoff}}$ = (113$\pm$8)~keV,
Norm = 0.040$\pm$0.005, and a reduced $\chi^2$ of 0.7 for 12 d.o.f.

The ISGRI spectrum is in agreement with spectra found by Del Santo et al. (\cite{delsanto05})
(see above) but the agreement with SPI is at best marginal.
In order to reconcile the two results, we fix the exponential cutoff in the model for the SPI data
at the value determined by ISGRI. We obtain a reduced $\chi^2$ of
1.9 for 9 d.o.f. and best fit parameters $\Gamma$ = 1.80$\pm$0.06, Norm = 0.28$\pm$0.06.
The residuals suggest an additional hard component. But our sensitivity
is not good enough to warrant a more detailed study.
The agreement of the SPI results with the fluxes reported by B08 
is very good (see Table \ref{tab-summary2}).

\subsection{IGR J17464-3213}
\label{sec-IGRJ17464}

\begin{figure}
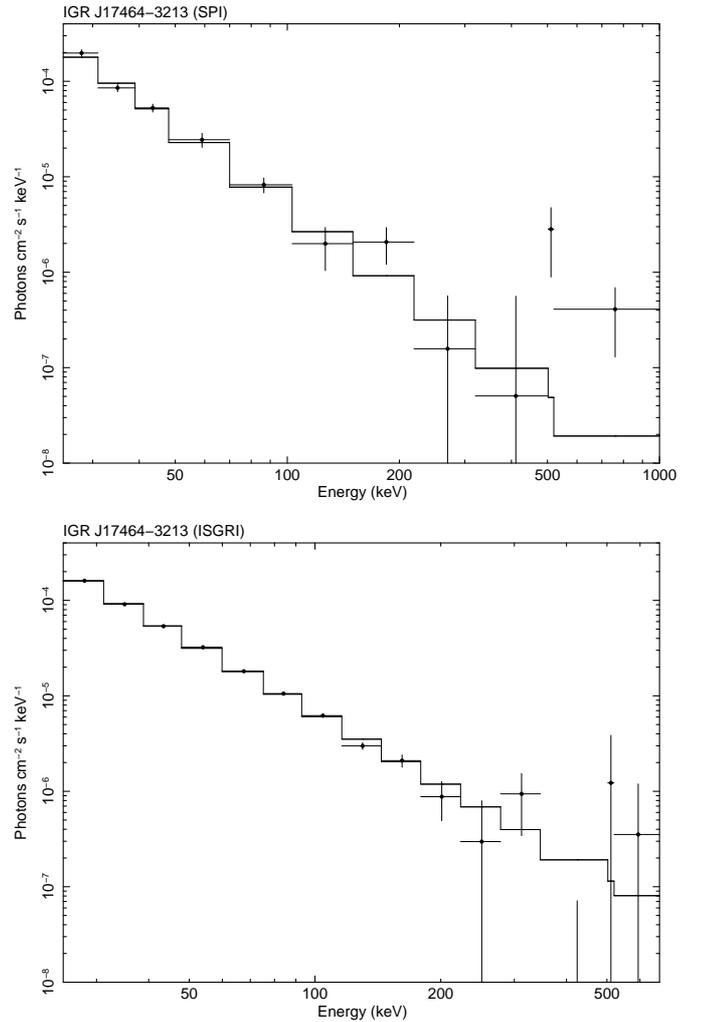

  \centering
  \includegraphics[angle=-90,width=9cm]{1284416a.eps}
  \includegraphics[angle=-90,width=9cm]{1284416b.eps}
  \caption{ The unfolded spectra of IGR J17464-3213 (a.k.a. H~1743-322) 
   derived from dataset 9 (see table \ref{tab-sourcedata}) with 
  statistical errors: {\bf (a)} (top) SPI spectrum with powerlaw fit, 
  {\bf(b)} (bottom) ISGRI spectrum with same model as (a). 
   The fit parameters are given in Sect. \ref{sec-IGRJ17464}.
   The significance of the emission above 200 keV is 3~$\sigma$.
}
  \label{fig-IGRJ17464}
  \end{figure}

The transient IGR~J17464-3213 was discovered by {\it INTEGRAL} in 2003  (Revnivtsev et al. \cite{revnivtsev03a})
and soon identified as the counterpart of the {\it HEAO~1} X-ray source H~1743-322.
The source was subsequently studied by several authors at X-ray, optical and radio wavelengths,
in particular with {\it INTEGRAL} (see Joinet et al. \cite{joinet05} for a recent review). 
Evidence for jets was found and the source is now regarded to be a microquasar (Corbel et al. \cite{corbel05}).
Timing and spectral properties of the object suggest that it is a LMXB containing a black hole
of about 10~M$_{\odot}$.

The {\it RXTE}/ASM lightcurve of IGR~J17464-3213 (e.g. Capitanio et al. \cite{capitanio06a})
shows that the source can be in strong outburst for months at a time. In between outbursts it
returns to a very low level of emission and remains there for months. Using data from {\it RXTE} and {\it INTEGRAL}
up to 150~keV, Joinet et al. (\cite{joinet05}) have studied the complex spectral variability of the object
(typical of BHCs): 
the source exhibits softening during flux increases followed by  phases of constant spectrum in spite 
of intensity changes by factors 2-3. Above a few ten keV, the spectrum is described by a powerlaw
with index $\Gamma$ varying between 1.0 and 2.6, and a not very well constrained exponential cutoff
around 100~keV. 

The ISGRI lightcurve (Courvoisier et al. \cite{isgri-lightcurves}) for IGR~J17464-3213 shows 
three strong outburst periods
and otherwise a low level of emission such that our time-averaged spectrum should be a superposition
of the low-hard and the high-soft state. 
When examining our SPI spectrum, we find evidence for increased  uncertainties in the
lowest two energy bins possibly due to to insufficient source region modelling at the lowest energies. 
To be on the safe side, we increase the statistical errors by 5~\% of the flux for the
entire spectrum. Fitting the spectrum with a powerlaw (Fig. \ref{fig-IGRJ17464}a), then leads to a reduced $\chi^2$ of
1.1 for 9 d.o.f. and best fit parameters $\Gamma$ = 2.82$\pm$0.14, Norm = 2.1$\pm$1.1 . 

Fitting a powerlaw to the ISGRI spectrum of (Fig. \ref{fig-IGRJ17464}c)
gives a perfect reduced $\chi^2$ of 0.9 and best fit parameters $\Gamma$ = 2.49$\pm$0.02
and Norm = 0.63$\pm$0.04. There is no evidence for a cutoff.
The powerlaw index is still in agreement with our SPI measurement.

\subsection{GRS 1758-258}    
\label{sec-GRS1758}

\begin{figure}
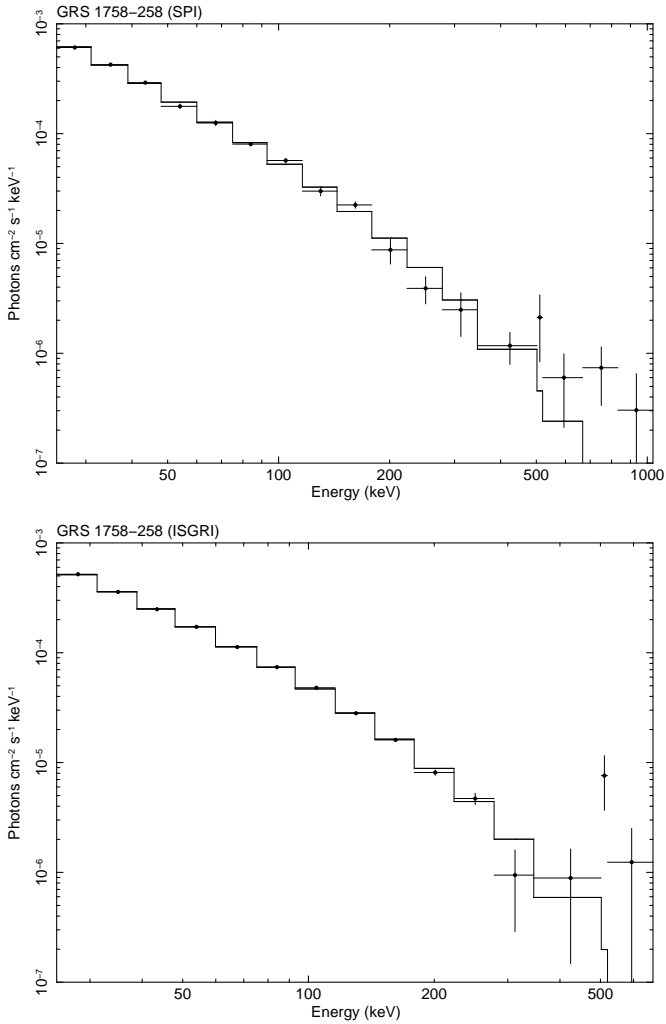

  \centering
  \includegraphics[angle=-90,width=9cm]{1284417a.eps}
  \includegraphics[angle=-90,width=9cm]{1284417b.eps}
  \caption{ The unfolded spectra of GRS~1758-258 
   derived from dataset 10 (see table \ref{tab-sourcedata}) with 
  statistical errors: {\bf (a)} (top) SPI spectrum, and 
  {\bf(b)} (bottom) ISGRI spectrum, both fit by powerlaw with exponential cutoff. 
   The fit parameters are given in Sect. \ref{sec-GRS1758}.
}
  \label{fig-GRS1758}
  \end{figure}

The LMXB GRS~1758+258 is one of the well known microquasars and regarded to be
similar to 1E~1740.7-2942. It is also a black hole candidate. For a recent review 
see e.g. Pottschmidt et al. (\cite{pottschmidt06}). The source is bright and
can show outburst which are however mostly too short for SPI's sensitivity. 
B08 and our analysis assume it to be a constant emitter. 
In the low-hard state, which the
source mostly assumes, the high energy spectrum can roughly be described by a hard 
powerlaw with an exponential cutoff around 100~keV. In terms of physical models,
thermal Comptonisation models and models involving black body emission from a disk 
have been applied. The spectrum above 200~keV was hardly studied.   

We fit our SPI spectrum initially with a powerlaw with exponential cutoff and obtain
a reduced $\chi^2$ of 1.7 for 14 d.o.f. (Fig. \ref{fig-GRS1758}a) and well constrained
best fit parameters $\Gamma$ = 1.54$\pm$0.07, $E_{\mathrm{cutoff}}$ = (178$\pm$28)~keV,
Norm = 0.12$\pm$0.03, in agreement with previous measurements.

The same spectral fit applied to the ISGRI data results in best fit parameters
$\Gamma$ = 1.36$\pm$0.02, $E_{\mathrm{cutoff}}$ = (123$\pm$5)~keV,
Norm = 0.059$\pm$0.004, and reduced $\chi^2$ = 1.9 for 12 d.o.f.,
in marginal agreement with the SPI result.

In both cases, the fit quality is not very good which may be the result of
assuming the source to be constant (but we chose to follow B08 consistently). 
In the SPI spectrum, there is
a hint of a hard tail in the residuals. In fact, we achieve a fit improvement 
(reduced $\chi^2$ = 1.6 for 13 d.o.f.) if we freeze the cutoff energy and add a
powerlaw as a second component. However, the chance probability of the improvement
is 0.2. 

\subsection{Ginga 1826-24}   
\label{sec-Ginga1826}

\begin{figure}
  \centering
  \includegraphics[angle=-90,width=9cm]{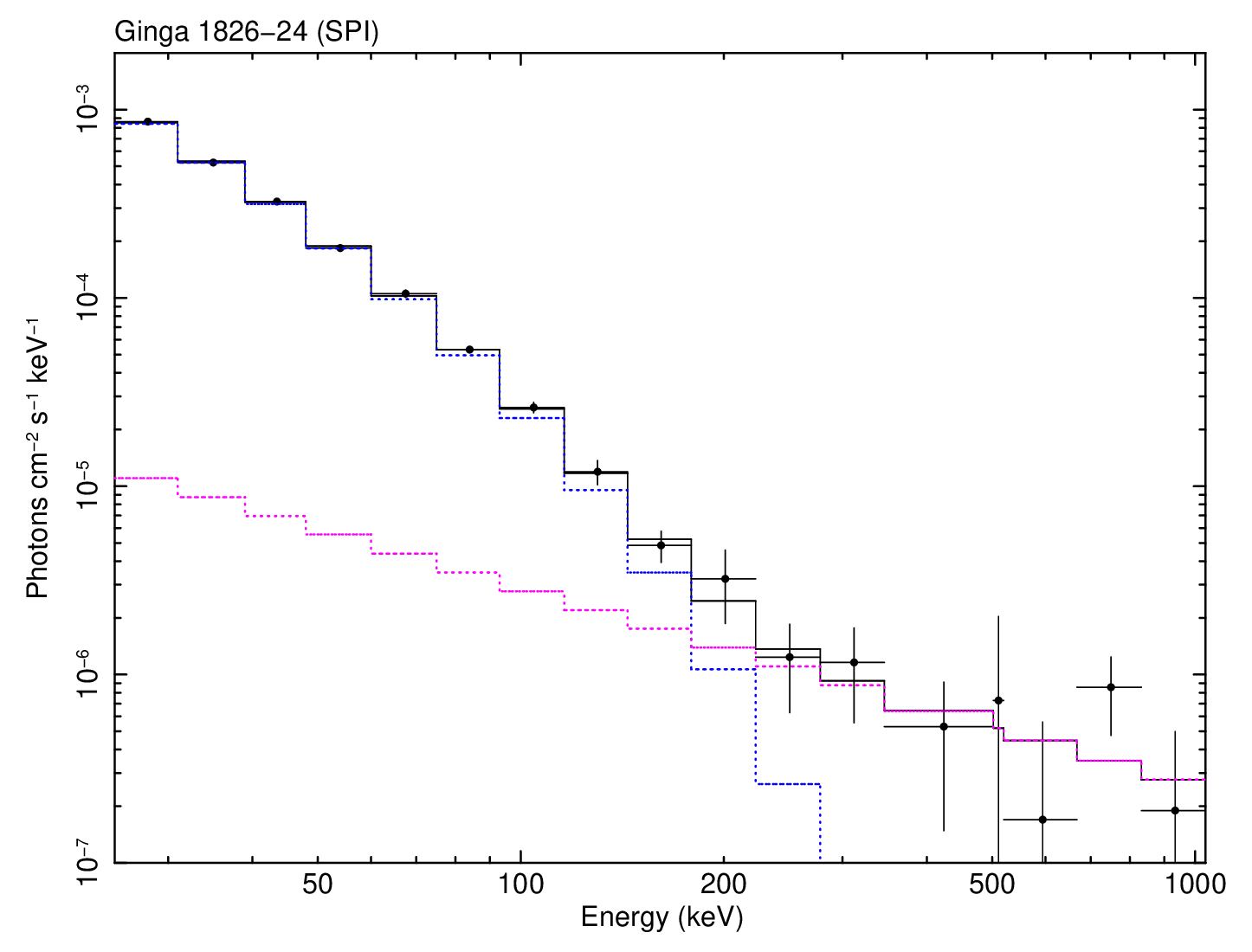}
  \includegraphics[angle=-90,width=9cm]{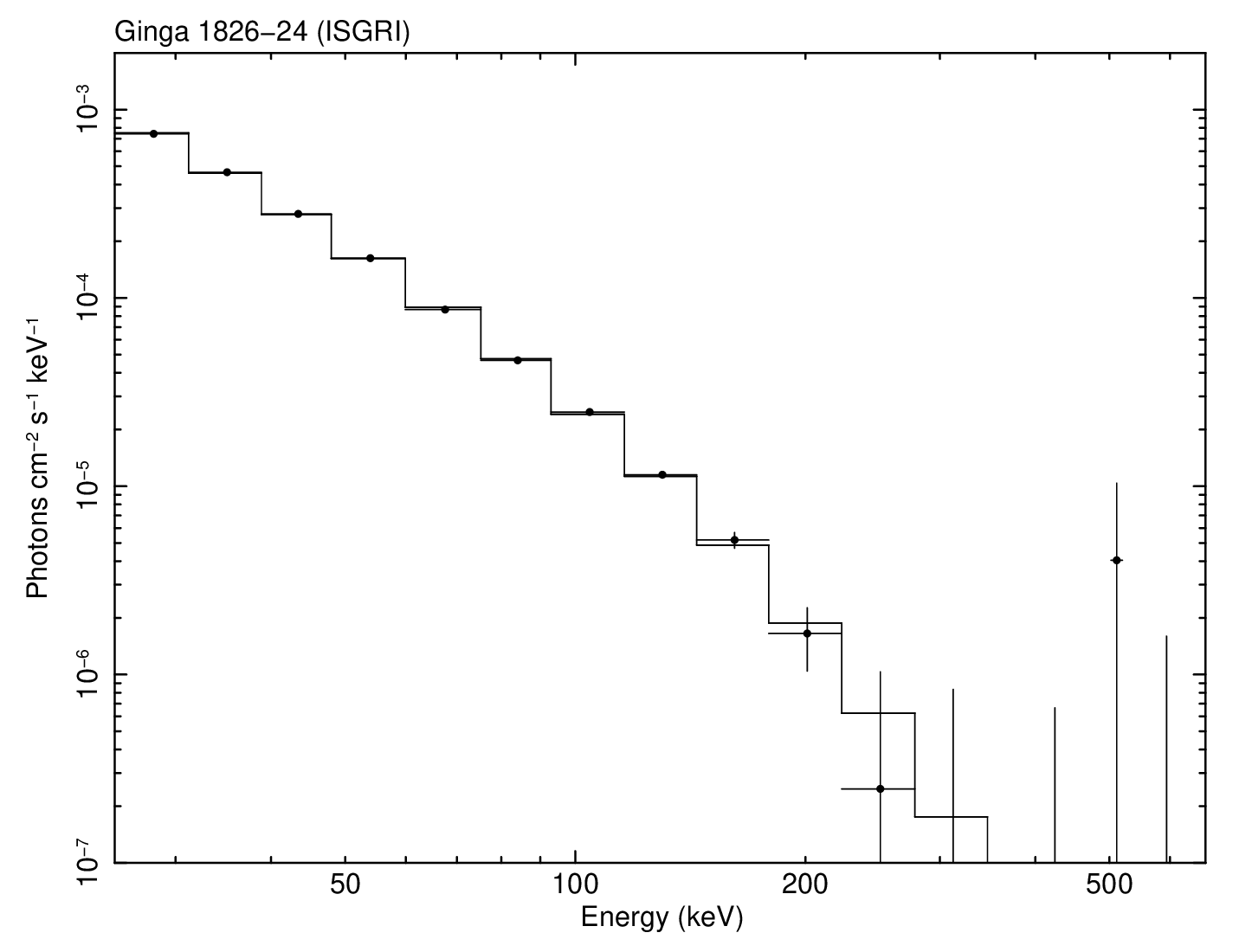}
  \caption{ The unfolded spectra of Ginga~1826-24
   derived from dataset 10 (see table \ref{tab-sourcedata}) with 
  statistical errors: 
  {\bf(a)} (top) SPI spectrum fit with two-component model consisting of a  powerlaw with exponential cutoff 
  and a second powerlaw, 
  {\bf(b)} (bottom) ISGRI spectrum fit with powerlaw with exponential cutoff. 
   The fit parameters are given in Sect. \ref{sec-Ginga1826}.
}
  \label{fig-Ginga1826}
  \end{figure}

The LMXB Ginga 1826-24 was discovered in 1988 and initially found similar in spectrum and variability
to the BHC Cyg~X-1. Not much later, the relatively low cutoff energy ($\approx$ 58~keV) of its powerlaw spectrum
above a few ten keV and the observation of X-ray bursts led to the classification
as a NS. The exceptional precision of the reoccurrence of the bursts (period $\approx$ 3.5~h) also led to the
nickname ``clocked burster''. For a recent review of the source properties see Thompson et al. (\cite{thompson05}).
The time-averaged high-energy spectrum up to about 300~keV was measured by Strickman et al. (\cite{strickman96a})
and modelled as a powerlaw with exponential cutoff. They found $\Gamma$ = 1.59$\pm$0.02 and
$E_{\mathrm{cutoff}}$ = (57$\pm$8)~keV.

Fitting our SPI spectrum of Ginga~1826-24 with a powerlaw with exponential cutoff, we obtain
a reduced $\chi^2$ of 1.2 for 14 d.o.f. and best fit parameters
$\Gamma$ = 1.7$\pm$0.1, $E_{\mathrm{cutoff}}$ = (60$\pm$7)~keV, Norm = 0.36$\pm$0.10
in good agreement with Strickman et al. (\cite{strickman96a}).

The residuals show evidence for a hard tail above 200~keV.
Accommodating the possibility of such a tail by adding a powerlaw component to the model,
we obtain from the fit to the SPI data (Fig. \ref{fig-Ginga1826}a) a reduced $\chi^2$ of 0.5 for 12 d.o.f. and
best fit parameters Norm$_1$ = 0.24$\pm$0.09, $\Gamma_2$ = 0.8$\pm$0.69,
Norm$_2$ (at 200~keV) = $(2.1\pm1.8)\times10^{-6}$cm$^{-2}$s$^{-1}$keV$^{-1}$.
The second powerlaw component becomes dominant above ca. 300~keV.

Applying the same modelling to the ISGRI data (Fig. \ref{fig-Ginga1826}b), we obtain for a simple powerlaw
with exponential cutoff a reduced $\chi^2$ of 1.5 for 12 d.o.f. and best fit parameters
$\Gamma$ = 1.78$\pm$0.04, $E_{\mathrm{cutoff}}$ = (69$\pm$3)~keV, Norm = 0.42$\pm$0.05.
The residuals don't show any posibility for improvement.

We conclude that our time averaged spectra agree well with previously
published results. SPI and ISGRI data are compatible with each other.
The evidence for a hard tail above 200 keV is weak even though B08
still find 25 mCrab at 200-600~keV. We find less than half of that flux.

\subsection{GRS 1915+105}    
\label{sec-GRS1915}

\begin{figure}
  \centering
  \includegraphics[angle=-90,width=9cm]{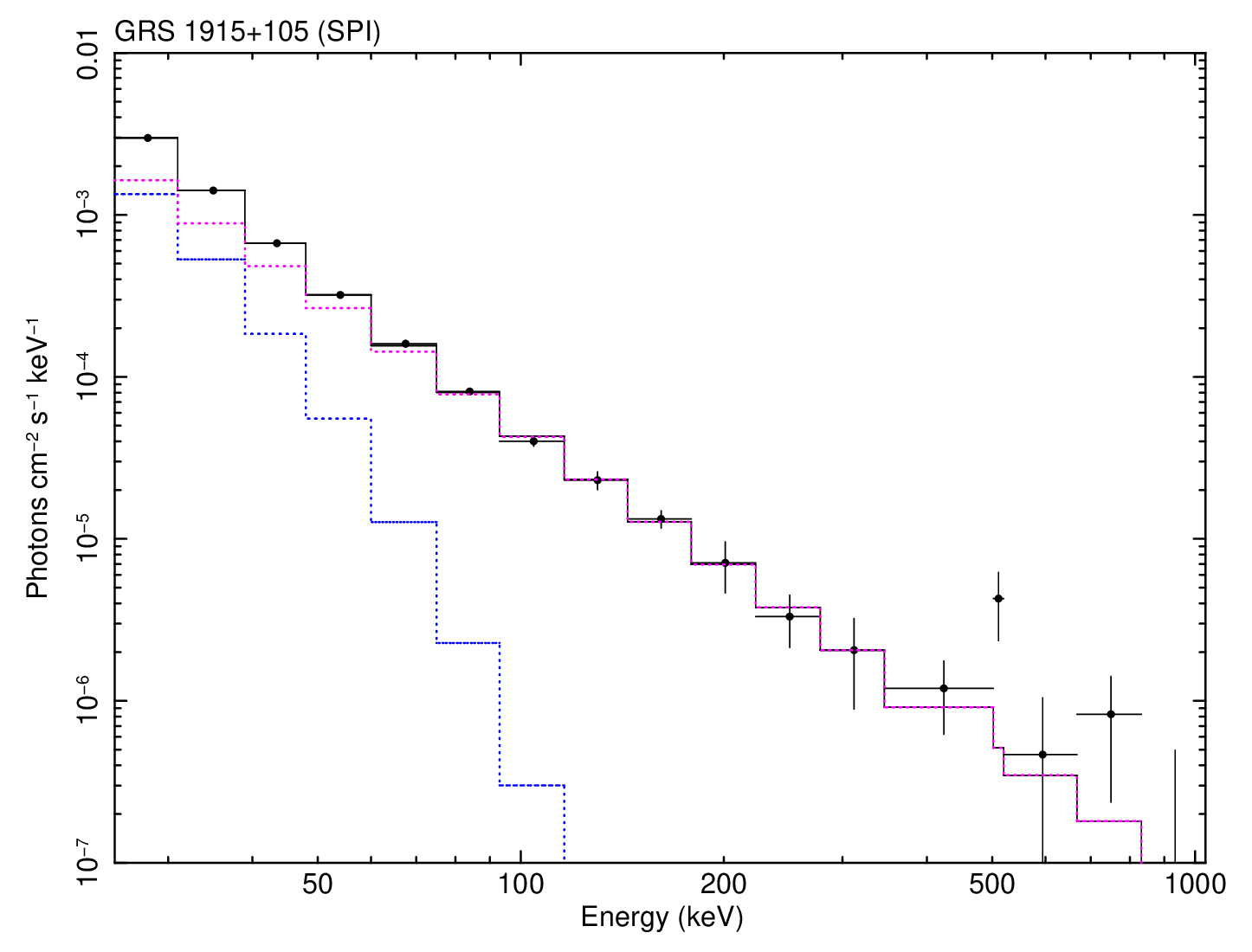}
  \includegraphics[angle=-90,width=9cm]{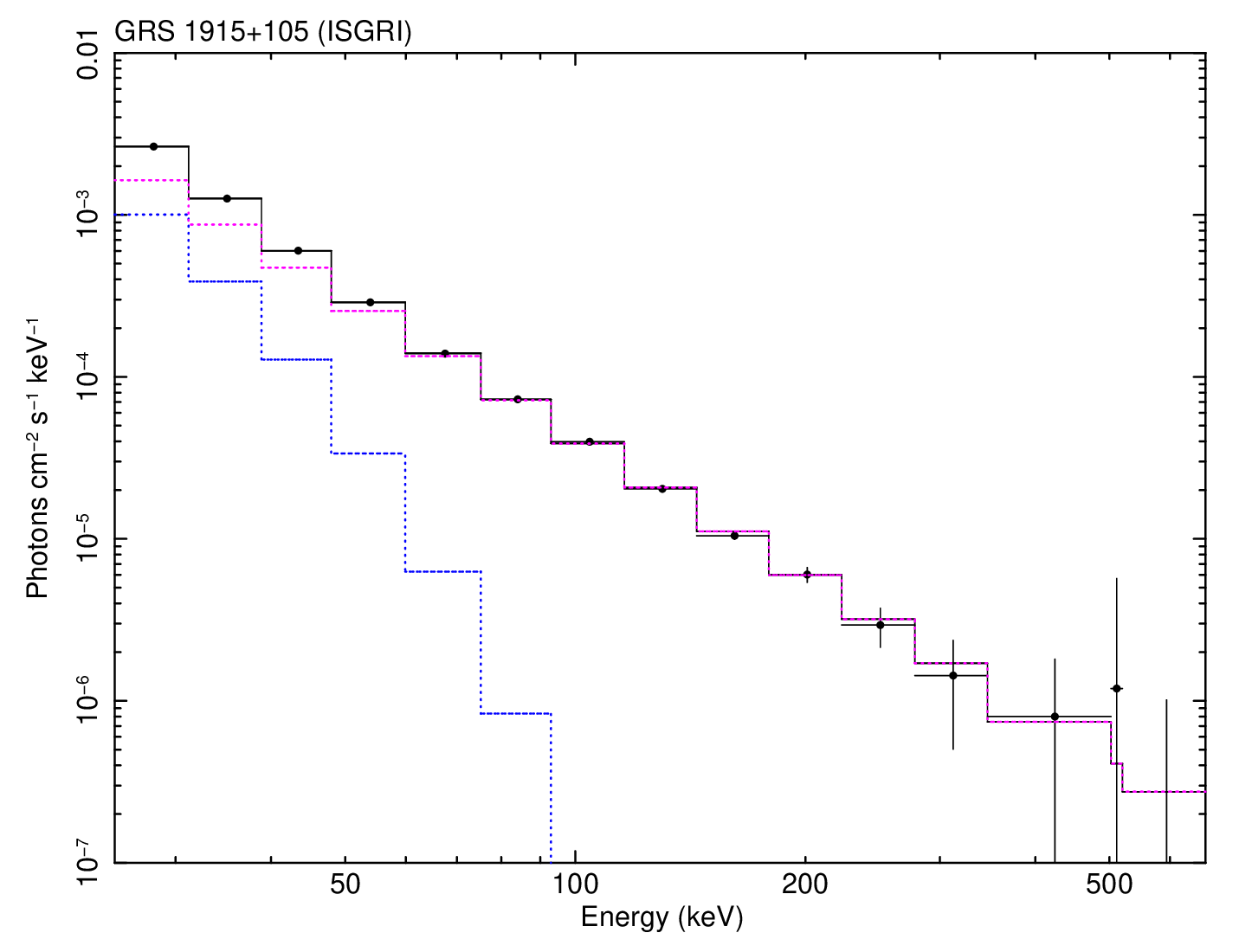}
  \caption{ The unfolded spectra of GRS~1915+105 
   derived from dataset 11 (see table \ref{tab-sourcedata}) with 
   statistical errors: {\bf (a)} (top) SPI spectrum fit with two-component 
   model consisting of powerlaw with exponential cutoff and second powerlaw, 
  {\bf(b)} (bottom) ISGRI spectrum fit by same model as used in (a). 
   The fit parameters are given in Sect. \ref{sec-GRS1915}.
}
  \label{fig-GRS1915}
  \end{figure}

The LMXB GRS~1915+105 is one of the most studied galactic black holes. For a review see e.g. Fender \& Belloni (\cite{fender04}) or
Rodriguez et al. (\cite{rodriguez08}). The object is a microquasar, the first to exhibit superluminal motion in its jets,
with a relatively large black hole mass of $\approx$ 14 M$_\odot$, a large binary period of 33.5~d,  
and an extremely complex time variability which is attributed to accretion disk instabilities (Greiner et al. \cite{greiner96}, Done et al. \cite{done04}).

GRS~1915+105 was extensively monitored by {\it INTEGRAL}. A detailed study of two years of the INTEGRAL observations was published
in  Rodriguez et al. (\cite{rodriguez08, rodriguez08b}). They analyse ISGRI spectra up to 300~keV and distinguish 
several spectral states which they fit with a number of multi-component models involving for the high energy
part thermal Comptonisation (Xspec model COMPTT) with seed photon temperatures between 0.7 and 2.1~keV
and a powerlaw with index between 2.0 and 2.9 .

Our time-averaged spectrum of the source is a superposition of all spectral states. As for the previous sources
we initially try to fit a powerlaw with exponential cutoff. This results in a high reduced $\chi^2$ of 2.4 for 14 d.o.f.,
a photon index of 3.30$\pm$0.02, and a cutoff energy far beyond the upper end of our energy range.
The residuals indicate that a more sophisticated model is needed.

Next we try a powerlaw with exponential cutoff with an additional powerlaw.
This results in a reduced $\chi^2$ of 0.6 for 12 d.o.f. and badly constrained best fit parameters
$\Gamma_1$ = 1.7$\pm$1.7, $E_{\mathrm{cutoff}}$ = (13$\pm$9)~keV, $\Gamma_2$ = 2.8$\pm$0.2,
Norm$_1$ = 3.7$\pm$16, and Norm$_2$ (at 200~keV) = $(1.4\pm0.1)\times 10^{-5}$cm$^{-2}$s$^{-1}$keV$^{-1}$, 
i.e. a dominant pure powerlaw
with a weak additional low-energy component becoming negligible at a few ten keV.
 
Applying the same model to the ISGRI data results in  $\Gamma_2$ = 2.87$\pm$0.05 and Norm$_2$ (at 200 keV) = 
$1.17\pm0.04\times 10^{-5}$cm$^{-2}$s$^{-s}$keV$^{-1}$ for the dominant component and
$\Gamma_1$ = 0.8$\pm$1.0, $E_{\mathrm{cutoff}}$ = 8$\pm$2~keV, Norm$_1$ = 0.09$\pm$0.25 for the weak low-energy
component. The reduced $\chi^2$ is 0.7 for 10 d.o.f.
The agreement with the SPI result is good.

\subsection{Cyg X-1} 
\label{sec-CygX-1}

\begin{figure}
  \centering
  \includegraphics[angle=-90,width=9cm]{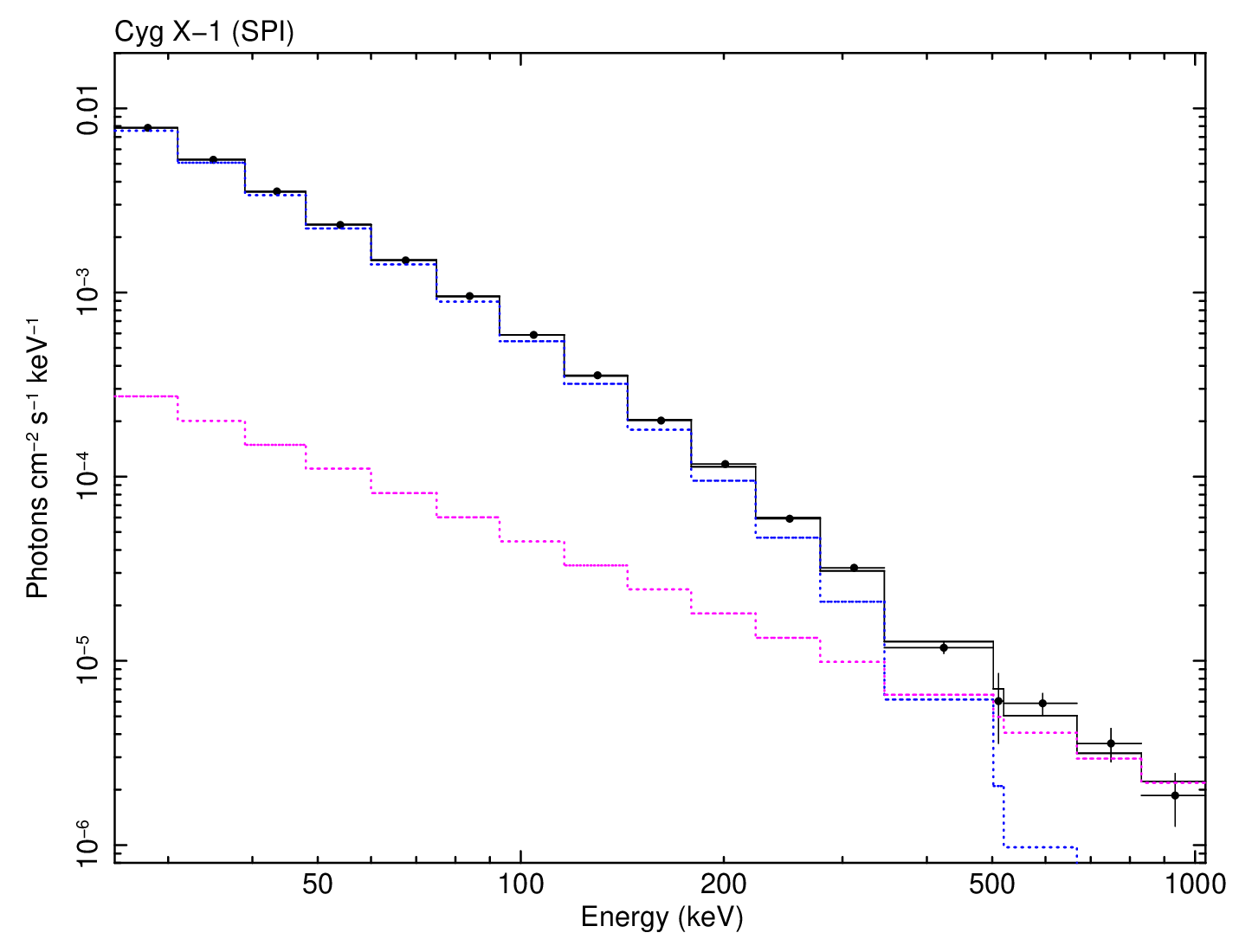}
  \includegraphics[angle=-90,width=9cm]{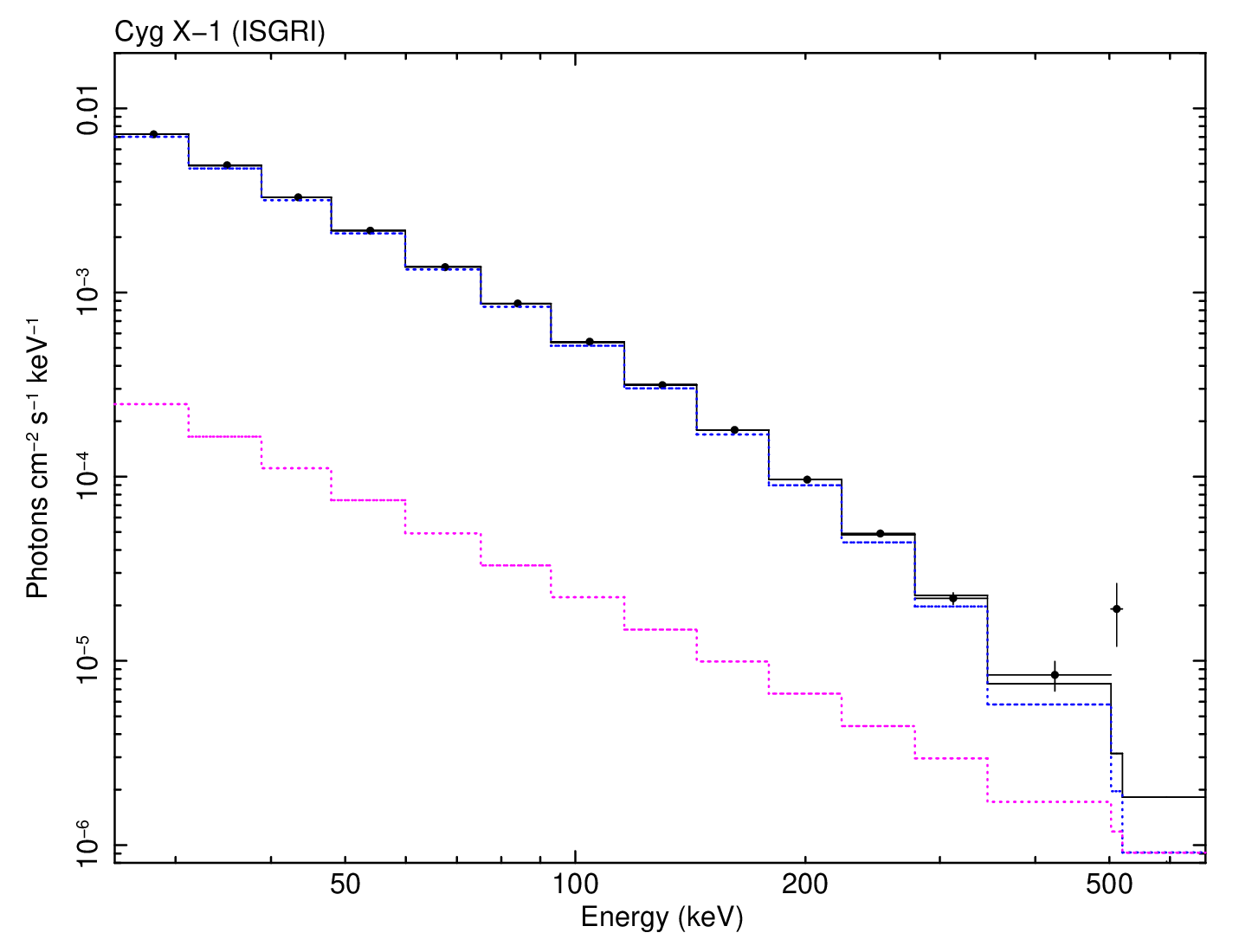}
  \caption{ The unfolded spectra of Cyg~X-1
   derived from dataset 12 (see table \ref{tab-sourcedata}) with 
  statistical errors: {\bf (a)} (top) 
 SPI spectrum fit with two-component model consisting of a  powerlaw with exponential cutoff 
  and a second powerlaw, 
  {\bf(b)} (bottom) ISGRI spectrum fit by same model as in (a) . 
   The fit parameters are given in Sect. \ref{sec-CygX-1}.
}
  \label{fig-CygX-1}
  \end{figure}

The HMXB Cyg~X-1 consists of a black hole candidate
of high but uncertain mass and an OB supergiant
(see Gies et al. \cite{gies03}, Ziolkowski \cite{ziolkowski05},`<
and references therein).  It is one of the brightest hard X-ray sources
and has been studied extensively by many authors in all wavebands.
For a brief review see e.g. Szostek \& Zdziarski (\cite{szostek07}) and
Shaposhnikov \& Titartchuk (\cite{shaposhnikov06}) and references therein. 
The radio emission shows evidence of a moderately relativistic jet. The object
has therefore recently been called a microquasar (e.g. Russell et al. \cite{russell07}).

Long-term monitoring has revealed two main spectral states like in other
BH binaries, the low-hard state and the high-soft state, between which the
source transits. It spends the dominant fraction of time in the low-hard state
(Zhang et al. \cite{zhang97}, Hjalmarsdotter et al. \cite{hjalmarsdotter08}).
Our time-averaged spectrum of Cyg~X-1 should therefore be more similar to the
low hard state which is roughly a hard powerlaw ($\Gamma$ between 1.4 and 2.1)
with a break at $E >$~50~keV (Cadolle Bel et al. \cite{cadollebel06}).

The X-ray spectra of individual states have been modelled up to 300~keV
by several authors with a variety of combinations of disk black body and thermal Comptonisation
with additional Fe fluorescence line and reflection components (see again Cadolle Bel et al. \cite{cadollebel06}).

Lacking the soft X-ray information and averaging over all source states,
we fit our SPI spectrum of Cyg~X-1 initially with a simple powerlaw with exponential cutoff and 
obtain a very large reduced $\chi^2$ of 5.3 for 14 d.o.f.
The best fit parameters ($\Gamma$ = 1.65$\pm$0.01, $E_{\mathrm{cutoff}}$ = (180$\pm$5)~keV,
Norm = 2.3$\pm$0.1) agree well with the expectations from previous measurements and, up to about
300~keV, the model describes the data reasonably well but there is a clear and very significant
hard tail above 300 keV.

This hard tail in the low-hard state spectrum of Cyg~X-1 was already observed up to about 5~MeV  with COMPTEL 
by McConnell et al. (\cite{mcconnell00}). They did not observe an indication of a cutoff.
It was also already studied with {\it INTEGRAL} by Cadolle Bel et al.(\cite{cadollebel06}) 
and Malzac et al. (\cite{malzac}).

Adding a powerlaw to our model to accommodate the hard tail, but leaving all parameters free, 
we obtain a much better fit (Fig. \ref{fig-CygX-1}a) with reduced $\chi^2$ 0.9 for 12 d.o.f.
and residuals which don't show any sign of systematic deviations.
The best fit parameters are $\Gamma_1$ = 1.3$\pm$0.2,  $E_{\mathrm{cutoff}}$ = (95$\pm$15)~keV,
Norm$_1$ = 0.43$\pm$0.35, $\Gamma_2$ = 2.11$\pm$0.1, Norm$_2$ (at 200 keV) = 
$(9.9\pm1.8)\times 10^{-5}$~cm$^{-2}$s$^{-1}$keV$^{-1}$.

Applying the same model to the ISGRI data (Fig. \ref{fig-CygX-1}b), we obtain a reduced $\chi^2$ of 1.5 for 10 d.o.f.
and best fit parameters $\Gamma_1$ = 1.44$\pm$0.14,  $E_{\mathrm{cutoff}}$ = (115$\pm$18)~keV,
Norm$_1$ = 0.9$\pm$0.5, $\Gamma_2$ = 2.2$\pm$0.14, Norm$_2$ (at 200~keV) = $(3.6\pm2.3)\times 10^{-5}$~cm$^{-2}$s$^{-1}$keV$^{-1}$,
i.e. the best fit parameters are in agreement with those from SPI but are not as well constrained because
the ISGRI spectrum is lacking high energy sensitivity.

To further compare with the COMPTEL data, we fit a powerlaw to our
SPI data above 500~keV. We find an index of 1.9$\pm$0.5 which is much harder than
 the index 3.3$\pm$0.4 measured by  McConnell et al. (\cite{mcconnell00}) in the neighbouring
energy range. One explanation of this difference could be source variability,
which has indeed been suggested to be present as a variable broad line feature around 1 MeV 
(McConnell et al. \cite{mcconnell89,mcconnell00}).

We conclude that even though our spectra have extremely high statistical significance,
it is possible to model them with a  phenomenological model with only five parameters
consisting of a cutoff powerlaw and second powerlaw.
The cutoff powerlaw can be interpreted as a contribution from thermal Comptonisation
as already suggested by many previous authors (see above) while the hard tail
is most likely a product of the jet.

\subsection{Cyg X-3}         
\label{sec-CygX-3}

\begin{figure}
  \centering
  \includegraphics[angle=-90,width=9cm]{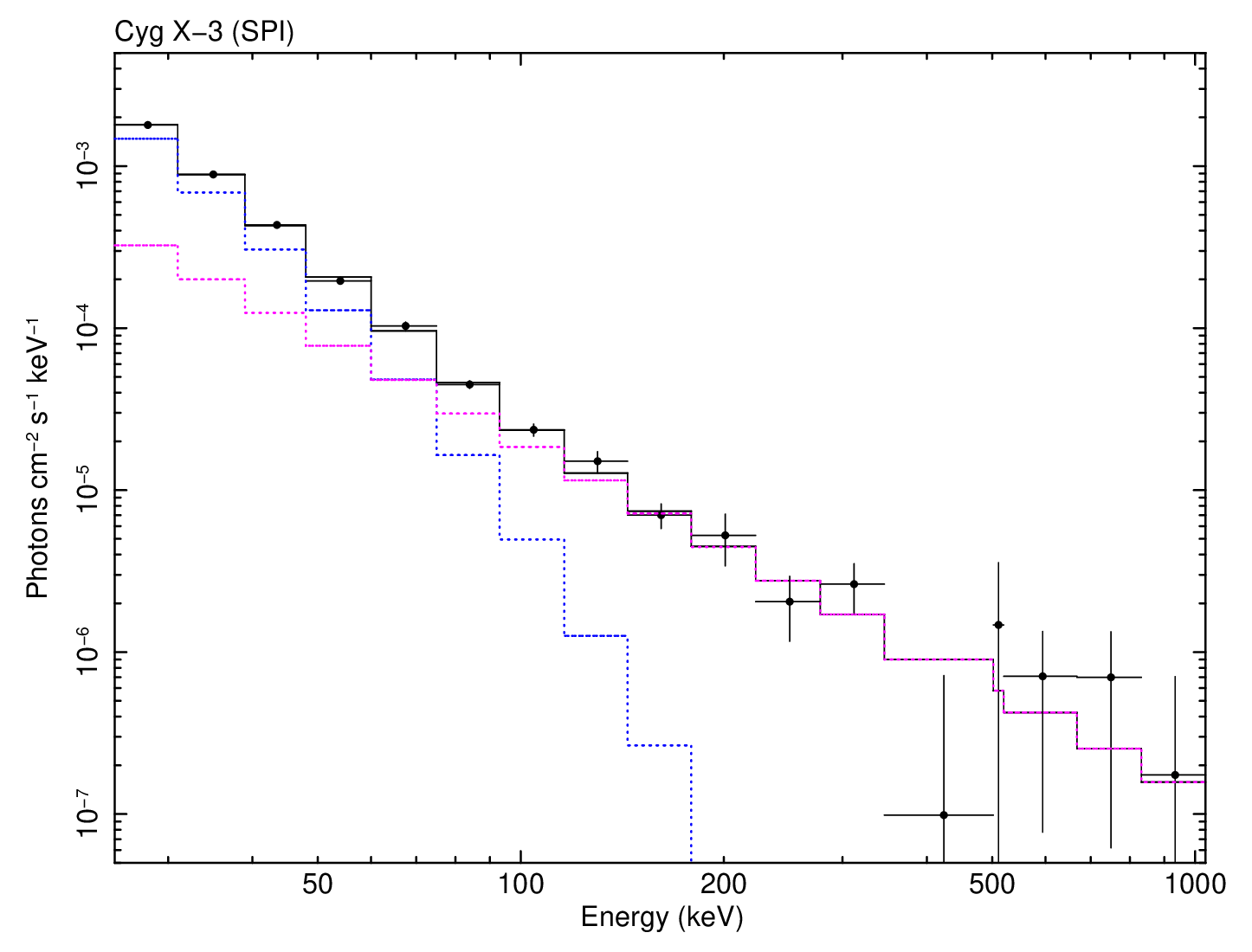}
  \includegraphics[angle=-90,width=9cm]{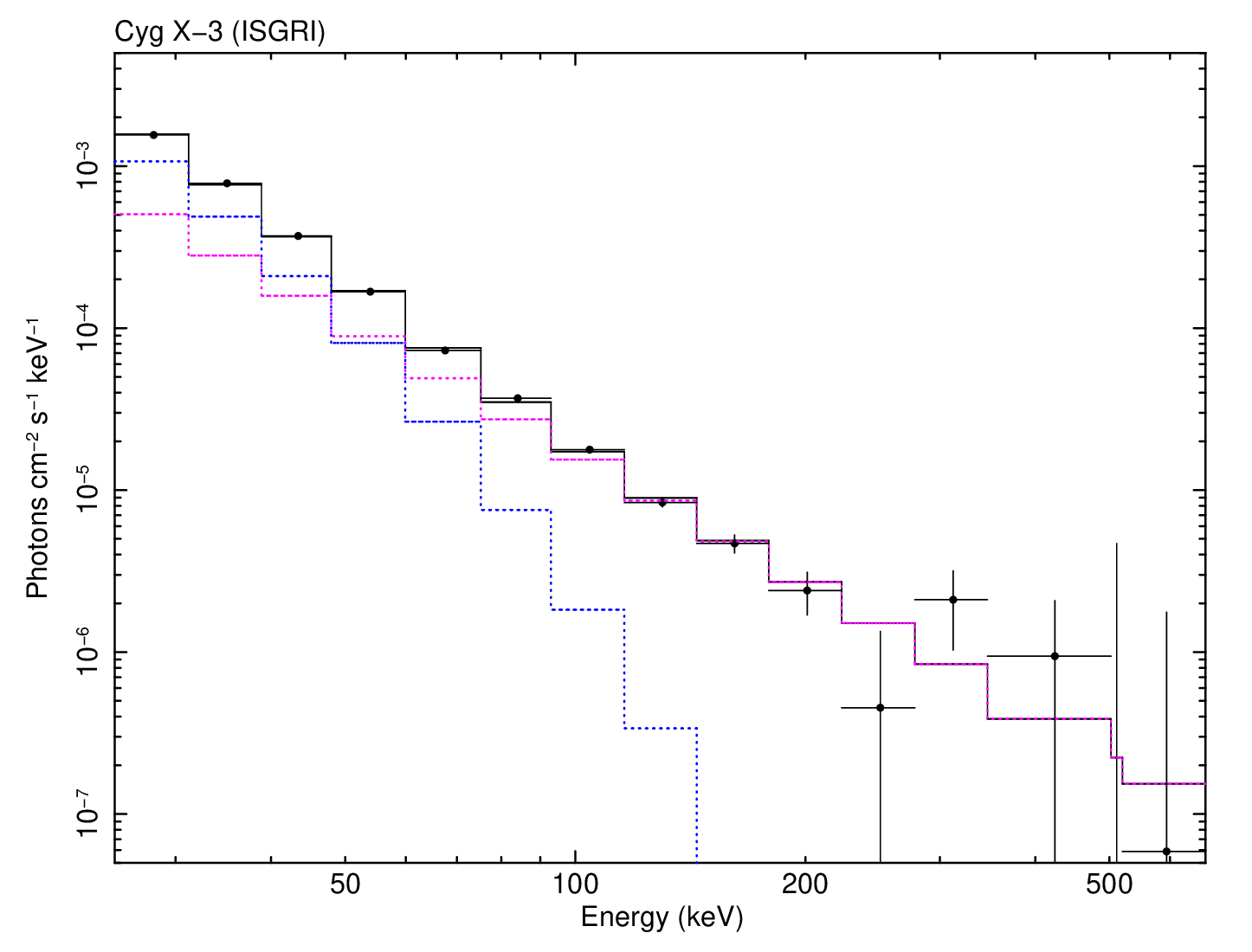}
  \caption{ The unfolded spectra of Cyg~X-3
   derived from dataset 13 (see table \ref{tab-sourcedata}) with statistical errors:
  {\bf(a)} (top) SPI spectrum fit with two-component model consisting of a  powerlaw with exponential cutoff 
  and a second powerlaw, 
  {\bf(b)} (bottom) ISGRI spectrum fit by same model as in (a) . 
   The fit parameters are given in Sect. \ref{sec-CygX-3}.
}
  \label{fig-CygX-3}
  \end{figure}

Cyg~X-3 is a HMXB with an unusually short binary period of 4.8~h and a Wolf-Rayet star as
an unusual stellar companion. Despite its brightness and the intensive study at all wavebands over
four decades, recently published articles still declare many of its properties as uncertain, in
particular the nature of the compact object contained in it. Szostek \& Zdziarski (\cite{szostek08}) and
Hjalmarsdotter et al. (\cite{hjalmarsdotter08}), H08 in the following, give a brief review of the literature. They also
argue that the compact object in Cyg~X-3 could be a quite massive black hole.
The object has also been called a microquasar because of its radio jets (Marti et al. \cite{marti07}).
At lower energies, the intense stellar wind from the companion star leads to strong absorption features.
Above 25~keV, however, the effect on the spectral shape is small (H08).

Cyg~X-3 has a binodal distribution of soft X-ray count rates. The distribution of its spectral hardness,
however, only shows one peak at low hardness with a long asymmetric tail to higher hardness (H08).
The authors note that the soft X-ray hardness in the case of Cyg~X-3 may be a bad indicator of 
the state of the source because of its strong absorption. The source probably
still has a separate hard state at higher energies which it occupies roughly two thirds of the time.

The time-averaged spectra we are studying here should therefore be a superposition of spectra from all states,
but more similar to the harder spectra.

We initially fit our SPI spectrum of Cyg~X-3 with a powerlaw with exponential cutoff.
This results in a relatively high reduced $\chi^2$ of 1.8, an index $\Gamma$ = 3.26$\pm$0.04
and a cutoff beyond the upper end of our energy range.
The residuals suggest a hard tail beginning at about 100~keV.
Adding a powerlaw as second component, we obtain a reduced $\chi^2$ of 0.8 for 12 d.o.f. (Fig. \ref{fig-CygX-3}a)
but the fit parameters are poorly constrained ($\Gamma_1$ = 2.0$\pm$0.7, $E_{\mathrm{cutoff}}$=(20$\pm$10)~keV,
Norm$_1$ = 3.5$\pm$6.6, $\Gamma_2$ = 2.4$\pm$0.3, Norm$_2$ (at 200~keV) = $(9.0\pm1.1)\times10^{-6}$~cm$^{-2}$s$^{-1}$keV$^{-1}$)
because the best fit cutoff energy
is outside our energy range.
The powerlaw component without cutoff constitutes the dominant part of the emission
above a few 10~keV.
There is no evidence for
a cutoff at higher energies, but our data points become insignificant above 350 keV.

An unusually low cutoff energy implying a most likely plasma temperature
of only 4~keV was also found in the {\it INTEGRAL} data analysis by  H08 when assuming a 
thermal Comptonisation model.

When inspecting our ISGRI spectrum of Cyg~X-3, we find evidence for enlarged systematic errors
which are probably caused by the crowded field around the object including the nearby and extremely bright
Cyg~X-1. We increase the statistical errors by 0.5~\% of the flux.
With this change, the fit of the same two-component model as for the SPI data with
$E_{\mathrm{cutoff}}$ fixed to 20~keV results in a reduced $\chi^2$ of 1.7 for 11 d.o.f. and best
fit parameters $\Gamma_1$ = 2.00$\pm$0.08, Norm$_1$ = 3.4$\pm$0.7, $\Gamma_2$ = 2.7$\pm$0.2, Norm$_2$
(at 200~keV) = $(5.3\pm0.5)\times10^{-6}$~cm$^{-2}$s$^{-1}$keV$^{-1}$
(Fig. \ref{fig-CygX-3}b) in agreement with the SPI results.

\subsection{Systematic errors of flux measurements in crowded fields}
\label{sec-control1}

\begin{figure}
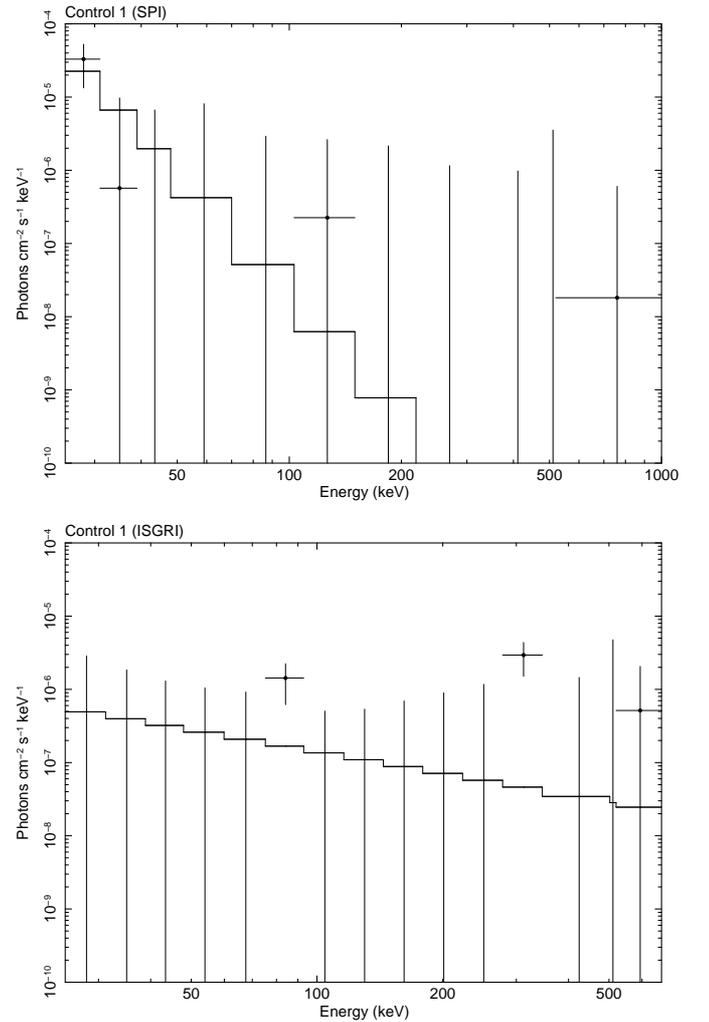

  \centering
  \includegraphics[angle=-90,width=9cm]{1284422a.eps}
  \includegraphics[angle=-90,width=9cm]{1284422b.eps}
  \caption{ The unfolded spectra for the control position ``Control1'' at RA = 259.5$^\circ$, Dec = -40.35$^\circ$ 
   derived from dataset 9 (see Table \ref{tab-sourcedata}) with 
  statistical errors as for all other sources: {\bf (a)} (top) SPI spectrum with powerlaw fit, 
  {\bf(b)} (bottom) ISGRI spectrum with powerlaw fit . 
   The fit parameters are given in Sect. \ref{sec-control1}.
}
  \label{fig-control1}
  \end{figure}

As described in Sect. \ref{sec-syserr} we derive our systematic error estimate from the study of our
Crab spectrum, i.e. from a source which is not in a crowded field as opposed to many of the X-ray binaries studied in this work.
We need to verify to which extent our analysis is subject to additional
systematic errors when applied to sources in crowded fields. In order to assess this possibility,
we have randomly added a source position ``Control1'' to the input catalog for one of our most complex 
fields (dataset 9) near GX~354-0
and 4U~1700-377 (see Table \ref{tab-sourcedata}). This position constitutes the worst case for
our analysis as it is close to variable bright sources in a location in the galactic plane
less than 15$^\circ$ from the galactic center where also significant diffuse emission is present.  
Furthermore, it is about 10~arcmin away from 1RXS~J171824.2-402934, an X-ray burster 
capable of very rare 4 minute bursts with peak intensities up to 1.3 Crab and a thermal spectrum peaking at 5~keV
detectable up to 28~keV (Kaptein et al. \cite{kaptein00}).
As a control case, this position therefore also gives an estimation of the influence of
variable and on average faint sources which are not included in our input catalog. 

The resulting SPI spectrum (Fig. \ref{fig-control1}a) does not contain any point with a flux significantly different
from zero. All points have less than 2~$\sigma$ significance.
The only indication of any systematic signal which could fake the presence of a source
is in the first bin near 30~keV where a level of 3 mCrab is reached.
If we naively fit the points with a powerlaw, we obtain a very steep spectrum with
best fit parameters  $\Gamma$ = 5.5, Norm = 2075.

Repeating the same procedure for the corresponding ISGRI spectrum (Fig. \ref{fig-control1}b) results
in best fit parameters $\Gamma$ = 1.0, Norm = 1.3$\times$10$^{-5}$.  
Also here all points are consistent with zero. The most significant point is
near 300~keV and has 2~$\sigma$ with a nominal flux of approx. 80~mCrab.

The fluxes derived from these powerlaws can serve as
additional estimations of systematic errors on the flux values measured by us.
They are given in Table \ref{tab-summary2} in the row labelled
``Estim. Syst. Error'' with negative sign in order to indicate that the values
need to be subtracted from the other flux measurements in the table.

\section{Summary and conclusions}
\label{secconclusions}

\begin{table*}[b] 
  \caption[]{Summary of the spectra obtained in this study (as shown in the figures of Sect. \protect\ref{secresults}). 
    Part Ia: spectral fit parameters for sources other than X-ray binaries.}
  \label{tab-summary1a}
  \setlength{\tabcolsep}{1mm}
  \begin{tabular}{|l|l|c|c|c|c|c|c|}
Object & Instru- & \multicolumn{6}{c|}{Spectral Model Fit Results$^a$}\\
      &  ment     &$\Gamma_1$&$E_{\mathrm{cutoff}}$&$E_{\mathrm{break}}$& $\Gamma_2$ &Norm & $\chi^2$/d.o.f.\\
      &            &         & $[$keV$]$           & $[$keV$]$           &            &$[$keV$^{-1}$cm$^{-2}$s$^{-1}]$ & \\
\hline
      &SPI         &2.11$\pm$0.01& -             &81$\pm$11               &2.20$\pm$0.01 &11.6$\pm$0.5 & 0.5 \\ 
\raisebox{1.5ex}[-1.5ex]{Crab}&ISGRI &2.119$\pm$0.006& - &96$\pm$4 &2.36$\pm$0.02            & 10.7$\pm$0.3 & 1.2 \\ 
\hline 
                                     & SPI   & 2.13$\pm$0.10 & - & - & - & 0.12$\pm$0.05 & 0.55 \\ 
\raisebox{1.5ex}[-1.5ex]{Vela Pulsar}& ISGRI & 2.04 $\pm$ 0.04&- & - & - & 0.056 $\pm$ 0.008 & 1.4 \\ 
\hline       
                                     & SPI   &  2.10$\pm$0.17 & -  & - & - & 0.4$\pm$0.25 & 1.4\\ 
\raisebox{1.5ex}[-1.5ex]{NGC 4151}&    ISGRI & 2.03$\pm$0.06 & - & - & -  & 0.24$\pm$0.05 & 1.4 \\ 
\hline
                                     & SPI$^b$   & 1.7$\pm$0.17 & 150 & - & - & 0.06$\pm$0.04 & 0.4 \\ 
\raisebox{1.5ex}[-1.5ex]{NGC 4945}&    ISGRI$^b$ & 1.56$\pm$0.05 & 150 & - & - & 0.033$\pm$0.006 & 0.6 \\ 
\hline
                                     & SPI   &  1.89$\pm$0.05  & - & - & - & 0.20$\pm$0.04 & 1.2 \\ 
\raisebox{1.5ex}[-1.5ex]{Cen A}&       ISGRI   &  1.80$\pm$0.02  & - & - & - & 0.117$\pm$0.009 & 0.9  \\ 
\hline
                                     & SPI    & 1.5$\pm$0.2 & - & - & - & 0.004$\pm$0.004 & 1.0 \\ 
\raisebox{1.5ex}[-1.5ex]{Swift J1656.3-3302}&ISGRI$^c$  & - & - & - & - & - & - \\ 
\hline             
  \end{tabular}
  \begin{list}{}{}
  \item[$^a$] Spectral parameters not used in the model for a particular object are marked as ``-''. 
  \item[$^b$] Fit model also included the strong photoelectric absorption with $N_{\mathrm H}$ fixed to 5.5$\times10^{24}$cm$^{-2}$.
  \item[$^c$] ISGRI spectral fit failed, see Sect. \ref{sec-SwiftJ}.
  \end{list}
\end{table*}

\begin{table*} 
  \caption[]{Summary of the spectra obtained in this study (as shown in the figures of Sect. \protect\ref{secresults}). 
     Part Ib: spectral fit parameters for X-ray binaries. }
  \label{tab-summary1b}
  \setlength{\tabcolsep}{1mm}
  \begin{tabular}{|l|l|c|c|c|c|c|c|c|}
Object & Instru- & \multicolumn{7}{c|}{Spectral Model Fit Results$^a$}\\
      &  ment     &$\Gamma_1$ & $E_{\mathrm{cutoff}}$ & $\Gamma_2$&Norm & Norm$_2$ & $\chi^2$/d.o.f. & F-test\\
      &           &           & $[$keV$]$            &           &$[$keV$^{-1}$cm$^{-2}$s$^{-1}]$ &$[10^{-6}$keV$^{-1}$cm$^{-2}$s$^{-1}]$ & & prob.$^b$\\
\hline
                                            & SPI   & 1.36$\pm$0.09 & 206$\pm$50  & - & 0.031$\pm$0.010 & - &1.6 & - \\ 
\raisebox{1.5ex}[-1.5ex]{XTE J1550-564$^c$}& ISGRI  &1.00$\pm$0.05 & 108$\pm$9   & - & 0.009$\pm$0.002& - & 1.3 & - \\  
\hline
                                           & SPI   & 3.3$\pm$0.2 & - & -  & 33$\pm$21 & - & 0.6 & - \\ 
\raisebox{1.5ex}[-1.5ex]{4U 1630-47} &  ISGRI$^d$  &  2.3$\pm$0.2  & -  & - & 0.043$\pm$0.038 & -  & 1.0 & - \\ 
\hline
                                     & SPI    & 1.23$\pm$0.07  & 23.9$\pm$3.2    &  1.27$\pm$0.58 &  0.16$\pm$0.04 & 2.1$\pm$1.0 & 0.7 &  0.015\\
\raisebox{1.5ex}[-1.5ex]{OAO 1657-415$^c$}& ISGRI  &   1.42$\pm$0.31 & 24.7$\pm$3.6   & -  &  0.23$\pm$0.21 & - & 0.4 & - \\ 
\hline
                                     & SPI       & 1.58$\pm$0.11 & 100   & 0.97$\pm$0.63 & 0.081$\pm$0.026 & 3.8$\pm$2.4 & 0.8 & 0.0035\\
\raisebox{1.5ex}[-1.5ex]{GX 339-4$^c$}&   ISGRI  & 1.71$\pm$0.10 & 100   & 1.0         & 0.038$\pm$0.011 & 1.9$\pm$0.8 & 0.7 & -  \\ 
\hline
                                         & SPI    & 2.00$\pm$0.08  &  47.5$\pm$3.9  & - & 3.14$\pm$0.71 & - & 1.6 & - \\
\raisebox{1.5ex}[-1.5ex]{4U 1700-377$^c$}& ISGRI  & 2.31$\pm$0.17  &  74.1$\pm$14.1 & - & 6.4$\pm$3.2 & - & 0.9 & - \\ 
\hline
                                     & SPI   & 1.75$\pm$0.20     & -  & - & 0.016$\pm$0.013 & -  & 0.4 & -  \\
\raisebox{1.5ex}[-1.5ex]{IGR J17091-3624$^c$}&ISGRI  & 1.93$\pm$0.04    & -  & - & 0.034$\pm$0.005 & - & 1.2 & -\\ 
\hline
                                     & SPI   & 3.26$\pm$0.06   & - & - & 26.4$\pm$5.9 & - & 0.7 & -\\
\raisebox{1.5ex}[-1.5ex]{GX 354-0$^c$}&   ISGRI  & 3.53$\pm$0.02   & - & - & 43.3$\pm$3.0 & - & 1.0 & -\\ 
\hline
                                     & SPI   &  1.80$\pm$0.06  & 113.4  & - &  0.28$\pm$0.06 & - & 1.9 &  - \\
\raisebox{1.5ex}[-1.5ex]{1E 1740.7-2942$^c$}&ISGRI  &  1.44$\pm$0.04 & 113.4$\pm$7.6  & - & 0.040$\pm$0.004 & - & 0.7 & - \\ 
\hline
                                     & SPI     &  2.82$\pm$0.14   & - & - & 2.1$\pm$1.1 &  - & 1.1 & - \\
\raisebox{1.5ex}[-1.5ex]{IGR J17464-3213}&ISGRI  & 2.49$\pm$0.02  & - & - & 0.63$\pm$0.04 & -  & 0.9 & - \\ 
\hline
                                          & SPI  & 1.54$\pm$0.07 & 178$\pm$28  & - & 0.12$\pm$0.027  & -  & 1.7 &  - \\
\raisebox{1.5ex}[-1.5ex]{GRS 1758-258$^c$}&ISGRI & 1.36$\pm$0.02 & 123$\pm$5   & - &  0.06$\pm$0.004 & - & 1.9 & - \\ 
\hline
                                     & SPI   & 1.51$\pm$0.14 & 48.4$\pm$7.8   & 0.82$\pm$0.69 & 0.24$\pm$0.09 & 2.1$\pm$1.8 & 0.5 &  0.002 \\
\raisebox{1.5ex}[-1.5ex]{Ginga 1826-24$^c$}&ISGRI  &  1.78$\pm$0.04 & 68.5$\pm$3.5   &  - &  0.42$\pm$0.05 & - & 1.5 & - \\ 
\hline
                                     & SPI   &  1.7$\pm$1.7 & 12.5$\pm$9.2   &  2.76$\pm$0.18 & 4$\pm$15 & 1.4$\pm$0.1 & 0.6 &  9.2$\times$10$^{-5}$ \\
\raisebox{1.5ex}[-1.5ex]{GRS 1915+105$^c$}&ISGRI  & 0.3$\pm$1.1 & 7.6$\pm$2.1   & 2.87$\pm$0.05 & 0.09$\pm$0.25 & 1.17$\pm$0.04 & 0.7 & 7.8$\times$10$^{-9}$  \\ 
\hline
                                     & SPI   &  1.28$\pm$0.19 & 95.9$\pm$15.5   &  2.11$\pm$0.09 & 0.43$\pm$0.35 & 99$\pm$18 & 0.9 &  9.4$\times$10$^{-6}$ \\
\raisebox{1.5ex}[-1.5ex]{Cyg X-1$^c$}&    ISGRI  & 1.44$\pm$0.14 & 115$\pm$18  & 2.20$\pm$0.14   & 0.90$\pm$0.53 & 36$\pm$23 & 1.5  & - \\ 
 \hline
                                     & SPI   & 2.0$\pm$0.7 & 20$\pm$10  & 2.4$\pm$0.3 & 3.5$\pm$6.6 & 9.0$\pm$1.1 &0.8 & 3.5$\times$10$^{-3}$\\
\raisebox{1.5ex}[-1.5ex]{Cyg X-3$^c$}&    ISGRI  & 2.00$\pm$0.08 & 20  &  2.6$\pm$0.2 &  3.4$\pm$0.7 & 5.3$\pm$0.5 & 1.7 & - \\
\hline
  \end{tabular}
  \begin{list}{}{}
  \item[$^a$] Spectral parameters not used in the model for a particular object are marked as ``-''. 
  \item[$^b$] F-test probability that the fit improvement achieved by the inclusion of a hard tail in addition to a cutoff powerlaw was by chance,
                   i.e. that the hard tail is not real. Note that since ISGRI has lower sensitivity above 500~keV and none
                   above 700~keV, we can in most cases not test for the presence of the additional component but only check the compatibility
                   with the ISGRI data.
  \item[$^c$] Parameters of alternative fit models are given in Sect. \ref{secresults}. 
  \item[$^d$] Spectrum very probably flawed by source contamination from  IGR~J16358-472.
  \end{list}
\end{table*}

\begin{table*}
  \caption[]{Summary of the spectra obtained in this study. Part II: 
    Our measurements of the Crab flux with SPI and ISGRI assuming the spectral shape given
   in Table \ref{tab-summary1a} for energy redistribution.}
  \label{tab-crab}
  \begin{tabular}{|l|c|c|c|c|c|}
    \hline
    Instrument & \multicolumn{5}{c|}{Crab Nebula Flux $[10^{-3}$cm$^{-2}$s$^{-1}]$}\\
               & 25-50 keV & 50-100 keV & 100-200 keV & 200-600 keV & 600-1000 keV \\
\hline
    SPI        & 156.5$\pm$0.4 & 72.18$\pm$0.15 & 32.14$\pm$0.12 & 17.81$\pm$0.27 & 3.47$\pm$0.15 \\
    ISGRI      & 140.2$\pm$0.4 & 64.38$\pm$0.16 & 27.46$\pm$0.09 & 13.47$\pm$0.25 & - \\
\hline
    Ratio SPI/ISGRI & 1.116$\pm$0.004 & 1.121$\pm$0.004 & 1.170$\pm$0.006 & 1.322 $\pm$ 0.032 & -\\
\hline
  \end{tabular}
\end{table*}

\begin{table*}
  \caption[]{Summary of the spectra obtained in this study. Part III: Source fluxes $[$mCrab$]$.}
   \centering
  \label{tab-summary2}
  \setlength{\tabcolsep}{0.5mm}
  \renewcommand{\arraystretch}{1.5}
  {\tiny
  \begin{tabular}{|l|rrr|rrr|rrr|r|}
{\large Object Name}& \multicolumn{10}{c|}{\large Flux $[$mCrab$]$$^{a,b}$}\\
  & \multicolumn{3}{c|}{25-100 keV} & \multicolumn{3}{c|}{100-200 keV}& \multicolumn{3}{c|}{200-600 keV} & 600-1000 keV \\
      & \multicolumn{1}{c}{SPI} & \multicolumn{1}{c}{ISGRI} & \multicolumn{1}{c|}{SPI(B08)} 
      & \multicolumn{1}{c}{SPI} & \multicolumn{1}{c}{ISGRI} & \multicolumn{1}{c|}{SPI(B08)} 
      & \multicolumn{1}{c}{SPI} & \multicolumn{1}{c}{ISGRI} & \multicolumn{1}{c|}{SPI(B08)} & \multicolumn{1}{c|}{SPI} \\
\hline
Crab &1000.0$\pm$2.0&1000.0$\pm$2.0&1000.0$\pm$2.0&1000.0$\pm$3.4&1000.0$\pm$3.5&1000.0$\pm$3.7&1000.0$\pm$11.8&1000.0$\pm$18.6&1000.0$\pm$8.3&1000.0$\pm$43.2 
\\
Vela Pulsar &10.0$\pm$0.4&7.2$\pm$0.1&9.8$\pm$0.7&11.6$\pm$1.3&10.2$\pm$0.9&12.6$\pm$2.1&5.6$\pm$5.3&0.0$+$19.0&13.9$\pm$4.9&23.0$\pm$18.5
\\
NGC 4151 &35.4$\pm$2.4&32.7$\pm$0.7&31.3$\pm$2.1&43.8$\pm$7.4&34.6$\pm$2.5&22.9$\pm$6.8&28.2$\pm$36.4&0.0$+$132.5&46.9$\pm$14.1&101.4$\pm$111 
\\
NGC 4945 &18.7$\pm$1.1&16.4$\pm$0.3&18.3$\pm$1.1&21.4$\pm$3.7&21.5$\pm$1.2&14.3$\pm$4.0&0.5$\pm$27.1&0.0$+$53.2&20.2$\pm$8.3&5.8$\pm$53.8 
\\
Cen A &40.7$\pm$1.1&36.0$\pm$0.3&40.5$\pm$1.3&59.7$\pm$3.7&53.7$\pm$1.2&56.9$\pm$4.6&55.9$\pm$20.7&87.7$\pm$24.2&70.1$\pm$9.8&12.9$\pm$48.7 
\\
XTE J1550-564&36.1$\pm$0.6&41.6$\pm$0.3&23.5$\pm$0.7&55.2$\pm$1.7&70.2$\pm$0.9&30.2$\pm$2.5&48.8$\pm$8.4&51.5$\pm$12.1&18.8$\pm$5.5&6.7$\pm$24.4 
\\
4U 1630-47&36.5$\pm$1.6&1.8$\pm$0.2$^c$&30.5$\pm$1.3&19.7$\pm$6.0&0.8$\pm$0.3$^c$&21.7$\pm$2.4&2.2$\pm$23.3&19.9$\pm$13.2$^c$&21.4$\pm$5.3&0.0$+$164.2 
\\
Swift J1656.3-3302&3.7$\pm$2.2&-\ \ \  $^d$& 6.0$\pm$1.1&8.0$\pm$3.4&-\ \ \  $^d$&  5.0$\pm$1.9 &15.2$\pm$7.8&-\ \ \  $^d$&  17.6$\pm$4.0&24.6$\pm$12.9 
\\
OAO 1657-415&66.6$\pm$0.6&51.5$\pm$1.3&51.3$\pm$0.7&15.0$\pm$1.3&7.6$\pm$0.6&10.3$\pm$2.2&15.0$\pm$8.0&0.0$+$24.9&13.8$\pm$4.8&2.3$\pm$26.6 
\\
GX 339-4&34.1$\pm$0.5&11.5$\pm$0.2&38.4$\pm$1.4&32.6$\pm$1.4&12.1$\pm$0.5&35.6$\pm$2.4&30.1$\pm$13.8&0.0$+$30.3&30.0$\pm$5.2&66.1$\pm$27.4 
\\
4U 1700-377&171.7$\pm$0.8&162.3$\pm$3.6&146.3$\pm$1.4&36.0$\pm$1.0&40.9$\pm$1.6&35.5$\pm$2.2&13.6$\pm$8.1&23.4$\pm$17.2&6.4$\pm$4.8&0.0$+$62.7 
\\
IGR J17091-3624&5.2$\pm$0.8&6.1$\pm$0.1&8.9$\pm$0.6&7.2$\pm$5.0&6.3$\pm$0.3&5.6$\pm$2.0&18.5$\pm$8.8&2.7$\pm$13.4&9.8$\pm$4.4&8.7$\pm$19.6 
\\
GX 354-0&34.2$\pm$0.5&23.4$\pm$0.1&24.4$\pm$0.5&10.5$\pm$1.1&5.4$\pm$0.2&7.6$\pm$1.8&0.7$\pm$10.0&1.0$\pm$10.5&7.9$\pm$4.1&7.4$\pm$21.4 
\\
1E 1740.7-2942&52.8$\pm$0.9&32.7$\pm$0.1&51.1$\pm$1.2&43.0$\pm$2.3&35.0$\pm$0.3&48.3$\pm$1.6&18.5$\pm$10.9&22.1$\pm$6.1&39.2$\pm$3.6&7.8$\pm$34.9 
\\
IGR J17464-3213&13.3$\pm$0.6&15.1$\pm$0.1&15.6$\pm$0.7&6.9$\pm$1.1&9.9$\pm$0.2&11.2$\pm$1.6&2.2$\pm$4.4&6.8$\pm$6.3&9.9$\pm$3.6&41.6$\pm$23.1 
\\
GRS 1758-258&68.8$\pm$0.7&66.6$\pm$0.2&70.8$\pm$0.9&84.6$\pm$2.1&82.2$\pm$0.5&83.4$\pm$1.6&41.1$\pm$7.2&49.3$\pm$6.0&47.4$\pm$3.6&51.2$\pm$18.0 
\\
Ginga 1826-24&75.8$\pm$0.5&73.1$\pm$0.3&74.6$\pm$0.5&30.3$\pm$0.9&34.2$\pm$0.4&33.6$\pm$1.9&18.4$\pm$5.7&0.0$+$19.8&24.5$\pm$3.9&38.3$\pm$22.4
\\
GRS 1915+105&189.1$\pm$0.8&187.8$\pm$0.5&150.1$\pm$0.9&73.2$\pm$2.2&60.5$\pm$0.5&61.8$\pm$2.3&38.8$\pm$8.7&40.9$\pm$10.2&38.6$\pm$5.2&53.9$\pm$27.2 
\\
Cyg X-1&841.0$\pm$1.5&871.6$\pm$1.7&843.8$\pm$1.0&873.7$\pm$3.1&907.6$\pm$3.1&769.9$\pm$3.1&575.5$\pm$9.8&581.3$\pm$16.1&500.6$\pm$6.9&386.4$\pm$37.3
\\
Cyg X-3&117.7$\pm$0.6&111.7$\pm$0.6&101.2$\pm$0.8&34.9$\pm$1.4&25.3$\pm$0.5&25.0$\pm$2.9&26.4$\pm$7.6&21.0$\pm$21.8&14.8$\pm$6.2&52.6$\pm$31.0 
\\
Estim. Syst. Error$^e$      & -0.9 & -0.1 &  -\ \ \   & -0.1 & -0.4 &  -\ \ \  & -0.1 & -1.2 &  -\ \ \   & -0.1 \\
\hline
  \end{tabular}
  \begin{list}{}{}
  \item[$^a$] For the columns ``SPI'' and ``ISGRI'' fluxes are expressed as fractions of the Crab fluxes for the individual instrument given in Table \ref{tab-crab}.
            In the columns ``SPI (B08)'' we give the fluxes obtained by Bouchet et al. (\cite{bouchet08}) (B08) for the available energy ranges 
            (see B08 for their definition of the unit ``mCrab''). 
  \item[$^b$]  The statistical errors shown in this table do not take into account the uncertainty of the Crab flux
     on which the flux values are normalised.
  \item[$^c$] Spectrum very probably flawed by source contamination from  IGR~J16358-472; therefore not included in Fig. \ref{fig-agree-spi-isgri}.
  \item[$^d$] ISGRI spectral fit failed, see Sect. \ref{sec-SwiftJ}; therefore not included in Fig. \ref{fig-agree-spi-isgri}.
  \item[$^e$] See Sect. \ref{sec-control1}.
  \end{list}
}
\end{table*}

Tables \ref{tab-summary1a}, \ref{tab-summary1b}, \ref{tab-crab}, and \ref{tab-summary2} summarise our 
spectral measurements.
They provide a catalog of the time-averaged spectra of the brightest known soft gamma-ray sources.
They also give an impression of the variety of possible spectral shapes for different
source types,  especially for X-ray binaries (XRBs) which represent more than half of the sources.
Furthermore, the tables permit to compare the response of SPI and ISGRI. 
In particular Tables \ref{tab-summary1a} and \ref{tab-summary1b} permit to study
the agreement of the corresponding measurements in terms of spectral shape while
Table \ref{tab-summary2} can be used to compare the flux normalisation of the
two instruments in several energy bands. In the latter table we also give the
results from B08 for those energy ranges where they
are available.

\begin{figure}[t]
  \includegraphics[width=8cm]{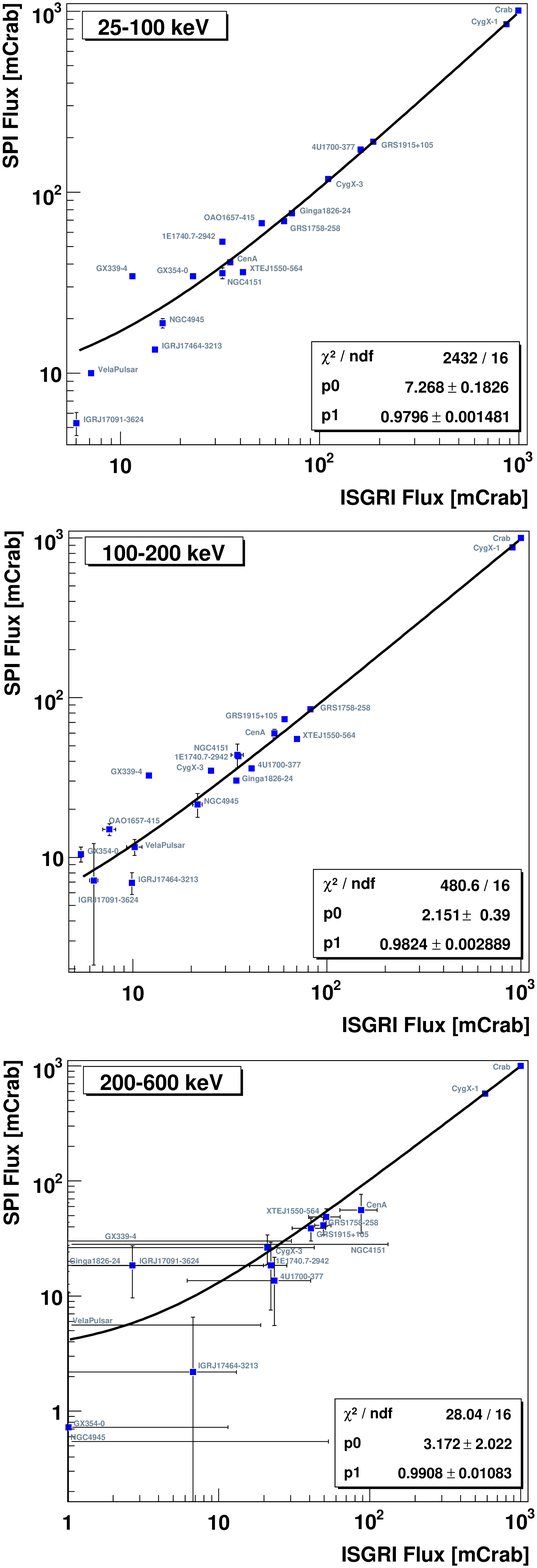}
  \caption{ Comparison of our flux values obtained from SPI with those from ISGRI for the
     energy bands {\bf (a)} (top) 25-100 keV, {\bf (b)} (middle) 100-200 keV, 
     {\bf (c)} (bottom) 200-600 keV. In each band, a linear function was fit to the points and fit
     parameters are shown in the figures. The flux values are also given in Table \ref{tab-summary2}.
     Only the statistical errors are shown. They take into account the uncertainty of the Crab flux. 
     See Sect. \ref{sec-agree-spi-isgri}.
  }
  \label{fig-agree-spi-isgri}
\end{figure}

\begin{figure}[t]
  \includegraphics[width=8cm]{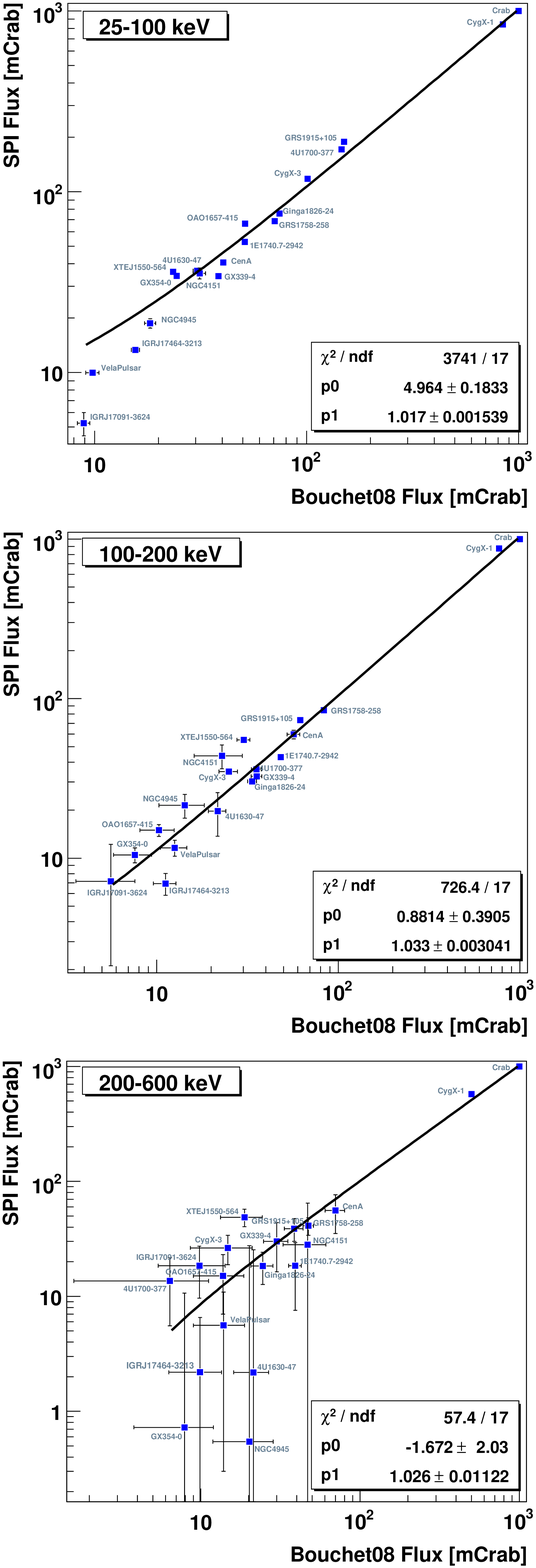}
  \caption{ Comparison of our flux values obtained from SPI with those by B08 (also obtained with SPI, largely using the same data
     but with a different analysis method) for the
     energy bands {\bf (a)} (top) 25-100 keV, {\bf (b)} (middle) 100-200 keV, 
     {\bf (c)} (bottom) 200-600 keV. In each band, a linear function was fit to the points and fit
     parameters are shown in the figures. The flux values are also given in Table \ref{tab-summary2}.
     Only the statistical errors are shown. They take into account the uncertainty of the Crab flux. 
     See Sect. \ref{sec-agree-spi-bouchet}.
  }
  \label{fig-agree-spi-bouchet}
\end{figure}

\subsection{Agreement of SPI and ISGRI spectra}
\label{sec-agree-spi-isgri}

The study of 20 bright sources with SPI and ISGRI using nearly identical datasets permits us
to carry out a quantitative comparison of the response of the two instruments and test the
quality of the spectra obtained with our new {\it spimodfit}-based analysis method.

The comparison of the absolute flux normalisation should be carried out using a steady source
such as the Crab. Our Crab flux measurements are given in Table \ref{tab-crab}.
While the flux ratio SPI/ISGRI is consistent with a constant value of about 1.12 up to 200~keV,
we observe a clear deviation at higher energies suggesting systematic differences in
flux calibration of additional 18~\% at 200 - 600~keV (see Sect. \ref{sec-Crab}).

If we correct for the differences in absolute flux calibration by normalising on the Crab flux,
we obtain the flux values in Table~\ref{tab-summary2}. 
Plotting the SPI values versus the ISGRI values from this table, we obtain Figs. \ref{fig-agree-spi-isgri}a, b, and c.
In all three energy ranges up to 600~keV, these plots show clearly a good agreement between the flux values
from the two instruments for all sources with fluxes higher than a few ten~mCrab. Only for the faintest sources
and only below 100 keV do we see obvious systematic deviations from the linear relationship. 
Here the SPI fluxes tend to be lower than
those from ISGRI. The origin of this is probably a combination
of two causes: (1) systematic differences in the handling of source variability by the analysis software, and (2)
systematic differences in the background determination.

\subsection{Agreement of our SPI measurements with those by B08}
\label{sec-agree-spi-bouchet}

To further verify our analysis method, we compare our SPI fluxes with those published by B08 who give
their flux values as fraction of their Crab flux. All values are given in Table \ref{tab-summary2}.
Plotting our values versus those from B08 (Figs. \ref{fig-agree-spi-bouchet}a, b, and c) we see that
the flux values agree well in all energy bands for sources brighter than 20~mCrab. Below this flux level,
our flux values seem to be systematically slightly lower. At least some of the differences can be attributed
to source variability as our dataset is not exactly the same as that used by B08. 

B08 defined the annihilation (positronium) emission between 300~keV and 511~keV as an individual component
in their source model and extracted its intensity as a result of their fit. Since this emission is extended 
(to first order a 8$^\circ$ FWHM Gaussian centered on the galactic centre) on a scale similar to the SPI field-of-view, 
it is in our analysis absorbed in the time-variable isotropic component of our background model which does not have
a separate annihilation component.
This is confirmed by the fact that our fluxes in the range 200~keV - 600~keV agree very well with 
those by B08 (see Fig.~\ref{fig-agree-spi-bouchet}), also for sources near the galactic centre.

\subsection{Emission above 200 keV}

We detect all of the 20 sources in our sample independently in the 25-100~keV and in the 100-200~keV band
both with SPI and with ISGRI. Above 200~keV we find evidence for emission of more than 3~$\sigma$ in 8 cases.
In the 600-1000~keV band, we find such evidence only for two sources.

XRBs constitute 70\% of our source sample.  
Table \ref{tab-summary1b} summarises our measurements of the spectra of these 14 XRBs.
They are not a random sample but the 14 brightest known X-ray binaries at 200~keV.
Four of them are HMXBs, the remaining are LMXBs. Ten of them contain black hole candidates (eight of which are classified
as microquasars), the remaining four contain neutron stars.

\subsection{Conclusions}

   We present spectra for the 20 brightest soft gamma-ray 
   sources averaged over timescales of years and give flux values in four energy bands between 25~keV and 1 MeV. 

   For SPI, these measurements were derived using a new tool ({\it spimodift} by Halloin \& Strong (\cite{spimodfit}))
   which can coherently treat data from the whole SPI mission by performing a single, large maximum likelihood fit.
  
   The agreement between our SPI and ISGRI measurements is good if we  correct for systematic 
   calibration differences by normalizing on the Crab spectrum.
   Our SPI flux measurements also agree well with those by Bouchet et al. (2008) (B08).
   Our Crab spectrum shows the also previously observed break near 100~keV where the spectral
   index softens from 2.1 to 2.2. 
  
   Up to 200~keV, all 20 sources in our sample
   are detected independently in two adjacent bands. At 200-600~keV we detect eight sources,
   and at 600-1000~keV we detect two sources.
   The spectra we find agree well with the results from previous publications.
   For six of the 14 XRBs in our sample (not counting the borderline case GRS~1758-258), 
   we find evidence for a hard powerlaw-component which becomes
   dominant above the cutoff energy of the thermal Comptonization component, i.e. above about hundred to
   a few hundred keV. For GX~339-4, GRS~1915+105, Cyg~X-1, and Cyg~X-3 such hard tails 
   have already  been noted by previous studies. For OAO~1657-415 and Ginga~1826-24, 
   our study provides the first weak indication of such emission. 

\begin{acknowledgements}
      We would like to thank R. Diehl, J. Greiner, and G. Sala (MPE, Garching), and E.~Jourdain and J.P.~Roques (CESR/CNRS, Toulouse)
      for useful discussions. DP was supported in part by the German Bundesministerium f\"{u}r 
      Bildung, Wissenschaft, Forschung und Technologie (BMBF/DLR) under contract No. FKZ 50 OG 0502.
\end{acknowledgements}


\begin{thebibliography}{}

  \bibitem[2004]{revofpartphys} Alvarez-Gaum\'{e}, L. et al. (eds.), 2004, ``Review of Particle Physics'',
      Phys. Lett. B, 592

  \bibitem[2007]{xspec} Arnaud, K., Dorman, B., \& Gordon, C. 2007, ``Xspec - an X-ray fitting package, User's Guide for version 12.3.1'', HEASARC, NASA/GSFC publications

  \bibitem[2006]{audley06} Audley, M.D., Nagase, F., Mitsuda, K., et al. 2006, MNRAS, 367, 1147

  \bibitem[2008]{barnstedt} Barnstedt, J., Staubert, R., Santangelo, A., et al. 2008, A\&{}A, 486, 293

  \bibitem[2005]{beckmann05} Beckmann, V., Shrader, C. R., Gehrels, N., et al. 2005, ApJ, 634, 939

  \bibitem[2006]{bosch06} Bosch-Ramon, V., Romero, G.E., Paredes, J.M., et al. 2006, A\&{}A, 457, 1011

  \bibitem[1991]{bouchet91} Bouchet, L., Mandrou, P., Roques, J.P., et al. 1991, ApJ, 383, L45

  \bibitem[2005]{bouchet05} Bouchet, L., Roques, J.P., Mandrou, P., et al. 2005, ApJ, 635, 1115

  \bibitem[2008]{bouchet08} Bouchet, L., Jourdain, E., Roques, J.P., et al. 2008, ApJ, 679, 1315


  \bibitem[2006]{cadollebel06} Cadolle Bel, M., Sizun, P., Goldwurm, A., et al. 2006, A\&{}A, 446, 591

  \bibitem[2006a]{capitanio06a} Capitanio, F., Bazzano, A., Ubertini, P., et al. 2006, Adv. in Space Research, 38, 2816

  \bibitem[2006]{capitanio06} Capitanio, F., Bazzano, A., Ubertini, P., et al. 2006, ApJ, 643, 376

  \bibitem[2008]{chaty08} Chaty, S., Rahoui, F., Foellmi, C., et al. 2008, A\&{}A, 484, 783

  \bibitem[2005]{corbel05} Corbel, S., Kaaret, P., Fender, R.P., et al. 2005, ApJ, 632, 504 

  \bibitem[2003]{thierry03} Courvoisier, T. J.-L., Walter, R., Beckmann, V., et al. 2003, A\&{}A, 411, L53

  \bibitem[2008]{isgri-lightcurves} Courvoisier, T. J.-L., et al. 2008, INTEGRAL Source Results Version 2, 
      available from the INTEGRAL Science Data Centre website http://isdc.unige.ch/

  \bibitem[2005]{delsanto05} Del Santo, M., Bazzano, A., \& Zdziarski, A.A., et al. 2005, A\&{}A, 433, 613

  \bibitem[1995]{dermer95} Dermer, C.D. \& Gehrels, N. 1995, ApJ, 447, 103

  \bibitem[1996]{done96} Done, C., Madjeski, G.M., \& Smith, D.A., 1996, ApJ, 463, L63

  \bibitem[2004]{done04} Done, C., Wardzinski, G. \& Gierlinski, M. 2004, MNRAS, 349, 393

  \bibitem[2006]{falanga06} Falanga, M., G\"{o}tz, D., Goldoni, P., et al. 2006, A\&{}A, 458, 21

  \bibitem[2004]{fender04} Fender, R. \& Belloni, T. 2004, Annu. Rev. Astron. Astrophys., 42, 317

  \bibitem[2005]{filippova05} Filippova, E.V., Tsygankov, S.S., Lutovinov, A.A., \& Sunyaev, R.A. 
    2005, Astronomy Letters, 30, 824

  \bibitem[2003]{gies03} Gies, D.R., Bolton, C.T., Thomson, J.R., et al. 2003, ApJ, 583, 424 

  \bibitem[2003]{goldwurm} Goldwurm, A., David, P., Foschini, L., et al. 2003, A\&{}A, 411, L223

  \bibitem[1995]{grabelsky95} Grabelsky, D.A., Maltz, S.M., Purcell, W.R., et al. 1995, ApJ 441, 800

  \bibitem[1996]{greiner96} Greiner, J., Morgan, E.H., \& Remillard, R.A. 1996, ApJL, 473, L107

  \bibitem[2007]{spimodfit} Halloin, H. \& Strong, A., 2007, ``spimodfit user manual'', OSA documentation, 
    available from the INTEGRAL Science Data Centre website http://isdc.unige.ch/

  \bibitem[2008]{hjalmarsdotter08} Hjalmarsdotter, L., Zdziarski, A.A., Larsson, S., et al. 2008, MNRAS, 384, 278

  \bibitem[2008]{itoh08} Itoh, T., Done, C., Makishima, K. et al. 2008, Publ. Astron. Soc. Japan, 60, S251

  \bibitem[1993]{johnson93} Johnson, W.N., Kurfess, J.D., Purcell, W.R., et al. 1993, A\&{}A Suppl. Ser., 97, 21

  \bibitem[2005]{joinet05} Joinet, A., Jourdain, E., Malzac, J., et al. 2005, ApJ, 629, 1008

  \bibitem[2007]{joinet07} Joinet, A., Jourdain, E., Malzac, J., et al. 2007, ApJ, 657, 400

  \bibitem[2008]{jourdain08} Jourdain, E. 2008, Gotz, D., Westergaard, N.J., et al.2008, , arXiv:0810.0646, POS (integral08) 144 

  \bibitem[2007]{kaptein00} Kaptein, R.G., in't Zand, J.J.M., Kuulkers, E., et al. 2000, A\&{}A, 358, L71

  \bibitem[2005]{kirsch05} Kirsch, M.G., Briel, U.G., Burrows, D., et al. 2005, in Sigmund, O.H.W., et al. (eds.) Proc. SPIE, 5898, 589803-1

  \bibitem[2007]{krivonos07} Krivonos, R., Revnivtsev, M., Churazov, E., et al. 2007, A\&{}A, 463, 957

  \bibitem[2007]{kubota07} Kubota, A., Dotani, T., Cottam, J., et al. 2007, Publ. Astron. Soc. Japan, 59, S185


  \bibitem[1994]{maisack94} Maisack, M., Kendziorra, E., Pan, H.C., et al. 1994, A\&{}A, 283, 841

  \bibitem[2006]{malzac} Malzac, J., Petrucci, P. O., Jourdain, E., et al. 2006, A\&{}A, 448, 1125

  \bibitem[2006]{mangano06} Mangano, V., Bocchino, F., Cusumano, G., et al. 2006, Advances in Space Research, 37, 1984

  \bibitem[2007]{marti07} Marti, J., Perez-Ramirez, D., Luque-Escamilla, P., et al. 2007, Astrophys. Space Sci., 309, 309

  \bibitem[2008]{masetti08} Masetti, N., Mason, E., Landi, R., et al. 2008, A\&{}A, 480, 712

  \bibitem[1989]{mcconnell89} McConnell, M.L., Forrest, D.J., Owens, A., et al. 1989, ApJ, 343, 317

  \bibitem[2000]{mcconnell00} McConnell, M.L., Ryan, J.M., Collmar, W., et al. 2000, ApJ, 543, 928


  \bibitem[2006]{okajima06} Okajima, T., Tueller, J., Markwardt, C., et al. 2006, ATel 799

  \bibitem[2006]{pottschmidt06} Pottschmidt, K., Chernyakova, M., Zdziarski, A.A., et al. 2006, A\&{}A, 452, 285

  \bibitem[2003a]{revnivtsev03a} Revnivtsev, M.G., Chernyakova, M., Capitanio, F., et al. 2003, ATel 132, 1

  \bibitem[2003b]{revnivtsev03} Revnivtsev, M.G., Gilfanov, M., Churazov, E., \& Sunyaev, R. 2003, ATel, 150, 1

  \bibitem[2004]{revnivtsev04} Revnivtsev, M.G., Sunyaev, R.A., Gilfanov, M.R., et al.
      2004, Astronomy Letters, 30, 527

  \bibitem[2006]{revnivtsev06} Revnivtsev, M.G., Sazonov, S., Gilfanov, M., Churazov, E., \& Sunyaev, R. 2006, A\&{}A, 452, 169

  \bibitem[2008a]{rodriguez08} Rodriguez, J., Hannikainen, D.C., Shaw, S.E., et al. 2008, ApJ, 675, 1436

  \bibitem[2008b]{rodriguez08b} Rodriguez, J., Shaw, S.E., Hannikainen, D.C., et al. 2008, ApJ, 675, 1449

  \bibitem[2003]{roques} Roques, J.P., Schanne, S., von Kienlin, A., et al. 2003, A\&{}A, 411, L91

  \bibitem[2006]{rothschild06} Rothschild, R.E., Wilms, J., Tomsick, J., et al. 2006, ApJ, 641, 801

  \bibitem[2007]{russell07} Russell, D.M., Fender, R.P., Gallo, E., \& Kaiser, C.R. 2007, MNRAS, 376, 1341

  \bibitem[2000]{schoenfelder00} Sch\"{o}nfelder, V., Bennett, K., Blom, J.J., et al. 2000,
      A\&{}A Suppl. Ser., 143, 145

  \bibitem[2006]{shaposhnikov06} Shaposhnikov, N. \& Titarchuk, L. 2006, ApJ, 643, 1098

  \bibitem[1998]{smith98} Smith, D.A. 1998, IAU Circ., 7008, 1
 
  \bibitem[1998]{steinle98} Steinle, H., Bennett, K., Bloemen, H., et al. 1998, A\&{}A, 330, 97

  \bibitem[1996a]{strickman96a} Strickman, M., Skibo, J., Purcell, W., Barret, D., \& Motch, C. 1996,  A\&{}A Suppl. Ser., 120, 217 

  \bibitem[1996b]{strickman96} Strickman, M., de Jager, O., \& Harding, A. 1996, A\&{}A Suppl. Ser., 120, 449

  \bibitem[2005]{sturner05} Sturner, S.J. \& Shrader, C.R. 2005, ApJ, 625, 923

  \bibitem[2007]{szostek07} Szostek, A. \& Zdziarski, A.A. 2007, MNRAS, 375, 793

  \bibitem[2008]{szostek08} Szostek, A. \& Zdziarski, A.A. 2008, MNRAS, 386, 593 

  \bibitem[2005]{thompson05} Thompson, T.W.J., Rothschild, R.E., Tomsick, J.A., \& Marshall, H.L. 2005, ApJ, 634, 1261

  \bibitem[2005]{tomsick05} Tomsick, J.A., Corbel, S., Goldwurm, A., \& Kaaret, P. 2005, ApJ, 630, 413

  \bibitem[2008]{tomsick08} Tomsick, J.A., Kalemci, E., Kaaret, P., et al. 2008, ApJ, 680, 593

  \bibitem[1994]{titarchuk94} Titarchuk, L. 1994, ApJ, 434, 570

  \bibitem[2003]{ubertini03} Ubertini, P., Lebrun, F., Di Cocco, G. et al. 2003, A\&{}A 411, L131-L139

  \bibitem[2000]{ulrich00} Ulrich, M.H. 2000, A\&{}A Rev, 10, 135

  \bibitem[2005]{vandermeer05} van der Meer, A., Kaper, L., di Salvo, T., et al. 2005, A\&{}A, 432, 999

  \bibitem[2003]{vedrenne03} Vedrenne, G., Roques, J.-P., Sch\"{o}nfelder, V., et al. 2003,
      A\&{}A, 411, L63

  \bibitem[2007]{wu07} Wu, Y.-X., Liu, C.-Z.,\& Li, T.-P. 2007, ApJ, 660, 1386

  \bibitem[1997]{zhang97} Zhang, S.N., Cui, W., Harmon, B.A., et al. 1997, ApJ, 477, L95

  \bibitem[2005]{ziolkowski05} Ziolkowski, J. 2005, MNRAS, 358, 851



\end{thebibliography}
\end{document}